\documentclass[a4paper,USenglish,autoref,pdfa]{lipics-v2021}
\pdfoutput=1
\usepackage[noadjust]{cite} 
\usepackage{ebproof/ebproof}
\ebproofset{
rule code={\hbox{\hspace*{-6pt}\rule{\dimexpr\hsize+12pt}{0.4pt}}},
label separation=1em,
}
\newcommand\topalignimage[1]{\raisebox{-\dimexpr\totalheight-.5\ht\strutbox}{#1}}
\newcommand\centeralignimage[1]{\raisebox{-.5\dimexpr\totalheight-.5\ht\strutbox}{#1}}
\newcommand\variable[1]{\textsf{\textit{#1}}}
\newcommand\varell{\variable{\large$\ell$}}
\newcommand\literal[1]{\texttt{#1}}
\newcommand\tss[1]{\textsubscript{#1}}
\newcommand\ellsub[1]{\ensuremath{\varell\!\variable{\tss{#1}}}}
\newcommand\plty[1]{\ensuremath{p_{#1}}}

\DeclareUnicodeCharacter{03B5}{$\varepsilon$}
\DeclareUnicodeCharacter{2200}{$\forall$}
\DeclareUnicodeCharacter{2264}{$\leq$}
\DeclareUnicodeCharacter{2265}{$\geq$}

\title{Automatic Layout of Railroad Diagrams}

\author{Shardul Chiplunkar}{School of Computer and Communication Sciences, EPFL, Lausanne, Switzerland \and \url{https://etaoin-shrdlu.xyz/}}{shardul.chiplunkar@epfl.ch}{https://orcid.org/0000-0002-0803-2133}{}
\author{Cl\'{e}ment Pit-Claudel}{School of Computer and Communication Sciences, EPFL, Lausanne, Switzerland \and \url{https://pit-claudel.fr/clement/}}{clement.pit-claudel@epfl.ch}{https://orcid.org/0000-0002-1900-3901}{}

\authorrunning{S. Chiplunkar and C. Pit-Claudel}
\Copyright{Shardul Chiplunkar and Cl\'{e}ment Pit-Claudel}



\supplementdetails[cite={dagstuhl-artifact-25789},subcategory={Source Code}]{Software}{https://github.com/epfl-systemf/librrd/releases/tag/ecoop2026-artifact}

\acknowledgements{The authors thank Viktor Kunčak, Emir Demirović, and Hugo Herbelin for many insightful discussions, and the anonymous peer reviewers for their helpful feedback. Both authors are members of the EPFL-Inria REMPAR associate team.}

\funding{This work was funded in part by the Swiss National Science Foundation (SNSF) under grant no.~10003649. This material is based upon work supported by the Air Force Research Laboratory (AFRL) and Defense Advanced Research Projects Agencies (DARPA) under Contract No.~FA8750-24-C-B044. Any opinions, findings, and conclusions or recommendations expressed in this material are those of the author(s) and do not necessarily reflect the views of the AFRL and DARPA.}

\newcommand{\TODO}[1]{}

\nolinenumbers

\EventEditors{Robbert Krebbers and Alexandra Silva}
\EventNoEds{2}
\EventLongTitle{40th European Conference on Object-Oriented Programming (ECOOP 2026)}
\EventShortTitle{ECOOP 2026}
\EventAcronym{ECOOP}
\EventYear{2026}
\EventDate{June 29--July 3, 2026}
\EventLocation{Brussels, Belgium}
\SeriesVolume{372}
\ArticleNo{2}

\supplementdetails[linktext={\\https://doi.org/10.4230/DARTS.12.1.13}, subcategory={ECOOP 2026 Artifact Evaluation approved artifact}]{Software}{https://doi.org/10.4230/DARTS.12.1.13}
\begin{document}

\def\Snospace~{\S{}}
\renewcommand\sectionautorefname{\Snospace}
\renewcommand\subsectionautorefname{\Snospace}
\renewcommand\subsubsectionautorefname{\Snospace}

\maketitle

\begin{abstract}
Railroad diagrams (also called ``syntax diagrams'') are a common, intuitive visualization of grammars, but limited tooling and a lack of formal attention to their layout mostly confines them to hand-drawn documentation.
We present the first formal treatment of railroad diagram layout along with a principled, practical implementation.
We characterize the problem as compiling a \emph{diagram language} (specifying conceptual components and how they connect and compose) to a \emph{layout language} (specifying basic graphical shapes and their sizes and positions).
We then implement a compiler that performs \emph{line wrapping} to meet a target width, as well as vertical \emph{alignment} and horizontal \emph{justification} per user-specified policies.
We frame line wrapping as optimization, where we describe principled dimensions of optimality and implement corresponding heuristics.
For front-end evaluation, we show that our diagram language is well-suited for common applications by describing how regular expressions and Backus-Naur form can be compiled to it.
For back-end evaluation, we argue that our compiler is practical by comparing its output to diagrams laid out by hand and by other tools.
\end{abstract}

\maketitle

\begin{figure}[p]
\centering

\begin{subfigure}[t]{0.58\linewidth}
\includegraphics[width=\linewidth]{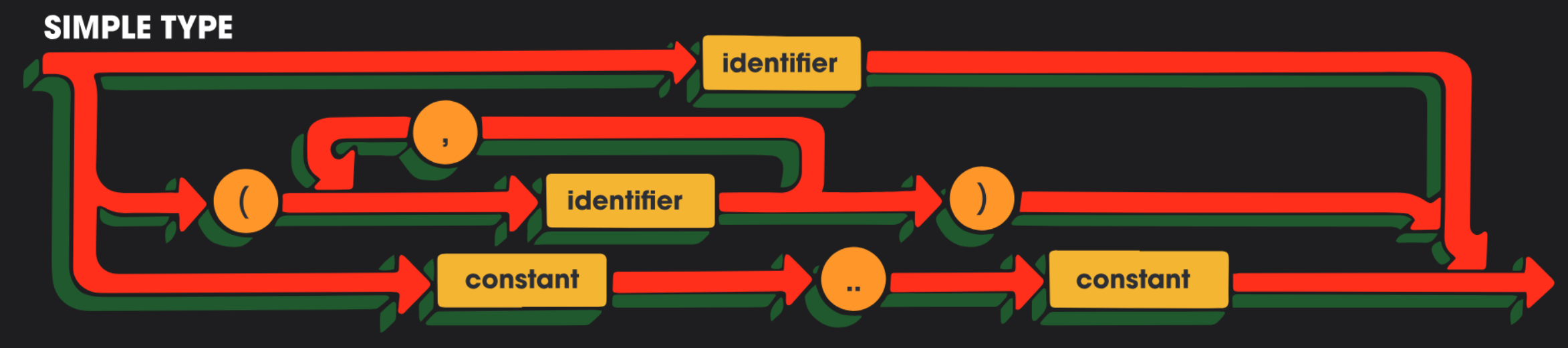}
\caption{Apple Pascal syntax chart, 1979~\cite{KamifujiRaskinApplePascalSyntax1979}.}
\label{fig:wild:apple}
\end{subfigure}\hfill\begin{subfigure}[t]{0.39\linewidth}
\includegraphics[width=\linewidth]{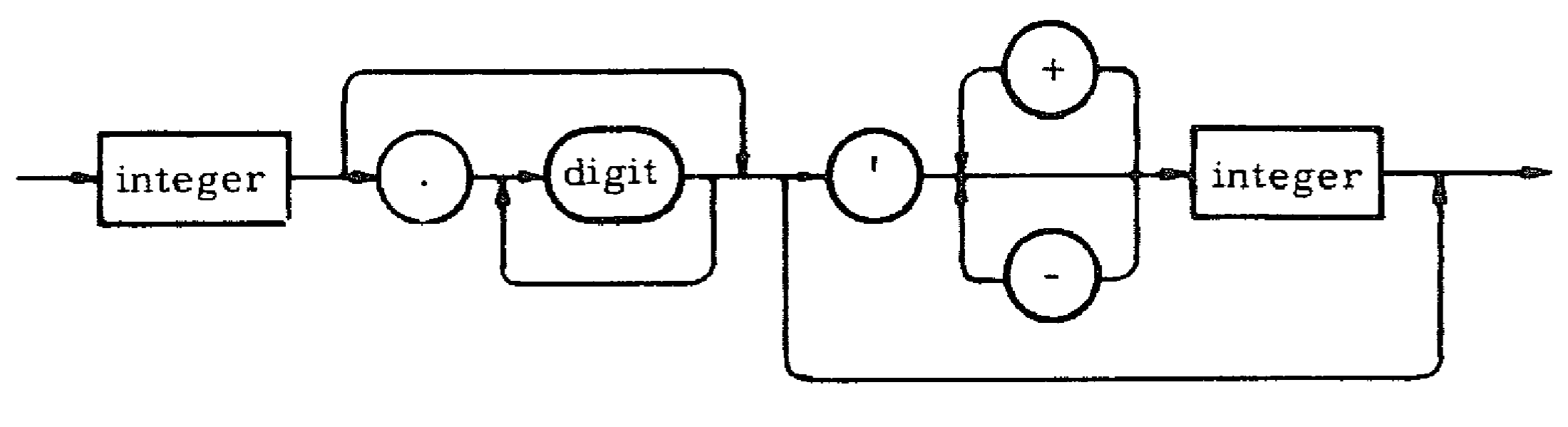}
\caption{Pascal manual, 1970~\cite{WirthProgrammingLanguagePascal1970}.}
\label{fig:wild:wirth}
\end{subfigure}\\[\baselineskip]

\begin{subfigure}[b]{0.4\linewidth}
\includegraphics[width=\linewidth]{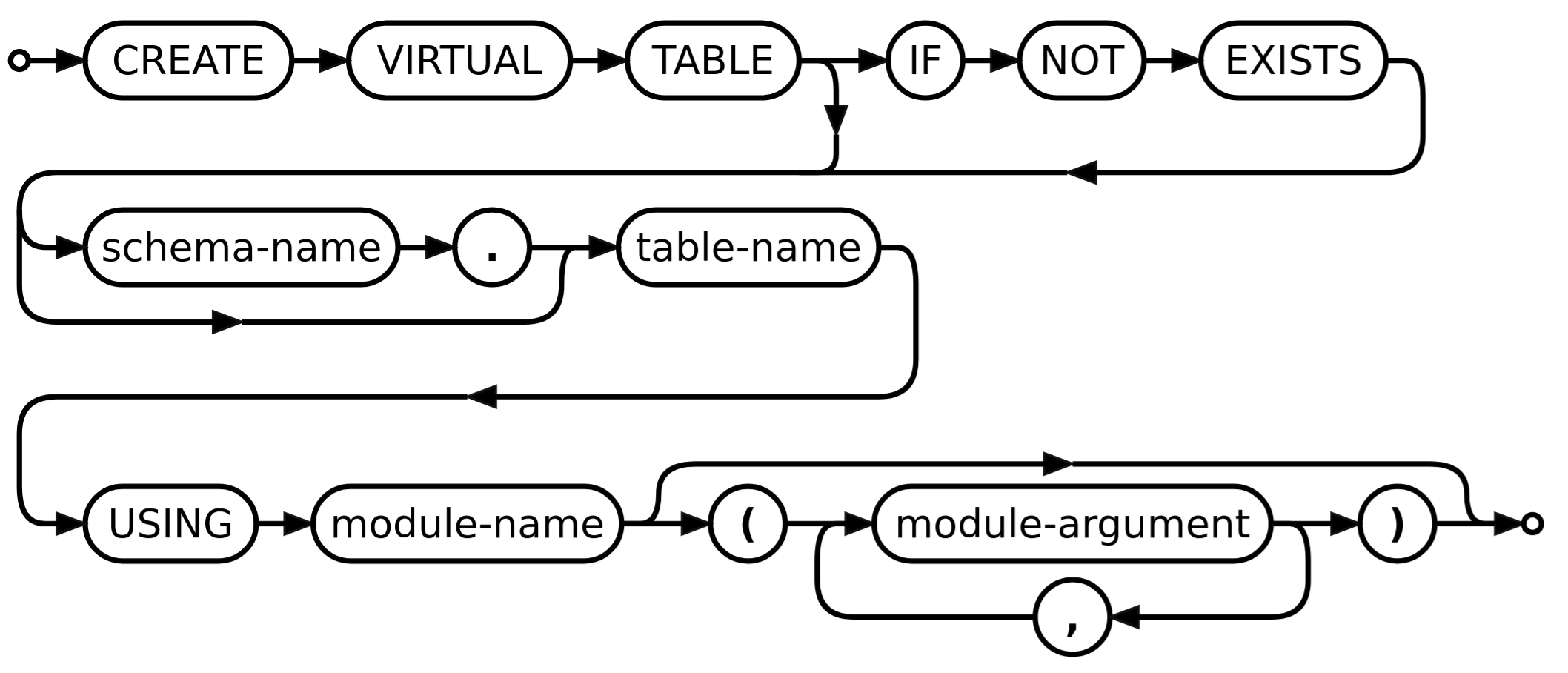}
\vspace{1.5mm}
\caption{SQLite documentation, 2024~\cite{SQLitecontributorsSyntaxDiagramsSQLite2024}.}
\label{fig:wild:sqlite}
\end{subfigure}\hfill\begin{subfigure}[b]{0.58\linewidth}
\includegraphics[width=\linewidth]{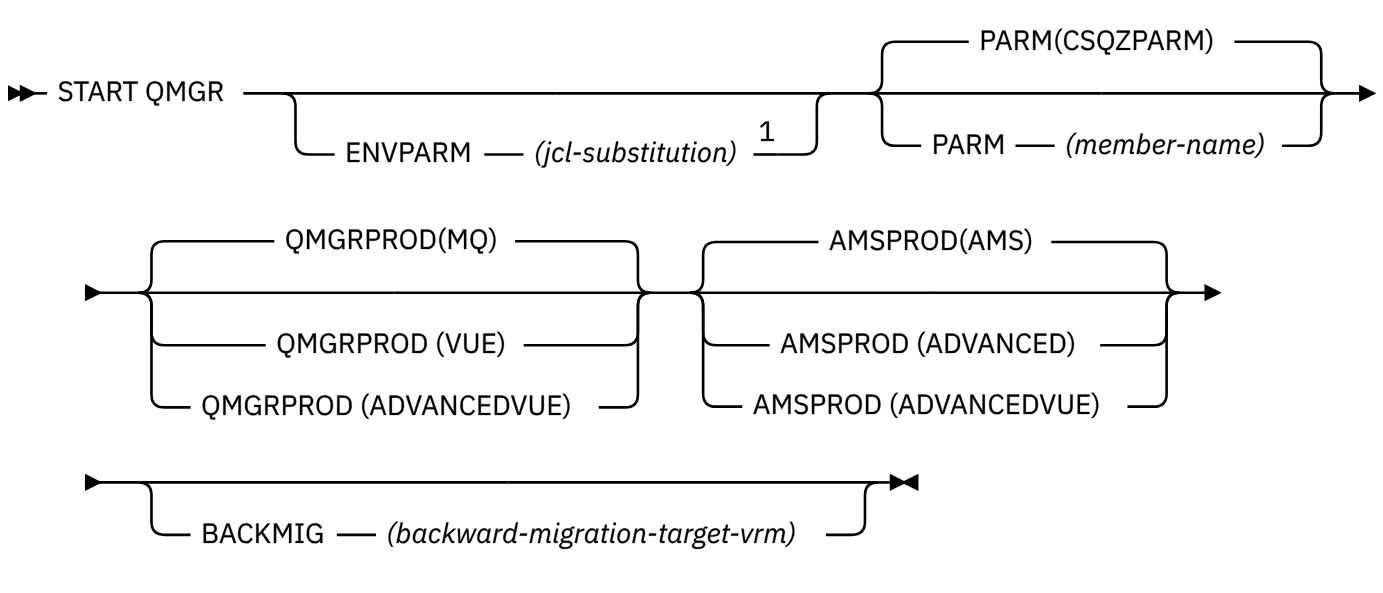}
\caption{IBM MQ documentation, 2025~\cite{IBMCorporationIBMMQReference2025}.}
\label{fig:wild:ibm}
\end{subfigure}

\caption{Examples of railroad diagrams taken from published works.
Others pictured later include~\cite{EcmaJSONDataInterchange2017} in \autoref{fig:wild:ecma} and~\cite{CrockfordJSONObjectSyntax2001} in \autoref{fig:counterexamples:semantic-nesting}.
All were laid out by hand.}
\label{fig:wild}
\end{figure}

\begin{figure}[p]
\begin{minipage}[c]{0.37\linewidth}
\centering
\includegraphics[height=3.31\baselineskip]{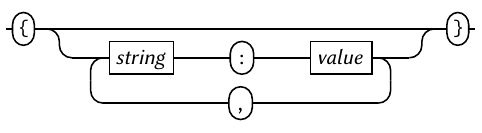}

\vspace{1.2mm}
\includegraphics[height=3.31\baselineskip]{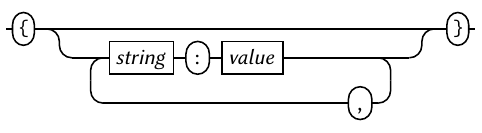}

\vspace{2mm}
\includegraphics[height=2.95\baselineskip]{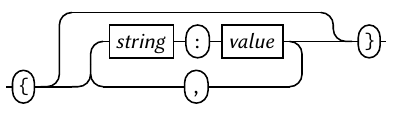}

\vspace{4mm}
\includegraphics[width=0.95\linewidth]{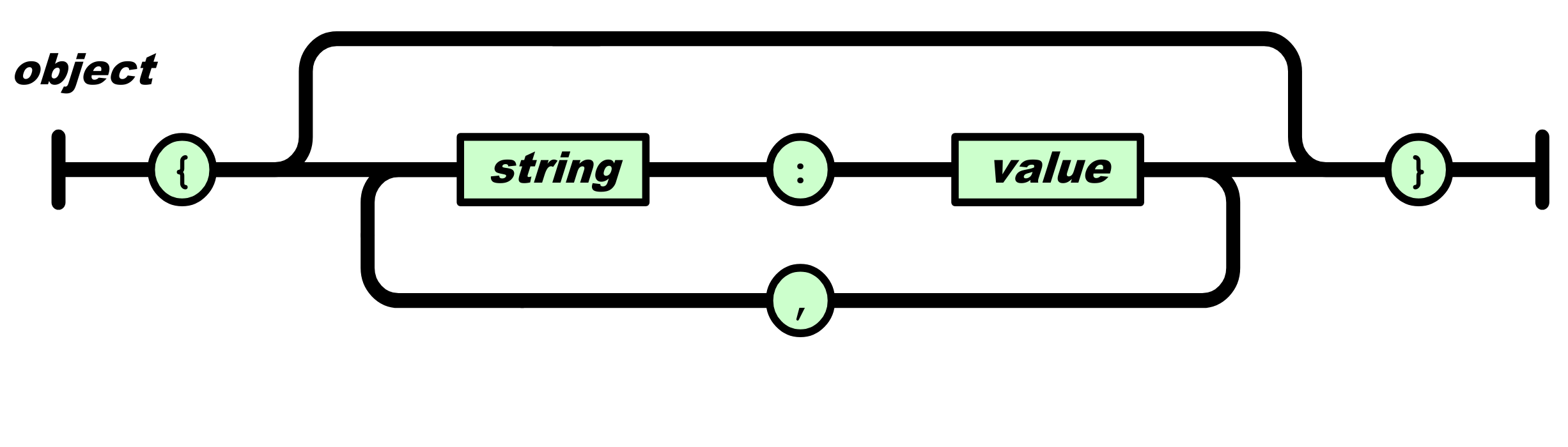}
\vspace{-2mm}

\caption{A railroad diagram has many possible layouts. (Bottommost from~\cite{EcmaJSONDataInterchange2017}; the rest from our tool.)}
\label{fig:wild:ecma}
\end{minipage}\hfill\begin{minipage}[c]{0.59\linewidth}
\centering
\topalignimage{\includegraphics[scale=0.7]{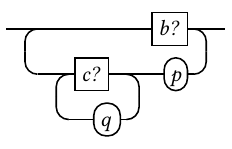}}
\hspace{1mm}
\raisebox{-2em}{$\equiv$}
\hspace{1mm}
\topalignimage{\includegraphics[scale=0.7]{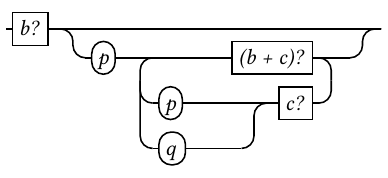}}
\caption{Railroad-style layout is versatile.
Above, we illustrate a theorem that~\cite[\S3.4]{KozenKleeneAlgebraTests1997} states as algebra: $(bp(cq)^\ast\overline{c})^\ast\overline{b} \equiv bp((b + c)(cq + \overline{c}p))^\ast\overline{b + c} + \overline{b}$, where $b, c$ are Boolean \emph{tests} and $p, q$ are arbitrary KAT terms.
In our diagrammatic notation, a boxed $b?$ selects the lower path after it iff $b$ is true.}
\label{fig:kat}
\end{minipage}
\end{figure}

\begin{figure}[p]
\centering

\begin{tabular}{>{\quad\hspace{0.5em}}c<{\qquad\qquad\hspace{0.5em}}c<{\qquad\quad\hspace{1em}}c}
align & wrap & justify \\[2pt]
\includegraphics{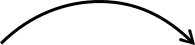} & \includegraphics{figures/arrow.png} & \includegraphics{figures/arrow.png} \\[2pt]
\end{tabular}

\centeralignimage{\includegraphics[scale=0.72]{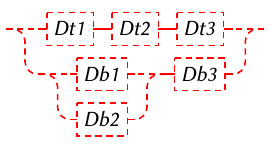}}
\hspace{0.3em}
\centeralignimage{\includegraphics[scale=0.72]{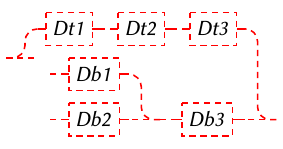}}
\hspace{0.3em}
\centeralignimage{\includegraphics[scale=0.72]{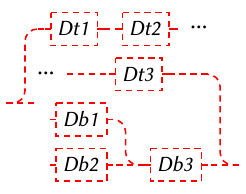}}
\hspace{0.8em}
\centeralignimage{\includegraphics[scale=0.72]{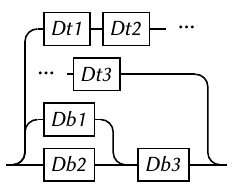}}
\caption{Our algorithm for compiling a diagram to a layout.
\autoref{sec:glossary} explains the terminology and graphical conventions, and
\autoref{sec:threesteps} describes the process in detail.}
\label{fig:compilation}
\end{figure}

\section{Introduction}
\label{sec:introduction}

Railroad diagrams, also called \emph{syntax diagrams}, are a class of schematic visualizations of formal grammars.
Common uses include documenting general-purpose and domain-specific languages (\autoref{fig:wild}) and teaching and illustrating regular languages~\cite{BellGilbertLearningRecursionSyntax1974,AvalloneRegexper2020,HinzeSelfcertifyingRailroadDiagrams2019,ChengParreauxSimpleRecipeWriting2026}.
While some sources explain how to read the diagrams, many do not -- including \mbox{\cite{WirthProgrammingLanguagePascal1970}}, despite being the first (to our knowledge) published instance.\footnote{See~\cite{SteeleItsTimeNew2017} for further evidence of this being the first usage.} This suggests that railroad diagrams have been considered intuitive or common enough to need no explanation since at least~1970.

There are many decisions to make when drawing a railroad diagram.
Some concern its semantic content, such as whether to represent recursion with a visual loop or a named self-reference.
Some are purely stylistic, such as the choice of colors.
The rest are about \emph{layout}:
the visual arrangement (positions, sizes) of the basic shapes that constitute the diagram.
(See \autoref{fig:wild:ecma}.)
These layout decisions are what we focus on in the present work.

Automating the layout of structured objects for display is a well-studied problem.
Examples include wrapping paragraphs of text~\cite{KnuthPlassBreakingParagraphsLines1981,KenningaOptimalLineBreak2003b,PeelsEtAlDocumentArchitectureText1985,BirdTransformationalProgrammingParagraph1986,deMoorGibbonsBridgingAlgorithmGap1999},
pretty-printing code~\cite{PorncharoenwaseEtAlPrettyExpressivePrinter2023,BernardyPrettyNotGreedy2017,HughesDesignPrettyprintingLibrary1995,OppenPrettyprinting1980,PughSinofskyNewLanguageindependentPrettyprinting1987,SwierstraChitilLinearBoundedFunctional2009,WadlerPrettierPrinter2003},
arranging content on web- and physical pages~\cite{W3CCSSFlexibleBox2018,CiancariniEtAlHighqualityPaginationPublishing2011,JacobsEtAlAdaptiveGridbasedDocument2003,JohariEtAlAutomaticYellowPagesPagination1997},
graph-drawing~\cite{TamassiaHandbookGraphDrawing2013,yWorksYFilesHTML2025,GansnerNorthOpenGraphVisualization2000}, and
visualizing data and statistics~\cite{WilkinsonGrammarGraphics2005,WickhamGgplot2ElegantGraphics2016,HeerBostockDeclarativeLanguageDesign2010,SatyanarayanEtAlVegaLiteGrammarInteractive2017,BostockEtAlD3DatadrivenDocuments2011,Hunter:2007}.
While automation makes layout less tedious, error-prone, and hard-to-update than doing it by hand, it loses some of the freedom to choose between valid alternative layouts.
Useful automation must identify, and often let the user specify, which choices are reasonable or desired:
it must formally characterize the layout problem.
Indeed, formal study underlies practical tools in almost all the research cited above.
For instance, pretty-printers from the seminal~\cite{OppenPrettyprinting1980} to the modern~\cite{PorncharoenwaseEtAlPrettyExpressivePrinter2023} have had formalism at their core, and text wrapping has even been cast as a model problem for formalism-driven programming~\cite{BirdTransformationalProgrammingParagraph1986,deMoorGibbonsBridgingAlgorithmGap1999}.

However, a class of layout problems that so far has not enjoyed the benefits of automation are those that exhibit a hierarchically nested reading order, like structured control flow graphs and electrical and hardware circuit diagrams, which we call \emph{1.5-dimensional} problems.
Our terminology stems from the observation that some layouts have 1-dimensional structure with a global, linear reading order, such as for text or code (first left-to-right, then top-to-bottom, in English), whereas others use available space more freely in 2~dimensions without a strict reading order, such as node-link graphs.
1.5-dimensional layouts are intermediate.
Like 1-dimensional layouts, they can be wrapped to fit a target size, but subcomponent layouts can split or merge while still independently participating in the reading order.
Meanwhile, they could be seen as stylized 2-dimensional layouts of directed graphs, but conventional graph layouts are much less rigid and structured, whether done by hand or by standard algorithms.
The fact that many 1.5-dimensional layouts are still created by hand or with \emph{ad hoc} adjustments to general-purpose graph layout tools\footnote{E.g., a 2021 dissertation about visualizing control flow graphs~\cite{DevkotaVisualizingControlFlow2021} states that visualizers ``commonly use general layered layout algorithms such as [GraphViz's] Dot'', even though that makes ``formulating the constraints for high-level requirements such as preserving program structures [\ldots{}] challenging''.
}
is evidence of the lack of a satisfactory automated alternative -- and of the potential value of a principled, formal approach.

\subsection{Contributions}
\label{sec:contributions}

In this paper, we study the layout of railroad diagrams as an emblematic instance of a 1.5-dimensional problem, whose challenges -- nested wrapping, alignment and justification, wide hand-drawn stylistic variation -- are representative of the class.
More precisely, we formalize and develop an algorithm for \emph{railroad layout}:
\begin{enumerate}
\item
We define a diagram language and a layout language (\autoref{sec:diagramlang}, \autoref{sec:layoutlang}) that lets us precisely specify railroad layout as compilation from the former to the latter (\autoref{sec:langrel}).
\item
We show that our formalism is realistic despite being simple: it captures most of the variation in hand-drawn railroad diagrams and guides the design of a practical algorithm.
\item
We design a three-step compilation algorithm, illustrated in \autoref{fig:compilation}, consisting of vertical \emph{alignment}, then \emph{wrapping} to meet a target width, and finally horizontal \emph{justification}.
\item
We frame wrapping as parametric optimization and develop practical heuristics~(\autoref{sec:wrapping}).
\item
We implement the compiler that we describe.
\end{enumerate}

First, we present our formalism~(\autoref{sec:formal}) and then our algorithm~(\autoref{sec:threesteps}), although the two evolved together in reality.
Then, we argue that our formalism is realistic by showing that common grammar notations are easy to translate to our diagram language~(\autoref{sec:frontends}) and that most manual layouts can be expressed in our layout language~(\autoref{sec:evalmanual}).
We further argue that our algorithm compares favorably to manual layout, going beyond existing tools~(\autoref{sec:evalauto}) while being performant enough for interactive use~(\autoref{sec:performance}).
Lastly, we discuss how prior work in layout and diagramming relates to ours~(\autoref{sec:related}).

\begin{figure}[t]
\centering\vspace{-3mm}

\begin{subfigure}[t]{0.4\linewidth}
\centering
\includegraphics[width=0.93\linewidth]{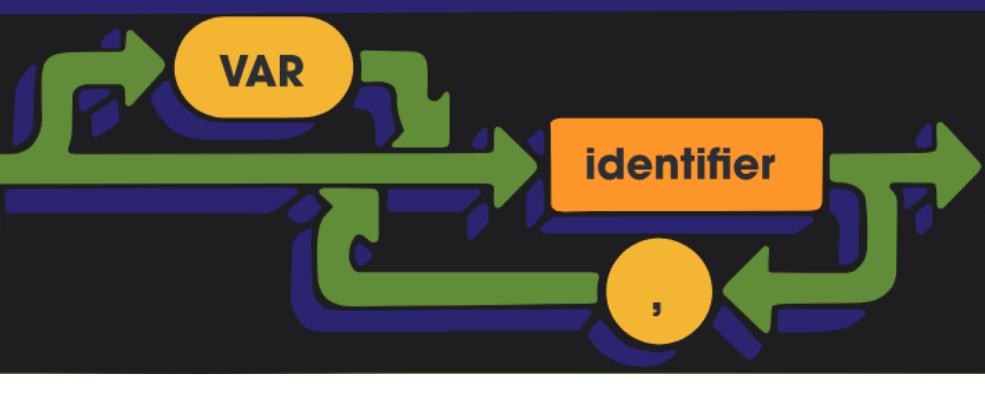}
\caption{Excerpt from Apple Pascal syntax chart~\cite{KamifujiRaskinApplePascalSyntax1979}.
This structure is ill-nested, and hence unrepresentable as a diagram, because the optional \literal{"VAR"} ends before the repeating \literal{[identifier]} starts.
(It would still be ill-nested if the \literal{","}-adjoint edges were reversed.)}
\label{fig:counterexamples:ill-nested}
\end{subfigure}
\hfill
\begin{subfigure}[t]{0.56\linewidth}
\centering
\includegraphics[width=0.77\linewidth]{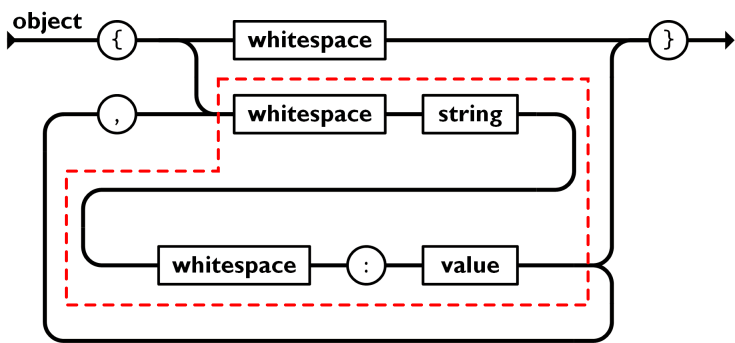}
\caption{Syntax of a JSON object~\cite{CrockfordJSONObjectSyntax2001}, with dashed red annotation.
The visual nesting of the indicated sublayout does not reflect the nesting of the corresponding subdiagram, as no rectangle can bound the layout of the five-token sequence \literal{([whitespace] [string] … [value])} without including the \literal{","} or the vertex after it.
Note that this \emph{diagram} is well-nested, but this particular \emph{layout} isn't.}
\label{fig:counterexamples:semantic-nesting}
\end{subfigure}\\[\baselineskip]

\begin{subfigure}[t]{0.68\linewidth}
\centering
\includegraphics[width=0.93\linewidth]{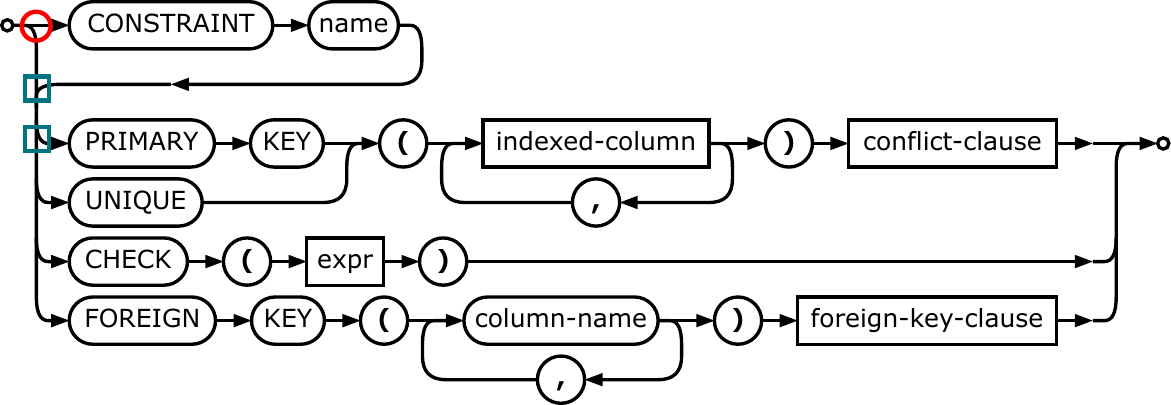}
\caption{Syntax of a SQLite \literal{table-constraint} term~\cite{SQLitecontributorsSyntaxDiagramsSQLite2024}, with red and blue annotations.
The first subdiagram of this sequence is an optional \literal{("CONSTRAINT" "name")} sequence, i.e.~a stack of those two tokens and the empty sequence~ε.
Hence, the vertical line between the red circle and the first blue square is a layout of the ε, violating the single semantic axis property.
(In contrast, the vertical line between the two blue squares is just ``visual syntax'' for the sequence and not a sublayout.)}
\label{fig:counterexamples:semantic-vertical}
\end{subfigure}
\hfill
\begin{subfigure}[t]{0.28\linewidth}
\centering
\includegraphics[width=0.6\linewidth]{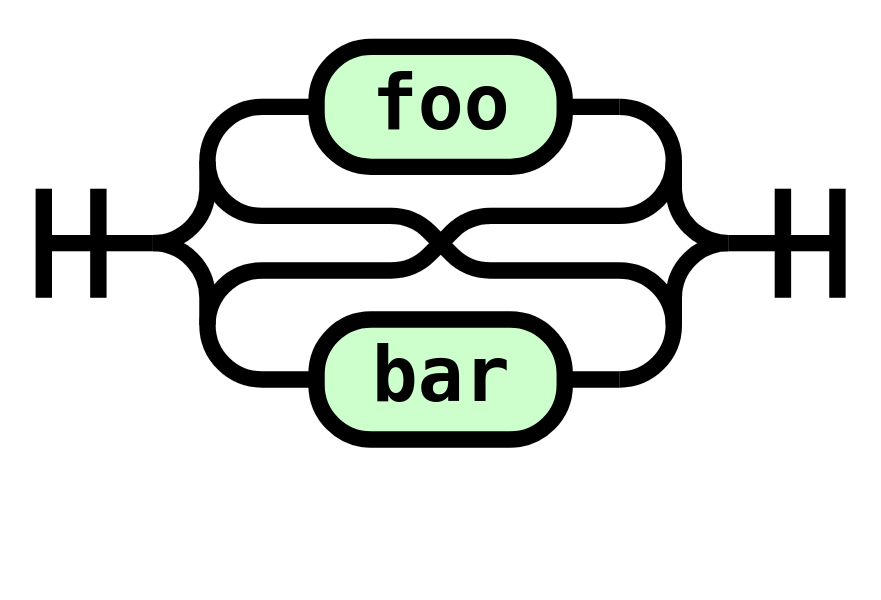}
\caption{An \literal{AlternatingSequence} construct from~\cite{AtkinsRailroaddiagramGenerator2024}, signifying the language of one or more instances of \literal{"foo"} or \literal{"bar"}, starting with either and alternating between them.}
\label{fig:counterexamples:unseen}
\end{subfigure}

\caption{Railroad diagrams outside our scope. We show our closest approximations in \autoref{fig:approx-counter}.}
\label{fig:counterexamples}
\end{figure}

The novelty of our solution lies in its practical, principled approach to automatic railroad layout, notably for wrapping.
When used as static documentation, railroad diagrams are often laid out and updated by hand (e.g.\ Figures~\ref{fig:wild} and~\ref{fig:counterexamples} except~\ref{fig:counterexamples:unseen}).
Manual layout is tedious and hard to update~\cite{BrazVisualSyntaxDiagrams1990}, especially when reflowing a wrapped layout or maintaining differently-wrapped layouts for different media.
Worse, it is prone to semantic errors and stylistic inconsistencies (e.g.\ Figures~\ref{fig:sqlite-inconsistency:both} and~\ref{fig:sqlite-inconsistency}) \cite{BrazVisualSyntaxDiagrams1990}.
Beyond addressing these problems with automation,
our tool can also reproduce many existing manual layouts and produce comparable alternatives for the rest, making it a practical replacement.
Further, it is unique in its principled approach aiming for convenient layout ``idioms'' and ``cost-benefit balance''~\cite{ChiplunkarPit-ClaudelDiagrammaticNotationsInteractive2023}, with exploratory, interactive theorem-proving and programming environments in mind.
For instance, few other tools perform automatic wrapping, and none let the user control it.

Formalism has long served as a tool for clarity of thought and computational expression in layout research, and we follow suit.
We hope that formal foundations will aid further study of railroad layout beyond our work and of its applications beyond syntax.
As an example, an extension of regular algebra gives Kleene Algebra with Tests (KAT), a powerful formalism for equational reasoning about programs and control flow~\cite{KozenKleeneAlgebraTests1997};
a corresponding slight extension to railroad diagrams can illustrate KAT terms
(e.g.~\autoref{fig:kat}).
Recent work on the formal semantics of railroad diagrams and similar diagrammatic calculi~\cite{HinzeSelfcertifyingRailroadDiagrams2019,PiedeleuZanasiFiniteAxiomatisationFinitestate2023,AlurEtAlRobustTheorySeries2023} as well as formally verified web layout~\cite{PanchekhaTorlakAutomatedReasoningWeb2016,PanchekhaEtAlModularVerificationWeb2019} and pretty-printing~\cite{PorncharoenwaseEtAlPrettyExpressivePrinter2023} suggests several directions for future work.
More broadly, structured layout remains an open problem for ubiquitous diagrammatic notations, and we hope railroad-style layout may play a part in a unifying solution.

\subsection{Glossary}
\label{sec:glossary}

\begin{description}
\item[diagram]
A conceptual specification in terms of components and how they connect and compose.
It has no height, width, or other visual properties, but it has structure.
\item[layout]
A graphical specification in terms of basic shapes and their sizes and positions.
It has visual properties like height and width, but also retains structure from which the diagram can be recovered.
It is the result of \emph{laying out} a diagram, possibly under constraints like width.
\autoref{fig:wild:ecma} illustrates how a diagram can have many different layouts.
\item[rendering]
A concrete realization of a layout, in a format that can be directly displayed (like a bitmap) or executed for display (like canvas drawing instructions).

A diagram cannot be directly rendered, but must first be laid out.
Yet, to have a visual reference for raw or partially laid-out diagrams,
we use layouts rendered with dashed red lines, sometimes partly built by hand, e.g.\ to depict ill-formed intermediate states in \autoref{fig:compilation}.
Proper layouts are in solid black and are all generated by our tool.

\item[1.5-dimensional layout problem]
When layouts exhibit a hierarchically nested reading order.
See the last paragraph before~\autoref{sec:contributions}.
Our definition aligns with previous, more limited uses of the term, discussed further in~\autoref{sec:related}.

\item[bounding box]
The smallest axes-aligned rectangle that completely encloses a shape.

\item[flexbox]
The CSS Flexible Box Layout Module~\cite{W3CCSSFlexibleBox2018}, which inspires some of our terminology but is not general enough for railroad layout.
We will return to it in~\autoref{sec:threesteps} and~\autoref{sec:related}.
\end{description}

\section{A formal account of railroad diagrams and their layout}
\label{sec:formal}

There is a lot of variation in what are called ``railroad diagrams'' by their creators, because the term has never been formally defined before, and because they are often hand-drawn.
Much of it can be ascribed to \ae{}sthetic choices that will naturally vary from author to author.
Yet, even after abstracting a unifying visual schema or grammar, some structural variation remains.
We choose to exclude a part of it from our scope to achieve succinct definitions that nonetheless capture a large portion of the variation seen in the wild, which we quantify in~\autoref{sec:evalmanual} and~\autoref{sec:evalauto}.
Our three exclusions are illustrated in \autoref{fig:counterexamples} and explained below.

First, we only consider \emph{well-nested} diagrams:
diagrams composed of \textit{n}-ary sequences of subdiagrams in the same direction, binary stacks of subdiagrams (potentially in opposite directions), and atomic tokens.
This almost describes \emph{two-terminal series-parallel (SP) graphs}, except that they are usually defined to be either undirected or acyclic~\cite{BrandstadtEtAlAlgebraicCompositionsRecursive1999,AlurEtAlRobustTheorySeries2023}, whereas we allow cycles.
(We further discuss SP graphs in~\autoref{sec:related} and justify why stacks are binary in~\autoref{sec:diagramlang}.)
This criterion naturally extends to layouts: the well-nesting of a diagram must be reflected in the visual nesting of its layout.
Specifically, we only consider layouts where the \emph{bounding box} of each sublayout (corresponding to the decomposition above) does not overlap with those of unrelated sublayouts.
The bounding boxes then have the same nesting structure as the diagram.
Figures~\ref{fig:counterexamples:ill-nested} and~\ref{fig:counterexamples:semantic-nesting} are examples of ill-nested diagrams and layouts.

Next, we require layouts to have a \emph{single semantic axis}:
all sublayouts (corresponding to the diagram decomposition) must be laid out horizontally.
\autoref{fig:counterexamples:semantic-vertical} is an example violation.

Lastly, we exclude constructs we have not found to be in common use, even if some railroad diagram tools can produce them.
\autoref{fig:counterexamples:unseen} is an example from a popular library.

(One additional condition is not essential but simplifies our presentation:
the backward component of a loop must not be above the forward component, as in \autoref{fig:wild:apple}.
Including such loops would be straightforward but would make our formalism longer and more confusing.)

\subsection{The diagram language}
\label{sec:diagramlang}

\begin{figure}[t]
\centering\vspace{-4mm}
\begin{subfigure}[t]{0.41\linewidth}
\centering\setlength{\tabcolsep}{3mm}
\begin{tabular}{cc}
\literal{"\variable{label}"}
& \literal{[\variable{label}]} \\[-1.5mm]
\topalignimage{\includegraphics[scale=0.85]{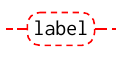}}
& \topalignimage{\includegraphics[scale=0.85]{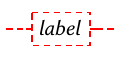}}
\\[8.6mm]
\literal{()} or ε
& \literal{(\variable{D1} \variable{D2} \variable{…})} \\[-1.5mm]
\topalignimage{\includegraphics[scale=0.85]{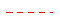}}
& \hspace{-2mm}\topalignimage{\includegraphics[scale=0.85]{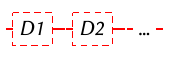}}
\\[9mm]
\literal{(+ \variable{Dtop} \variable{Dbot})}
& \literal{(- \variable{Dtop} \variable{Dbot})} \\[-1.5mm]
\topalignimage{\includegraphics[scale=0.85]{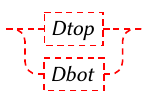}}
& \topalignimage{\includegraphics[scale=0.85]{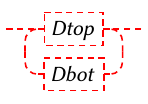}}
\end{tabular}
\bigskip\smallskip
\caption{Diagram language constructors and the special case ε.}
\label{fig:diagramlang}
\end{subfigure}\hfill\begin{subfigure}[t]{0.54\linewidth}
\centering
\setlength{\tabcolsep}{8mm}
\begin{tabular}{c}
\literal{(+ (+ \variable{D1} \variable{D2}) \variable{D3})} \\[0.7mm]
\centeralignimage{\includegraphics[scale=0.8]{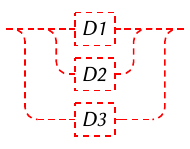}}
\hspace{1mm}\textrightarrow\hspace{1mm}
\centeralignimage{\includegraphics[scale=0.8]{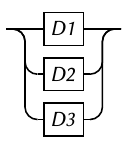}}
\\[15mm]
\literal{(+ (- \variable{D1} \variable{D2}) \variable{D3})} \\[0.7mm]
\centeralignimage{\includegraphics[scale=0.8]{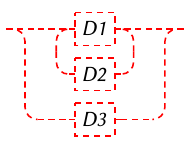}}
\hspace{1mm}\textrightarrow\hspace{1mm}
\centeralignimage{\includegraphics[scale=0.8]{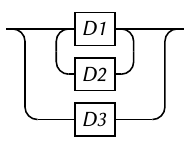}}
\end{tabular}
\caption{Unlike the positive stack, it would be invalid to collapse the sides of the negative stack, as the edges below \variable{D1} would not have a consistent direction.}
\label{fig:collapsing}
\end{subfigure}
\caption{The design of our diagram language.}
\end{figure}

We define our diagram language inductively in \autoref{def:diagramlang}.
The first two cases are terminal and nonterminal \emph{tokens} parameterized by a string \emph{label}.
The next case is a \emph{sequence} with zero or more subterms.
The last case is a \emph{stack} with exactly two subterms, parameterized by a \emph{polarity}, \mbox{\literal{+} or \literal{-}}.
An empty sequence \literal{()} is also denoted ε.
For the rest of this paper, a ``diagram'' is a term in this language.
The constructors are illustrated in \autoref{fig:diagramlang}.

Although laid-out stacks often appear to have more than two subdiagrams (e.g.\ the three-way stack in \autoref{fig:wild:apple}), we find that having stacks be only binary is the simplest model that accounts exactly for all possible layouts.
We use the same constructor for positive and negative stacks because both result in a layout with
(i)~vertically stacked sublayouts with brackets on the sides,
(ii)~no wrapping opportunities beyond what the sublayouts present, and
(iii)~possible \emph{collapsing} with a containing stack.
However, they differ in when such collapsing is possible, such as in \autoref{fig:collapsing}.
We thus need layout well-formedness rules to capture such differences, and in~\autoref{sec:layoutwf}, we state a clean set of such rules while treating all stack layouts as binary.
Consequently, to simplify the relation between diagrams and layouts, we treat (diagram-level) stacks as binary, too.
(The association order of nested binary positive stacks does not affect the rendering, unlike negative stacks (see~\autoref{sec:frontends}).)

In contrast to stacks, the association order of subdiagrams in a sequence has no bearing on the layout, so sequences are $n$-ary without ado.
Formally, the \emph{canonical form} of a diagram is the result of repeatedly splicing nested sequences, a trivially confluent rewriting:
any sequence \mbox{\literal{(\variable{D\tss{1}} \variable{…)}}} with a subdiagram \variable{D\tss{i}} which is itself a sequence \mbox{\literal{(\variable{D\tss{i\tss{1}}} \variable{…})}} is rewritten to \mbox{\literal{(\variable{D\tss{1}} \variable{…} \variable{D\tss{i\tss{1}}} \variable{…} \variable{…})}}, replacing \variable{D\tss{i}} with its subdiagrams.
We call two diagrams \emph{equivalent} if they have the same canonical form.

\subsection{The layout language}
\label{sec:layoutlang}

\begin{figure}[p]
\vspace{-2mm} \begin{minipage}[b]{0.48\linewidth}
\raggedright
\begin{tabular}[t]{r>{\sffamily\itshape}r>{\sffamily}r>{\ttfamily}l}
diagram
& d & := & "\variable{lbl}" \textsf{|} [\variable{lbl}] \\
&&  | & (\variable{d…}) \\
&&  | & (\variable{pol} \variable{d} \variable{d})
\end{tabular}
\caption{The diagram language.}
\label{def:diagramlang}

\bigskip\medskip
\begin{tabular}[b]{r<{\hspace{-2pt}}>{\sffamily\itshape}r>{\sffamily}r>{\ttfamily}l}
direction
& dir   & := & ltr \textsf{|} rtl \\
width
& w     & := & \textsf{\textbf{real} ≥ 0} \\
label, marker
& lbl, mk & := & \textsf{\textbf{string}} \\
terminal flag
& tm?    & := & \textsf{\textbf{boolean}} \\
polarity
& pol   & := & + \textsf{|} - \\
row number
& r     & := & \textsf{\textbf{integer} > 0} \\
proportion
& p     & := & \textsf{0 ≤ \textbf{real} ≤ 1}
\end{tabular}
\caption{Shared definitions.}
\label{def:commonlang}
\end{minipage}\hfill\begin{minipage}[b]{0.46\linewidth}
\centering
\begin{tabular}[t]{>{\sffamily}r>{\ttfamily}l}
& \hspace{-7.5mm}\textrm{layout \varell} \\
:= & (rail \variable{dir} \variable{w}) \\
| & (space \variable{dir}) \\
| & (station \variable{dir} \variable{lbl} \variable{tm?}) \\
| & (hconcat \variable{dir} \varell{} \varell\variable{…}) \\
| & (vconcat-inline \variable{dir} \variable{ts} \variable{ts} \variable{mk} \\
&\quad \varell{} \varell{} \varell\variable{…}) \\
| & (vconcat-block \variable{dir} \variable{ts} \variable{ts} \variable{pol} \varell{} \varell{}) \\[1mm]
& \hspace{-7.5mm}\textrm{tip specification \variable{ts}} \\
:= & vertical \\
| & (logical \variable{r}) \\
| & (physical \variable{p})
\end{tabular}
\caption{The layout language.}
\label{def:layoutlang}
\end{minipage}
\end{figure}

\begin{figure}[p]
\centering \renewcommand\arraystretch{2.3}\begin{minipage}[t]{0.28\linewidth}
\begin{tabular}[t]{m{22mm}c}
\literal{(space ltr)} & \\
\literal{(rail ltr 20)} & \centeralignimage{\includegraphics[scale=1.1]{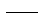}} \\
\literal{(station ltr} \par\quad \literal{\variable{label} \#t)} & \centeralignimage{\includegraphics[scale=1.1]{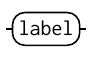}} \\
\literal{(station ltr} \par\quad \literal{\variable{label} \#f)} & \centeralignimage{\includegraphics[scale=1.1]{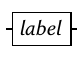}} \\[4mm]
\mbox{\literal{(hconcat ltr \variable{L1} \variable{L2})}} & \hspace{-6mm}\raisebox{-1.2\dimexpr\totalheight}{\includegraphics[scale=1.1]{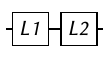}} \\[16mm]
\literal{(hconcat rtl} \mbox{\quad\literal{\variable{L1} (rail rtl 40) \variable{L2})}} & \hspace{-17mm}\raisebox{-1.5\dimexpr\totalheight}{\includegraphics[scale=1.1]{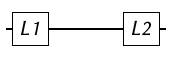}}\vspace{-8mm}
\end{tabular}
\end{minipage}\hfill\begin{minipage}[t]{0.61\linewidth}
\renewcommand\arraystretch{2.6}
\begin{tabular}[t]{m{55mm}>{\hspace{-5mm}}c}
\literal{(vconcat-inline ltr} \par\quad
\literal{(logical 1) (logical 1)} \literal{"…"} \par\qquad\quad \variable{L1} \par\qquad\quad
\variable{L2} \par\qquad\quad
\literal{(hconcat ltr} \par\qquad\qquad
\literal{\variable{L3} (rail ltr 20)))}
& \centeralignimage{\includegraphics[scale=1.1]{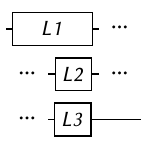}} \\
\literal{(vconcat-block ltr} \par\quad
\literal{(logical 1) (logical 1)} \par\quad
\literal{+ \variable{L1} \variable{L2})} & \centeralignimage{\includegraphics[scale=1.1]{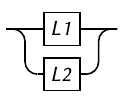}} \\
\literal{(vconcat-block ltr} \par\quad
\literal{(logical 1) (logical 1)} \par\quad
\literal{- \variable{L1} (hconcat rtl \variable{L3} \variable{L2}))} & \centeralignimage{\includegraphics[scale=1.1]{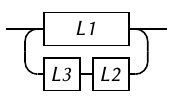}} \\
\verb|(vconcat-block ltr|\par
\verb|  (physical 0.2) (logical 3)|\par
\verb|  + |\variable{L1}\verb| (vconcat-block ltr|\par
\verb|         vertical vertical|\par
\verb|         + (vconcat-block ltr|\par
\verb|             vertical vertical|\par
\verb|             + |\variable{L2}\verb| |\variable{L3}\verb|) |\variable{L4}\verb|))| & \centeralignimage{\includegraphics[scale=1.1]{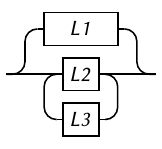}} \\
\verb|(vconcat-block ltr|\par
\verb|  (logical 2) (physical 1)|\par
\verb|  + |\variable{L1}\verb| (vconcat-block ltr|\par
\verb|         (logical 1) (logical 1)|\par
\verb|         - |\variable{L2}\verb| |\variable{L3}\verb|))| & \centeralignimage{\includegraphics[scale=1.1]{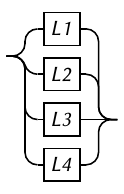}}
\end{tabular}
\end{minipage}
\vspace{-3mm}
\caption{Layout constructors, and the effect of tip specifications.}
\label{fig:layoutlang}
\end{figure}

We define our layout language inductively in \autoref{def:layoutlang} and illustrate it in \autoref{fig:layoutlang}.
The constructors, and their parameters beyond \emph{direction} (left-to-right or right-to-left), are:
\begin{itemize}
\item (atomic constructors) rails, with a nonnegative real \emph{width}; spaces; and stations, with a string \emph{label} and a \emph{terminal flag} indicating whether it is a terminal;
\item a horizontal concatenation of one or more subterms;
\item an inline vertical concatenation (``inline VC'') of two or more subterms, with a string \emph{marker} and left and right \emph{tip specifications}; and
\item a block vertical concatenation (``block VC'') of exactly two subterms, with a \emph{polarity} (positive or negative) and left and right \emph{tip specifications}.
\end{itemize}
A \emph{tip} is where a layout starts or ends, and where a containing layout can connect to enter or exit it.
Naturally, each layout has a tip on either side.
Either tip of a VC can be specified as:
\begin{itemize}
\item \literal{vertical}, for collapsing with a containing stack;
\item a \emph{logical row number}, a positive integer, to align with an inner sublayout after collapse; or
\item a \emph{physical proportion}, a real between 0 and 1, interpolating between the highest~(0) and lowest~(1) possible tips, which are not necessarily aligned with logical rows.
\end{itemize}
Any other layout has \literal{(logical 1)} tips on both sides by default.
Lastly, we define the \emph{start} and \emph{end sides} of an inline VC as left and right if the direction is left-to-right, else \emph{vice versa}.

\subsection{Layout well-formedness}
\label{sec:layoutwf}

Unlike diagrams, not all layouts in the language above are \emph{well-formed}.
For instance, a right-to-left horizontal concatenation must be constructed with right-to-left sublayouts, in visual (i.e., reverse) order (e.g.\ bottom left of \autoref{fig:layoutlang}).
In informal terms, our definition of well-formedness (to follow) aims to avoid ``nonsense'' layouts, where, say, a collapsed edge has no consistent direction, or sublayouts are visually disconnected.
Moreover, a well-formed layout leaves no ambiguity or context-dependence in its rendering.

\begin{figure}[tb]
\centering\vspace*{-4mm}
\begin{minipage}[b]{0.23\linewidth}
\centering \includegraphics[width=0.38\linewidth]{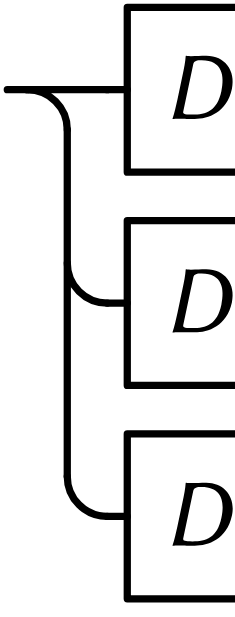}\hspace{0.9em}\includegraphics[width=0.38\linewidth]{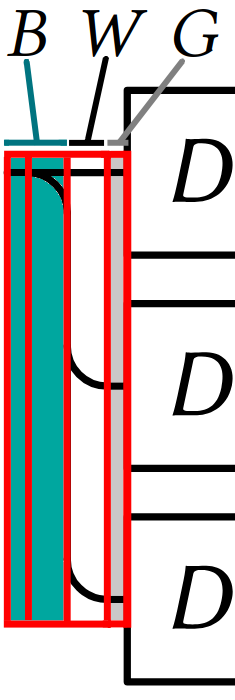}\hspace*{0.5em}
  \caption{The left side of a positive block VC with a \literal{(logical 1)} tip, without and with width annotations.
The blue~$B = 3S$ width is part of the VC.
The white~$W = 2S$ width comes from spaces at the left end of each sublayout; the bracket of the VC ``reaches into'' them.
The gray~$G = S$ width is intrinsic to the stations.}
\label{fig:bracket-widths}
\end{minipage}\hspace{0.04\linewidth}\begin{minipage}[b]{0.19\linewidth}
\centering
\includegraphics[width=0.87\linewidth]{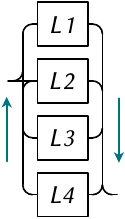}
\caption{A negative block VC with its left tip at its first logical row and its right tip at its last.
If the left tip were any higher, the edge between \variable{L1} and \variable{L2} would not have a consistent direction.}
\label{fig:negative-rows}
\end{minipage}\hspace{0.04\linewidth}\begin{minipage}[b]{0.5\linewidth}
\centering
\includegraphics[width=0.9\linewidth]{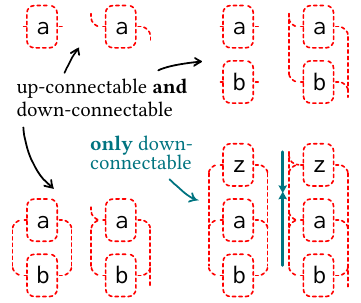}
\caption{Each pair of drawings shows a layout by itself and in the hypothetical context of a block~VC connecting from above (on the left) or below (right).
The top left represents a station~\literal{"a"} surrounded by (invisible) spaces; the top right, \mbox{\literal{(+ "a" "b")}}; the bottom left, \mbox{\literal{(- "a" "b")}}; and the bottom right, \mbox{\literal{(- (+ "z" "a") "b")}}, which is not up-connectable as the marked edge would not have a consistent direction.}
\label{fig:connectability}
\end{minipage}
\end{figure}

To state the well-formedness rules, we must first define three other layout properties: width, number of logical and connectable rows on either side, and up- and down-connectability on either side.
The definition of \textbf{width} below assumes a constant $S$, the unit width around curves and boxes.
\begin{itemize}
\item The width of a rail is its \literal{width} parameter.
\item The width of a space is $2S$.
\item Stations have implementation-dependent widths, plus $2S$.
\item The width of a horizontal concatenation is the sum of its sublayouts'.
\item The width of an inline VC is the sum of:
(i)~the width of its first sublayout;
(ii)~the width of its marker (implementation-dependent); and
(iii)~$3S$ if its start-side tip specification is a physical proportion other than~0 (i.e.~if it needs a bracket), and likewise for its end side if other than~1.
\item The width of a block VC is that of its first sublayout, plus $3S$ for each non-\literal{vertical} tip.
\end{itemize}
The multiples of $S$ account for vertical brackets in the rendering, e.g. as illustrated by \autoref{fig:bracket-widths} for a block VC.

Second, we define the \textbf{number of logical rows} on either side.
The \textbf{number of connectable rows} is equal to the logical one except where noted.
All tips and numbers of rows refer to the same universally quantified side:
\begin{itemize}
\item Spaces, rails, stations, and block VCs with a non-\literal{vertical} tip all have 1~row.
\item A horizontal concatenation has as many rows as its sidemost sublayout.
\item An inline VC has as many rows on its start side as its first sublayout on that side, and likewise for its end side with its last sublayout.
\item A positive block VC with a \literal{vertical} tip has as many rows as the sum of the number of connectable rows of its top and bottom sublayouts.
\item A negative block VC with a \literal{vertical} tip has as many logical rows as the sum of its top and bottom sublayouts, minus the number of connectable rows of its top sublayout, plus~1; and only 1~connectable row. (See \autoref{fig:negative-rows}.)
\end{itemize}

Third, we define \textbf{up- and down-connectability} on either side.
Up-connectability is meant to capture if a layout can be connected to ``from above'' when contained in block VCs, and likewise for down-connectability; see \autoref{fig:connectability}.
``Both-'' and ``neither-connectable'' have their natural meanings.
``Up-/down-'' means up- and down- separately.
On each side:
\begin{itemize}
\item Rails, stations, and block VCs with a non-\literal{vertical} tip are all neither-connectable.
\item A space is both-connectable.
\item A horizontal concatenation is up-/down-connectable only if its sidemost sublayout is.
\item An inline VC is up-/down-connectable on its start side only if its first sublayout is; and likewise on its end side with its last sublayout.
\item The up-/down-connectability of a block VC with a \literal{vertical} tip and polarity \variable{pol} depends on the nature of its top/bottom sublayout and is best described with tables.
Below, ``if top/bot'' means ``if the top/bottom sublayout has the property under consideration''.

\begin{center}
\begin{tabular}{ccc}
\textbf{\variable{pol}} & \textbf{top sublayout} & \textbf{up-ctbl.} \\
\literal{+} & \literal{-}ve block VC & no \\
\literal{+} & else & if top \\
\literal{-} & \literal{+}ve block VC & no (\autoref{fig:connectability}) \\
\literal{-} & else & if top
\end{tabular}
\qquad
\begin{tabular}{ccc}
\textbf{\variable{pol}} & \textbf{bot sublayout} & \textbf{down-ctbl.} \\
\literal{+} & \literal{-}ve block VC & no \\
\literal{+} & else & if bot \\
\literal{-} & \literal{-}ve block VC & no \\
\literal{-} & else & if bot
\end{tabular}
\end{center}
\end{itemize}
\newcommand\WFlabel[1]{\textsf{WF\tss{#1}}}
\newcommand\WF{\textsf{WF}}
\newcommand\WFc{\WF\tss{c}}
\newcommand\WFtip{\WF\tss{t}}
\newcommand\WFtop{\ensuremath{\widehat{\WF}}}
\begin{figure}[t]
\begin{minipage}[c]{0.45\linewidth}
\centering
\sffamily
\begin{prooftree}
\infer0[\WFlabel{r}]{\WF(\literal{(rail \variable{dir} \variable{w})})}
\end{prooftree}

\bigskip\smallskip
\begin{prooftree}
\infer0[\WFlabel{sp}]{\WF(\literal{(space \variable{dir})})}
\end{prooftree}

\bigskip\smallskip
\begin{prooftree}
\infer0[\WFlabel{st}]{\WF(\literal{(station \variable{dir} \variable{lbl} \variable{tm?})})}
\end{prooftree}

\bigskip\bigskip
\begin{prooftree}
\hypo{$∀\: \variable{i},\; \WF(\ellsub{i})$}
\hypo{\ellsub{i}.\literal{dir} = \variable{dir}}
\infer2[\WFlabel{c}]{\WFc(\variable{dir}\ \ellsub{i}\ldots)}
\end{prooftree}

\bigskip\bigskip
\begin{prooftree}
\hypo{\begin{matrix}
\WFc(\variable{dir}\ \ellsub{i}\ldots)
\\[1.5mm]
\forall\: 2 \leq \variable{i},
\: \ellsub{i}.\literal{left}.\literal{ctbl} = \literal{nr.}
\\
\forall\: \variable{i} \leq \variable{n} - 1,
\: \ellsub{i}.\literal{right}.\literal{ctbl} = \literal{nr.}
\end{matrix}}
\infer1[\WFlabel{hc}]{\WF(\literal{(hconcat \variable{dir} \ellsub{1}} \ldots{} \literal{\ellsub{n})})}
\end{prooftree}

\bigskip\bigskip
\begin{prooftree}
\hypo{\WF(\varell)}
\hypo{
\begin{matrix}
\varell{}.\literal{left}.\literal{ctbl} = \literal{nr.} \\
\varell{}.\literal{right}.\literal{ctbl} = \literal{nr.}
\end{matrix}
}
\infer2[\WFtop]{\WFtop(\varell)}
\end{prooftree}
\end{minipage}\begin{minipage}[c]{0.55\linewidth}
\centering\vspace*{-2mm}
\sffamily
\begin{prooftree}
\hypo{
\begin{matrix}
\variable{ts} = \literal{(logical \variable{r})} \implies \variable{r} \leq \variable{side}.\literal{num-rows} \\
\variable{ts} = \literal{vertical} \implies \variable{side}.\literal{ctbl} \neq \literal{neither}
\end{matrix}
}
\infer1[\WFlabel{t}]{\WFtip(\variable{side}\ \variable{ts})}
\end{prooftree}

\bigskip\bigskip
\begin{prooftree}
\hypo{
\begin{matrix}
\ellsub{top}.\literal{width} = \\\quad
\ellsub{bot}.\literal{width} \\[1.5mm]
\ellsub{top}.\literal{dir} = \variable{dir} \\
\ellsub{bot}.\literal{dir} = \variable{pol}\,(\variable{dir})
\end{matrix}
}
\hypo{
\begin{matrix}
\WF(\ellsub{top}) \quad \WF(\ellsub{bot}) \\[1.5mm]
\forall\:\variable{side},\:\WFtip(\variable{side}\ \variable{ts\tss{side}}) \\
\ellsub{top}.\variable{side}.\literal{ctbl} \succeq \literal{down} \\
\ellsub{bot}.\variable{side}.\literal{ctbl} \succeq \literal{up} \\
\end{matrix}
}
\infer2[\WFlabel{bvc}]{
\begin{matrix}
\hspace{-2em}\WF(\literal{(vconcat-block \variable{dir}} \\
\qquad\qquad
\literal{\variable{ts\tss{left}}\ \variable{ts\tss{right}} \variable{pol} \ellsub{top} \ellsub{bot})})
\end{matrix}
}
\end{prooftree}

\bigskip\bigskip\hspace*{1em}
\begin{prooftree}
\hypo{\begin{matrix}
\WFc(\variable{dir}\ \ellsub{i}\ldots) \qquad\hspace*{-0.5em}
\forall\:\variable{side},\:\WFtip(\variable{side}\ \variable{ts\tss{side}})
\\[1.5mm]
\forall\: 2 \leq \variable{i},
\: \ellsub{i}.\literal{start}.\literal{ctbl} = \literal{neither}
\\
\forall\: \variable{i} \leq \variable{n} - 1,
\: \ellsub{i}.\literal{end}.\literal{ctbl} = \literal{neither}
\\[1.5mm]
\ellsub{1}.\literal{width} = \ellsub{n}.\literal{width} = \variable{w} \geq \variable{m} = \variable{mk}.\literal{width} \\
\forall\: 2 \leq \variable{i} \leq \variable{n} - 1,
\: \ellsub{i}.\literal{width} = \variable{w} - \variable{m}
\end{matrix}}
\infer1[\WFlabel{ivc}]{
\begin{matrix}
\hspace{-2em}\WF(\literal{(vconcat-inline \variable{dir}} \\
\qquad\qquad
\literal{\variable{ts\tss{left}}\ \variable{ts\tss{right}} \variable{mk} \ellsub{1}} \ldots{} \literal{\ellsub{n})})
\end{matrix}
}
\end{prooftree}
\end{minipage}
\caption{Layout well-formedness.}
\label{fig:layoutwf}
\end{figure}
Finally, we can state the well-formedness rules, also presented as formulae in \autoref{fig:layoutwf}.
\begin{description}
\item[\WFlabel{r}, \WFlabel{sp}, \WFlabel{st}] Rails, spaces, and stations are well-formed.
\item[\WFlabel{c}] A horizontal or inline vertical concatenation \varell\ is well-formed if all sublayouts are well-formed and have the same direction as \varell.
\item[\WFlabel{hc}] In addition to \WFlabel{c}, a horizontal concatenation is well-formed if all sublayouts are neither-connectable, except possibly its sidemost sublayouts.
\item[\WFlabel{t}] A VC is well-formed if, on either side, if the tip is \mbox{\literal{(logical \variable{r})}}, then \variable{r} is no greater than the number of logical rows; or if it is \literal{vertical}, then the VC is either up- or down-connectable.
\item[\WFlabel{ivc}] In addition to \WFlabel{t}, an inline VC is well-formed if:
\begin{itemize}
\item All sublayouts are neither-connectable, except possibly the first sublayout on its start side or the last sublayout on its end side.
\item The first and last sublayouts have the same width $w \geq m$, and the other sublayouts have width $w - m$, where $m$ is the width of the marker.
\end{itemize}
\item[\WFlabel{bvc}] In addition to \WFlabel{t}, a block VC with direction \variable{dir} and polarity \variable{pol} is well-formed if:
\begin{itemize}
\item Both sublayouts are well-formed and have the same width.
\item The top sublayout is down-connectable on each side, and has direction \variable{dir}.
\item The bottom sublayout is up-connectable on each side, and has direction \variable{dir} iff \variable{pol} is \literal{+}.
\end{itemize}
\item[\WFtop] In addition to all the above, a top-level (i.e.~outermost) layout is well-formed if it is neither-connectable on both sides.
\end{description}

\subsection{Compilation or ``laying out'' relation}
\label{sec:langrel}

To \emph{lay out} a diagram is to compile a diagram \variable{d} to a layout \variable{l} such that
(i)~\variable{d} and \mbox{\literal{(diagram-of \variable{l\,})}} are equivalent, with equivalence as defined at the end of~\autoref{sec:diagramlang}, and the function \literal{diagram-of} as defined below; and
(ii)~\WFtop(\variable{l\,}) as defined above.

\medskip
\begin{tabular}{>{\ttfamily}p{0.45\linewidth}>{\textsf{:=}\quad\ttfamily}p{0.5\linewidth}}
(diagram-of (space \variable{dir})) & () \\
(diagram-of (rail \variable{dir} \variable{w})) & () \\
(diagram-of (station \variable{dir} \variable{label} \#t)) & "\variable{label}" \\
(diagram-of (station \variable{dir} \variable{label} \#f)) & [\variable{label}] \\
(diagram-of (hconcat \literal{ltr} \variable{l\tss{1}} … \variable{l\tss{n}})) & ((diagram-of \variable{l\tss{1}}) … (diagram-of \variable{l\tss{n}})) \\
(diagram-of (hconcat \literal{rtl} \variable{l\tss{1}} … \variable{l\tss{n}})) & ((diagram-of \variable{l\tss{n}}) … (diagram-of \variable{l\tss{1}})) \\
(diagram-of (vconcat-inline \variable{dir} \par\quad \variable{lts} \variable{rts} \variable{mk} \variable{l\tss{1}} … \variable{l\tss{n}})) & ((diagram-of \variable{l\tss{1}}) … (diagram-of \variable{l\tss{n}})) \\
(diagram-of (vconcat-block \variable{dir} \par\quad \variable{lts} \variable{rts} \variable{pol} \variable{l\tss{1}} \variable{l\tss{2}})) & (\variable{pol} (diagram-of \variable{l\tss{1}}) (diagram-of \variable{l\tss{2}}))
\end{tabular}

\begin{figure}[b]
\centering\begin{tabular}{>{\quad\hspace{0.5em}}c<{\qquad\qquad\hspace{0.5em}}c<{\qquad\quad\hspace{1em}}c}
align & wrap & justify \\[2pt]
\includegraphics{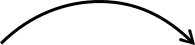} & \includegraphics{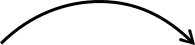} & \includegraphics{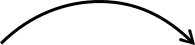} \\[2pt]
\end{tabular}

\centeralignimage{\includegraphics[scale=0.752]{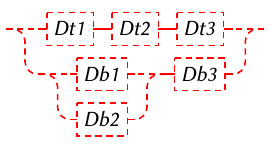}}
\hspace{0.3em}
\centeralignimage{\includegraphics[scale=0.72]{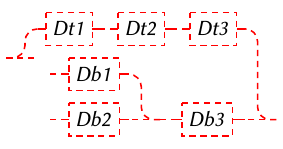}}
\hspace{0.3em}
\centeralignimage{\includegraphics[scale=0.72]{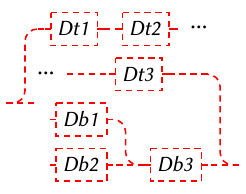}}
\hspace{0.8em}
\centeralignimage{\includegraphics[scale=0.72]{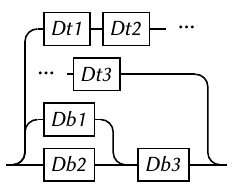}}
\renewcommand{\thefigure}{\ref{fig:compilation}}
\captionsetup{list=no}
\caption{(reproduced from page~\pageref{fig:compilation}) Our three-step algorithm for compiling a diagram to a layout.}
\addtocounter{figure}{-1}
\end{figure}

\section{A three-step layout algorithm}
\label{sec:threesteps}

In the previous section, we defined the problem of railroad layout.
In this section, we describe how we solve it.
The space of possible layouts for a diagram is large, with several degrees of freedom at each level of nesting;
our solution aims to navigate that space while balancing flexibility, efficiency, and ease of use.
In broad strokes, our algorithm has three steps (illustrated in \autoref{fig:compilation} below):
\begin{enumerate}
\item \emph{Alignment:}
determining whether the layout of each subdiagram collapses with its container, and if not, determining its vertical positioning;
controlled by the \literal{align-items} policy.
\item \emph{Wrapping:}
deciding how to wrap the sublayouts of each sequence across visual rows to meet a target width.
We discuss wrapping parameters in~\autoref{sec:wrapping}.
\item \emph{Justification:}
deciding how to distribute the width available inside a container among and around its sublayouts;
controlled by the \literal{justify-content} policy.
\end{enumerate}

Operationally, each step targets an aspect of the compilation relation from~\autoref{sec:langrel}.
Alignment is about choosing tip specifications and placing spaces such that the layout is well-formed.
Wrapping is about choosing a composition of horizontal and inline vertical concatenations for each sequence.
Justification is about placing rails (ε-sequences) to add up to the correct widths.
None of these choices affect diagram equivalence as defined at the end of~\autoref{sec:diagramlang}.

\begin{figure}[t]
\centering\vspace{-2mm}
\begin{minipage}[c]{0.38\linewidth}
\centering
\begin{tabular}{>{\sffamily\itshape}r>{\sffamily}r>{\ttfamily}l}
id & := & (station \variable{dir} \variable{lbl} \variable{tm?}) \\
&  | & (\variable{dir} \variable{id…}) \\
&  | & (vconcat-block \variable{dir} \\
&    & \quad \variable{pol} \variable{id} \variable{id})
\end{tabular}
\caption{The immediate diagram (\variable{id}) language.
\variable{dir}, \variable{lbl}, \variable{tm?}, \variable{pol} are as in \autoref{def:commonlang}.}
\label{tab:immediatelang}

\bigskip\bigskip\bigskip
\begin{tabular}{>{\sffamily\itshape}r>{\sffamily}r>{\ttfamily}l}
ad & := & (station \variable{dir} \variable{lbl} \variable{tm?}) \\
&  | & (\variable{dir} \variable{ts} \variable{ts} \variable{ad…}) \\
&  | & (vconcat-block \variable{dir} \\
&    & \quad \variable{ts} \variable{ts} \variable{pol} \variable{ad} \variable{ad}) \\
&  | & (space \variable{dir})
\end{tabular}
\caption{The aligned diagram (\variable{ad}) language.
\variable{dir}, \variable{lbl}, \variable{tm?}, \variable{ts}, \variable{pol} are as in \autoref{def:commonlang}.}
\label{fig:alignedlang}
\end{minipage}\hfill\begin{minipage}[c]{0.54\linewidth}
\raggedleft
\begin{tabular}[t]{l>{\sffamily\itshape}r>{\sffamily}r>{\ttfamily}l}
sequence wrap of \variable{d} \hspace{-14em} &&&\\
& sw\tss{d} & := & \variable{row\tss{d}} \\
&    &  | & (vconcat-inline \variable{dir} \\
&    &    & \quad \variable{ts} \variable{ts} \variable{mk} \variable{row\tss{d}} \variable{row\tss{d}} \variable{row\tss{d}…}) \\
& row\tss{d} & := & (hconcat \variable{dir} \variable{d} \variable{d…}) \\[2mm]
locally wrapped diagram \hspace{-14em} &&&\\
& lwd & := & (station \variable{dir} \variable{lbl} \variable{tm?}) \\
&     &  | & (vconcat-block \variable{dir} \\
&     &    & \quad \variable{ts} \variable{ts} \variable{pol} \variable{lwd} \variable{lwd}) \\
&     &  | & (space \variable{dir}) \\
&     &  | & \textbf{\textsf{ordered set}} \variable{sw\tss{lwd}…} \\[2mm]
global wrap \hspace{-14em} &&&\\
& gw & := & (station \variable{dir} \variable{lbl} \variable{tm?}) \\
&    &  | & (vconcat-block \variable{dir} \\
&    &    & \quad \variable{ts} \variable{ts} \variable{pol} \variable{gw} \variable{gw}) \\
&    &  | & (space \variable{dir}) \\
&    &  | & \variable{sw\tss{gw}} \\[2mm]
globally wrapped diagram \hspace{-14em} &&&\\
& gwd & := & \textbf{\textsf{ordered set}} \variable{gw…}
\end{tabular}
\caption{The wrapped diagram languages.
\variable{dir}, \variable{lbl}, \variable{tm?}, \variable{ts}, \variable{pol} are as in \autoref{def:commonlang}.}
\label{tab:wrappedlang}
\end{minipage}
\end{figure}

In this section, we build our layout algorithm as a progressive lowering from diagrams to layouts.
The first lowering is immediate:
per the compilation relation, tokens become stations;
stacks become block VCs, although tip specifications are not known yet;
and bottom subdiagrams of negative stacks sequences are reversed.
Assuming the top-level term is left-to-right, it is easy to compute subterm directions to satisfy the directional conditions of well-formedness.
This gives an \emph{immediate diagram} as in \autoref{tab:immediatelang}.

\begin{figure}[t]
\centering\vspace*{-6mm}
\begin{minipage}[b]{0.55\linewidth}
\centering
\begin{subfigure}{\linewidth}
\centering
\includegraphics[width=\linewidth]{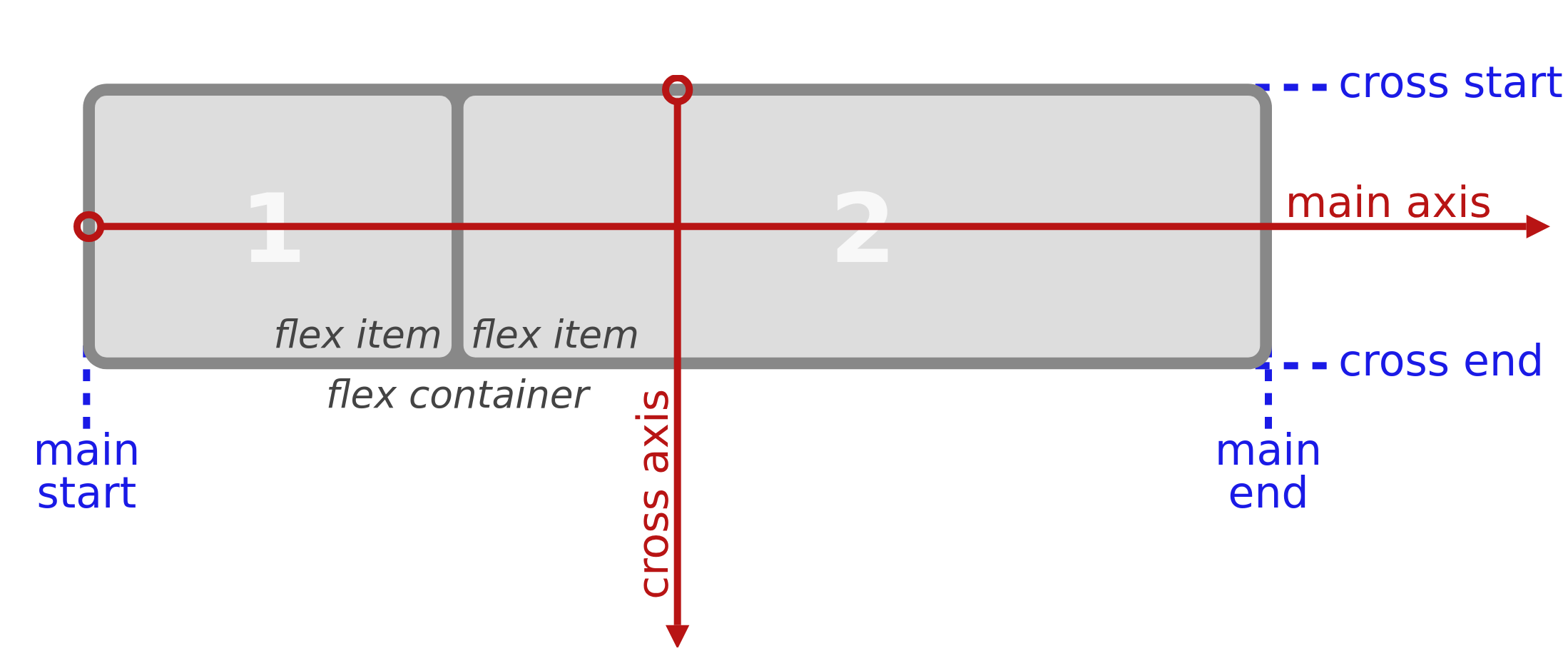}
\caption{Main and cross axes for English text.}
\label{fig:flexterms:terms}
\end{subfigure}

\bigskip\smallskip
\begin{subfigure}[b]{\linewidth}
\centering
\includegraphics[width=\linewidth]{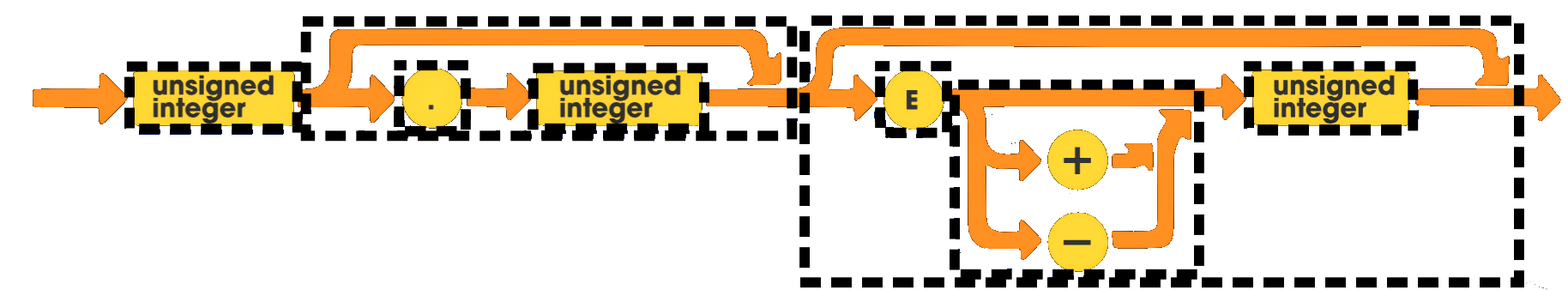}
\caption{Alignment in this railroad layout (from~\cite{KamifujiRaskinApplePascalSyntax1979}) is hard to explain in terms of bounding boxes (which we draw with dashed black lines).}
\label{fig:flexterms:notbox}
\end{subfigure}
\end{minipage}\hfill\begin{minipage}[b]{0.4\linewidth}
\centering
\begin{subfigure}{\linewidth}
\centering\small
\hspace*{-2mm}\begin{tabular}[b]{cc}
top & center \\
\topalignimage{\includegraphics[scale=0.53]{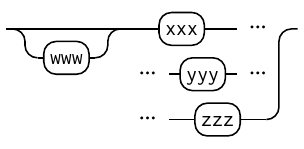}}
& \hspace{-1mm}\topalignimage{\includegraphics[scale=0.53]{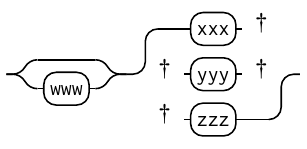}} \\[15mm]
bottom & baseline \\
\topalignimage{\includegraphics[scale=0.53]{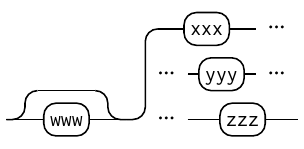}}
& \hspace{-1mm}\topalignimage{\includegraphics[scale=0.53]{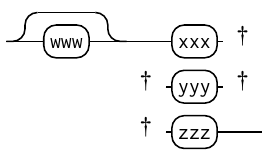}}
\end{tabular}
\caption{Alignment policies.}
\label{fig:flexterms:align}
\end{subfigure}

\bigskip
\begin{subfigure}[b]{\linewidth}
\raggedleft
\includegraphics[width=\linewidth]{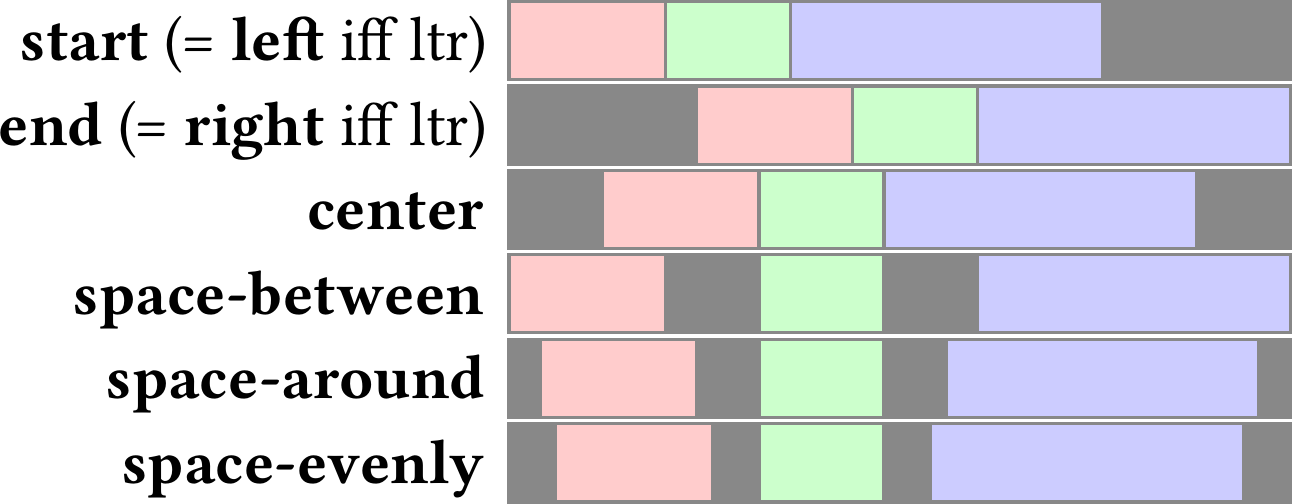}
\caption{CSS justification policies.}
\label{fig:flexterms:justify}
\end{subfigure}
\end{minipage}

\caption{Alignment and justification terminology. (Subfigures~\ref*{fig:flexterms:terms}, \ref*{fig:flexterms:justify} adapted from~\cite{W3CCSSFlexibleBox2018,W3CCSSBoxAlignment2025}.)}
\label{fig:flexterms}
\end{figure}

\subsection{Alignment and justification à la flexbox}

Layouts of linear content often have a \emph{main axis} aligned with the inherent linear order and an orthogonal \emph{cross axis}, each with a natural \emph{start} and \emph{end} side.\footnote{Although we explain our terminology without further citation below, it is heavily inspired by the CSS Flexible Box Layout Module~\cite{W3CCSSFlexibleBox2018}, or \emph{flexbox} for short. We discuss key differences in~\autoref{sec:related}.
}
E.g., the main axis of English text is left-to-right, and the cross axis is top-to-bottom (\autoref{fig:flexterms:terms});
the main axis of a vinyl is circular, and the cross axis is radial.
Railroad layouts have a horizontal main axis with a direction per their \variable{dir} parameter, and a top-to-bottom cross axis.
Then, alignment and justification are about the cross- and main-axis positioning of sublayouts, respectively.

Railroad alignment is best described in terms of the tips of each layout on each side, rather than the positions of their bounding boxes.
For example, the sublayouts of sequences in \autoref{fig:flexterms:notbox} cannot be explained as top-, center-, or bottom-aligned with respect to their bounding boxes.
It is simpler to say that each is laid out with a default choice of \literal{logical} tips, and the containing layouts are formed by positioning the sublayouts so as to align the tips.
An alignment policy, then, specifies for any diagram which tips its subdiagrams are to be laid out with under which circumstances.
Alignment accounts for collapsing: a subdiagram contained in a stack may be laid out with a \literal{vertical} tip to collapse with its container.
\autoref{fig:flexterms:align} illustrates the alignment policies we suggest.
``Baseline'' captures common patterns beyond the three simpler positional ones, such as choosing \mbox{\literal{logical 2}} tips for a stack whose top subdiagram is an empty sequence and bottom has only one row.
It also demonstrates that the left and right tips of a layout need not be at the same height.

Thus, alignment lowers an immediate diagram to an \emph{aligned diagram} as presented in \autoref{fig:alignedlang}, where each sequence and stack has tip specifications for its eventual layout, and spaces are explicit.
To ensure well-formedness, our compiler:
(i)~specifies \literal{vertical} tips and inserts spaces to make the sublayouts of block VCs correctly connectable (\hyperref[fig:layoutwf]{\WFlabel{bvc}}), as the recursive definitions of connectability have only spaces as their positive base case;
(ii)~avoids their negative base cases by conservatively never specifying \literal{vertical} tips for negative block VCs;\footnote{Creators sometimes prefer negative block VCs to not collapse even when possible, such as in \autoref{fig:wild:ecma}.}
and
(iii)~avoids inserting spaces in the middle of a sequence (\hyperref[fig:layoutwf]{\WFlabel{hc}}, \hyperref[fig:layoutwf]{\WFlabel{ivc}}) or at the ends of an outermost sequence (\hyperref[fig:layoutwf]{\WFtop}).

Next, wrapping turns each sequence into a set of possible compositions of horizontal and inline vertical concatenations (\emph{wraps}) and orders them by preference.
We describe this process in detail in the next section but assume its result, viz.~\autoref{tab:wrappedlang}, to explain justification below.
In particular, we assume the \literal{min-content} and \literal{max-content} widths are defined for each term:
the minimum width of its eventual layout if, respectively, all wrapping opportunities are taken, or none are.

\begin{figure}[b]
\centering\setlength{\tabcolsep}{0em}\vspace{-2mm} \begin{tabular}[t]{m{1.4em}>{\hspace{0.5em}}cm{0.7\linewidth}>{\hspace{6mm}}c}
\raggedleft (i)\par\hspace*{0pt}\par\hspace*{0pt}&\hspace{0pt}& All subdiagrams must have at least their \literal{min-content} widths (\mbox{$n_{\textsf{A}} = 10$}, \mbox{$n_{\textsf{B}} = 10$} in the illustration) and at least a \literal{gap}-width (a global parameter) rail between them.
& \centeralignimage{\includegraphics[height=15.4mm]{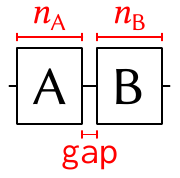}} \\[9mm]
\raggedleft (ii)\par\hspace*{0pt} && As the available width grows, all subdiagrams should grow to their \literal{max-content} widths ($x_{\textsf{A}} = 12$, $x_{\textsf{B}} = 20$) at equal rates.\footnotemark
& \centeralignimage{\includegraphics[height=16mm]{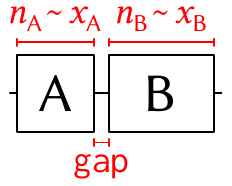}} \\[8mm]
\raggedleft (iii)\par\hspace*{0pt}\par\hspace*{0pt}\par\hspace*{0pt} && Only after all have reached their \literal{max-content} widths should any extra width ($E$) be used, in part for spacing ($\literal{fa} \cdot E$, where \literal{fa} is a global parameter \literal{flex-absorb}) and in part for letting subdiagrams grow further at equal rates, if they can.
& \centeralignimage{\includegraphics[height=16mm]{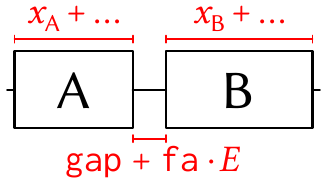}}
\end{tabular}
\caption{Justification aims to ensure three properties at each level of nesting.}
\label{fig:justification-aims}
\end{figure}

The last step of lowering to the layout language is justification.
It chooses one of the possible wraps and distributes extra width around and among items in each (wrapped) row, aiming to ensure three properties at each level of nesting, explained in \autoref{fig:justification-aims}.
The policies for distributing width around items (if any) are illustrated in \autoref{fig:flexterms:justify}.
This process satisfies width-related well-formedness criteria by creating horizontal concatenations and rails while preserving diagram equivalence.

Our compiler implements justification as a recursive algorithm on wrapped diagrams with an additional \variable{target-width} argument.
The top-level \variable{target-width} the user requests must be at least the \literal{min-content} of the top-level diagram to ensure that some layout is possible.
Then:
\begin{itemize}
\item For stations and spaces, the algorithm does nothing, as they are inflexible.
\item For each vertical concatenation, the algorithm
(i)~subtracts the fixed widths associated with tips and markers from \variable{target-width};
(ii)~ensures that each subdiagram is a horizontal concatenation (potentially by constructing one); and finally
(iii)~recurses on each (horizontal concatenation) subdiagram with the (updated) \variable{target-width}.
\item For each ordered set of wraps, the algorithm selects the first (i.e.~most preferred) wrap whose \literal{min-content} is no greater than \variable{target-width} (i.e.~which is guaranteed to fit).
\end{itemize}
\footnotetext{In~(ii), ``at equal rates'' means in proportion to the difference between the \literal{max-} and \literal{min-content} of each subdiagram.
In~(iii), it means in proportion to the \literal{max-content}.
In theory, different proportions could be used (e.g.~nonlinear), but in practice, the difference is barely noticeable (a few pixels at most).
}\smallskip
For each horizontal concatenation, the algorithm is more involved:
\begin{enumerate}[1.]
\item Let \variable{aw} and \variable{rw} be variables representing the \emph{absorbed} and \emph{remaining width} to be distributed around and among sublayouts, respectively.
Let \variable{sw\tss{i}} be variables representing the \emph{subwidths} assigned to each subdiagram~\variable{d\tss{i}}, for \mbox{1 ≤ \variable{i} ≤ \variable{n}}.
\begin{enumerate}[a.]
\item Initialize \variable{aw} to $(\variable{n} - 1) \cdot \literal{gap}$ (minimum spacing).
\item Initialize each \variable{sw\tss{i}} to \literal{\variable{d\tss{i}}.min-content} (minimum widths).
\item Initialize \variable{rw} to $\variable{target-width} - \variable{aw} - \Sigma\,\variable{sw\tss{i}}$.
\end{enumerate}
We will maintain the invariant $\variable{rw} + \variable{aw} + \Sigma\,\variable{sw\tss{i}} = \variable{target-width}$. By the end, $\variable{rw} = 0$.
\item Let the \emph{growth width} $\variable{g\tss{i}} = (\literal{\variable{d\tss{i}}.max-content} - \literal{\variable{d\tss{i}}.min-content})$.
Let the \emph{maximum growth width} $\variable{mg} = \min(\variable{rw}, \Sigma\,\variable{g\tss{i}})$.
Increment each \variable{sw\tss{i}} by $\variable{mg} \cdot \variable{g\tss{i}}/\Sigma\,\variable{g\tss{i}}$,
and decrement \variable{rw} by \variable{mg}.
\item Increment \variable{aw} by $\variable{rw} \cdot \literal{flex-absorb}$, and decrement \variable{rw} by the same amount.
\item Let \variable{d\tss{j}}, etc.\ denote only those subdiagrams that are concatenations (and hence can grow further).
If there are none, increment \variable{aw} by \variable{rw} and set \variable{rw} to zero.
Else, increment each \variable{sw\tss{j}} by $\variable{rw} \cdot (\literal{\variable{d\tss{j}}.max-content})/\Sigma(\literal{\variable{d\tss{j}}.max-content})$ and set \variable{rw} to zero.
\item Distribute \variable{aw} among rails between and around subdiagrams per \literal{justify-content}, \literal{gap}, and the current direction.
Recursively justify each subdiagram with target width \variable{sw\tss{i}}.
Return a new horizontal concatenation with said justified subdiagrams and rails.
\end{enumerate}

\begin{figure}[bt]
\centering\vspace{-2mm}
\centeralignimage{\includegraphics[scale=0.7]{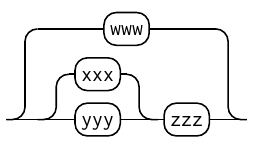}}
\quad$\xleftarrow[\text{space-evenly}]{\text{justification:}}$\quad
\centeralignimage{\includegraphics[scale=0.7]{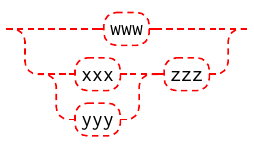}}
\quad$\xrightarrow[\text{start}]{\text{justification:}}$\quad
\centeralignimage{\includegraphics[scale=0.7]{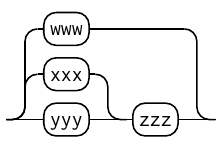}}
\caption{A sublayout cannot collapse with its container during alignment (here, bottom-alignment) if the justification policy (here, space-evenly) could potentially insert a rail in between.}
\label{fig:rrd-align-justify}
\end{figure}

\enlargethispage{\baselineskip} 
We make three remarks about the relationship between alignment and justification.
First, alignment applies to each sublayout of a container individually, whereas justification only makes sense as a collective property of all the sublayouts taken together, so the former policy is called \literal{align-items} and the latter \literal{justify-content}.
Second, if a sequence wraps across multiple rows, alignment and justification apply to each row independently, although policies may specify different behaviors for the first, middle, and last rows.
And third, alignment actually depends on the justification policy, because the side of a stack can collapse with its container only if the two are guaranteed to be directly adjacent, without the possibility of any rails in between (e.g.~\autoref{fig:rrd-align-justify}).
Now, justification must happen after wrapping because it works on each wrapped row, and wrapping must happen after alignment because it depends on widths, which depend on tip specifications; but the dependency is not circular, because alignment only depends on the justification policy, and not its realization.

The net result of the lowering is thus a well-formed layout, equivalent to the original diagram, that has width exactly \variable{target-width}, with alignment and justification per the \literal{align-items}, \literal{justify-content}, \literal{flex-absorb}, and \literal{gap} global parameters.

\subsection{Wrapping as optimization}
\label{sec:wrapping}

Many layout problems are stated in terms of hard constraints and softer optimization objectives.
\mbox{1-dimensional} problems like text wrapping and code pretty-printing treat layout width as a hard constraint, while trying to minimize properties like height,\footnote{This seems obvious but bears further thought.
It is the main axis that has a hard constraint, and the cross axis that is subject to minimization.
Layouts with different reading orders behave accordingly:
for instance, traditional Chinese calligraphy was laid out on horizontal scrolls, with a top-to-bottom main axis (limited by paper height) and a right-to-left cross axis (e.g.~\cite{ChengsuXizhiLantingjiXuScroll650}).
1-dimensional layouts are informally often measured along the cross axis -- books in pages, code in lines, scrolls in inches -- implicitly assuming the layout makes good use of a reasonable main-axis size.}
deviance from normative spacing, and drastic variation between lines~\cite{KnuthPlassBreakingParagraphsLines1981,PorncharoenwaseEtAlPrettyExpressivePrinter2023}.
In contrast, the constraints and objectives for 2-dimensional layout problems are typically isotropic properties, unlike size.
For instance, among Graphviz's 8~graph layout algorithms, 5~do not favor any axis over another; 1~(``circo'') distinguishes radial and circular axes, but minimizes edge crossings (independent of direction); and 2 (``dot'' and ``twopi'') are explicitly hierarchical layouts that impose a 1-dimensional structure of discrete ``ranks'' or ``levels'') on the graph~\cite{GansnerNorthOpenGraphVisualization2000}.

\begin{figure}[t]
\vspace{-3mm}
\begin{minipage}[b]{0.6\linewidth}
\centering
\setlength{\tabcolsep}{3mm} \hspace{-2mm}\begin{tabular}[t]{cc}\topalignimage{\includegraphics[scale=0.5]{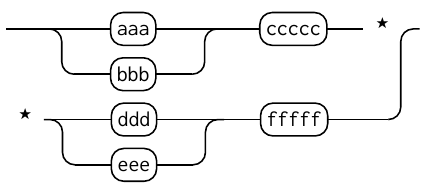}}
& \topalignimage{\includegraphics[scale=0.5]{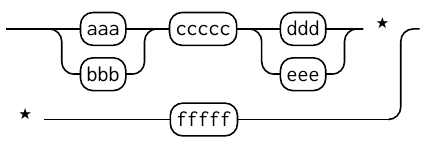}} \\[16mm]
\topalignimage{\includegraphics[scale=0.63]{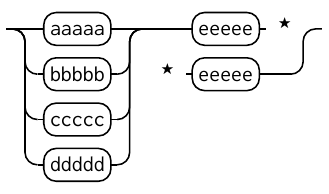}}
& \topalignimage{\includegraphics[scale=0.63]{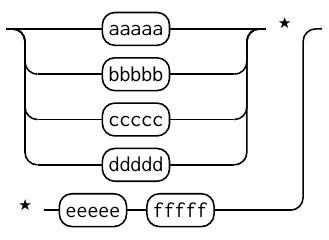}} \\[26mm]
\topalignimage{\includegraphics[scale=0.63]{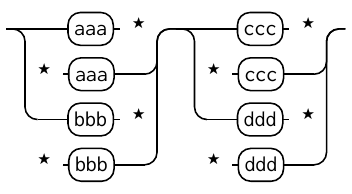}}
& \topalignimage{\includegraphics[scale=0.63]{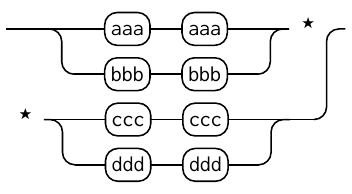}}
\end{tabular}
\caption{Each pair of drawings shows two layouts of the same diagram at the same width.
The first shows a text-like tradeoff between height and balance.
The second shows a code-like ``indentation'' choice.
The third shows a wrapping decision that neither (1-dimensional) layout can express.}
\label{fig:wrapexamples}
\end{minipage}\hfill\begin{minipage}[b]{0.36\linewidth}
\centering
\begin{tabular}{rl}
SW1 & \centeralignimage{\includegraphics[scale=0.45]{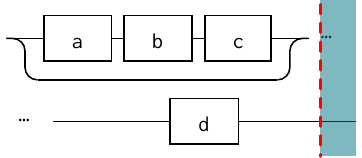}} \\[7mm]
WD1 & \centeralignimage{\includegraphics[scale=0.45]{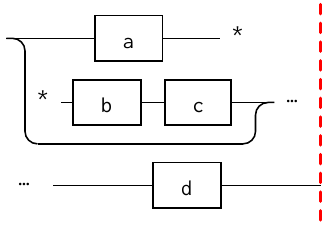}} \\[13mm]
SW2 & \centeralignimage{\includegraphics[scale=0.45]{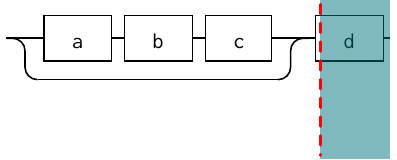}} \\[7mm]
WD2 & \centeralignimage{\includegraphics[scale=0.45]{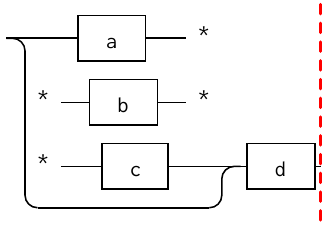}}
\end{tabular}
\caption{Between two sequence wraps (SW) that give wrapped diagrams (WD) of a given target width (dashed red), SW1 has lesser excess \literal{max-content} width (shaded blue), leading to less internal wrapping.}
\label{fig:wrappingobjectives}
\end{minipage}
\end{figure}

Railroad layouts, being 1.5-dimensional, fall somewhere in between.
Like 1-dimensional layouts, they follow a reading order and are wrapped in practice, so we treat target width as a hard constraint
(and in fact, an exact one, as railroad layouts stretch more and suffer less from sparse rows than text or code).
But \autoref{fig:wrapexamples} illustrates how the usual parameters of 1-dimensional layout -- width, height, and balance -- cannot explain the additional ``.5~dimension'' of railroad layout decisions.
Informally, layouts with less and shallower wrapping are preferred, while trying to make good use of available space.

One of our key contributions is to identify that encoding these preferences as optimization for just the wrapping step is sufficient for practical purposes, instead of phrasing the whole railroad layout problem as optimization.
This helps us avoid the instability typical of optimization-based 2-dimensional layouts and develop heuristics for \emph{local wrapping} (of every sequence) in addition to \emph{global wrapping} (of the whole diagram).
In computational terms, we express the result of wrapping as a total order over possible \emph{wraps}, i.e.~compositions of horizontal and inline vertical concatenations for each sequence, either per sequence (``local'') or for all sequences in the diagram at once (``global'').
This results in a \emph{locally} or \emph{globally wrapped diagram} as in \autoref{tab:wrappedlang}.
Standard techniques can then find the optimal (per the total order) wrap subject to a target width constraint.\footnote{Technically, a total order is stronger than we need:
an order in which any subset of wraps bounded by width has a least element would be sufficient.
However, we have not found the distinction to matter.}

Next, rather than try to define a single total order to cater to all styles and use cases, we describe a few principled optimality parameters that explain wrapping decisions we observed in the wild.
We begin with three preliminary definitions: the \emph{wrap specification} for a sequence, the \literal{min-content} and \literal{max-content} widths of a wrapped diagram, and the \emph{height} of a globally wrapped diagram.

A \emph{wrap specification} encodes a composition of horizontal and inline vertical concatenations for a sequence as a set of positive integer \emph{wrap points}, each specifying the index of the first subdiagram on a row (and hence each no less than~1 and no greater than the number of subdiagrams).
Naturally, \literal{1} must be a wrap point;
the wrap specification \literal{\{1\}} means there is only one row, i.e.\ a horizontal concatenation, and other specifications signify an inline VC containing horizontal concatenations.\footnote{Although it would be valid to nest inline VCs further, our encoding cannot express it, because we have not found it useful in practice and it can be achieved with nested sequences anyway.}
For example, for the sequence \variable{d\tss{1}}, \variable{d\tss{2}}, \variable{d\tss{3}},~\variable{d\tss{4}}, the wrap specification \mbox{\literal{\{1, 4\}}} means \variable{d\tss{1}}, \variable{d\tss{2}}, and~\variable{d\tss{3}} are on the first row, and \variable{d\tss{4}} is on the second.

The \literal{min-} or \literal{max-content} width of a wrapped diagram is the minimum width of its eventual layout if, respectively, all wrap points are taken, or none are.
\begin{itemize}
\item For stations and spaces, both are equal to the layout width as defined in~\autoref{sec:layoutlang}.
\item For a horizontal concatenation, \literal{min-content} is computed just like the layout width, except using the \literal{min-content}s of its subdiagrams instead of their layout widths, and accounting for the minimum \literal{gap} between subdiagrams; and likewise for \literal{max-content}.
\item For a vertical concatenation, \literal{min-content} is computed just like the layout width, except using the maximum \literal{min-content} among its subdiagrams instead of the layout width of the first sublayout; and likewise for \literal{max-content}.
\item For a set of wrapped diagrams, \literal{min-content} is the minimum value of \literal{min-content}, and \literal{max-content} is the \literal{max-content} of the element in which all wrap specifications are \literal{\{1\}}.
\end{itemize}
Note that we are not actually performing layout to compute these properties, but just using the same formulae.
Note also that a globally wrapped diagram is a set of global wraps, but global wraps contain no sets themselves (see \autoref{tab:wrappedlang}).
Hence, \literal{min-} and \literal{max-content} coincide for a global wrap and are equal to its layout width, which we will call just the \literal{content} width, because their formulae only differ when there is a set of possibilities.

Lastly, the \emph{height} of a global wrap is straightforward to compute.
Stations have implementation-dependent heights;
the height of a vertical concatenation is the sum of the heights of its subdiagrams, plus any space between rows;
and the height of a horizontal concatenation can be computed by vertically aligning the internal tips of its subdiagrams to form a ``baseline'', and taking the difference between the largest extent above and below that baseline of any subdiagram.

We can now explain the wrapping decisions we observed in manual layouts as finding an optimal global wrap with the objectives below.
$\variable{gw\tss{1}} <_g \variable{gw\tss{2}}$ means \variable{gw\tss{1}} is preferable to \variable{gw\tss{2}}.

\begin{description}
\item[Greater \literal{content} width.]
In other words, we prefer layouts that use more of the available width.
All else being equal, \mbox{$\literal{\variable{gw\tss{1}}.content} > \literal{\variable{gw\tss{2}}.content}$} implies \mbox{$\variable{gw\tss{1}} <_g \variable{gw\tss{2}}$}.
\item[Less and shallower wrapping.]
Let the \emph{depth}~$d_{\variable{t}}$ of each term~\variable{t} in a global wrap be defined as zero for the top-level term and one greater than the containing term for each subterm.
Let $\ell_{\variable{t}}$ be the length of the wrap specification if \variable{t} is a sequence wrap, else zero.
Let $\plty{d}(d_{\variable{t}})$ be a real-valued \emph{depth penalty} function and $\plty{\ell}(\ell_{\variable{t}})$ a real-valued \emph{wrap-length penalty} function.
Then, the \emph{wrap penalty} $\plty{w}(\variable{gw})$ is \mbox{$\Sigma_{\variable{t}}\,\plty{\ell}(\ell_{\variable{t}}) \cdot \plty{d}(d_{\variable{t}})$}.
We need $\plty{d}$ and $\plty{\ell}$ to be monotonic and positive.
All else being equal, \mbox{$\plty{w}(\variable{gw\tss{1}}) < \plty{w}(\variable{gw\tss{2}})$} implies \mbox{$\variable{gw\tss{1}} <_g \variable{gw\tss{2}}$}.
\item[Lower height.]
All else being equal, \mbox{$\literal{\variable{gw\tss{1}}.height} < \literal{\variable{gw\tss{2}}.height}$} implies \mbox{$\variable{gw\tss{1}} <_g \variable{gw\tss{2}}$}.
\end{description}
The last three criteria are phrased in terms of ``all else being equal'' assumptions.
Two practical ways to satisfy them all at once are
(i)~to make $<_g$ a lexicographic order, defining a priority order among the three, or
(ii)~to make $<_g$ a numerical order on a linear combination of \literal{content} width, wrap penalty, and height, where the first has a negative weight and the other two have positive weights.
Either option makes $<_g$ a total order.

\enlargethispage{\baselineskip}
\sloppy
Of course, it is impractical to enumerate all global wraps to choose one:
a diagram with $k$~sequences of $n$~subdiagrams each has $2^{k(n-1)}$ possible wraps.
This motivates our locally wrapped diagrams, for which a recursive layout process can alternate between justification and choosing a wrap for each sequence, avoiding the combinatorial explosion.
Below, we translate the global objectives into reasonable local versions.
Recall that \literal{min-content} and \literal{max-content} are now distinct, and the unresolved ambiguity in subterms means that we cannot yet compute the height of a diagram.
\begin{description}
\item[Lesser \literal{max-content} width beyond target width.]
For global wraps, we prefer \emph{greater} \literal{content} widths for using more of the available space, because we know they are no greater than the target width.
But for sequence wraps, the \literal{max-content} width is typically greater than the target width;
the smaller the excess, the closer its subdiagrams will be to their ``full'' \literal{max-content}s, reducing further nested wrapping.
(See \autoref{fig:wrappingobjectives}.)
\item[Lesser \literal{max-content} width for multi-row sequences.]
As the \literal{max-content} of a multi-row sequence is the maximum of its subdiagrams',
minimizing it leads to more balanced rows and better use of horizontal space, as moving any subdiagram from one row to another would only increase it.
(The first row of \autoref{fig:wrapexamples} illustrates this objective.)
\item[Less and shallower wrapping.]
As for global wraps.
\end{description}
As before, lexicographic order or numerical order on a linear combination of \literal{max-content} and the wrap penalty are practical choices.
(\literal{min-content} seems irrelevant to the order.
It roughly scales with the width of the widest or most deeply nested station in a diagram, not with any holistic property.
Moreover, in our experiments, we have not yet found a case in which \literal{min-content} distinguishes between two otherwise equal sequence wraps.)

\subsection{Implementation}
\label{sec:impl}

We implemented a prototype of the algorithm above in Scala (compiled with Scala.js).
It is available at
\href{https://github.com/epfl-systemf/librrd}{github.com/epfl-systemf/librrd}.
As a summary, the parameters to layout are:
\begin{itemize}
\item the target width (and whether it should be achieved by global or local wrapping);
\item \literal{align-items}, the policy for vertical alignment of sublayouts;
\item \literal{justify-content}, the policy for horizontal justification of sublayouts;
\item \literal{flex-absorb}, the proportion of any extra width a horizontal concatenation uses for justification before passing the rest on to its sublayouts; and
\item the minimum \literal{gap} between sublayouts of a horizontal concatenation.
\end{itemize}
Layouts are rendered to SVG with classes and hierarchical structure amenable to further CSS styling.
Our implementation is 1141 SLOC (excluding empty lines and comments), plus 314 SLOC for rendering.
A web UI is available at
\href{https://systemf.epfl.ch/etc/librrd/}{systemf.epfl.ch/etc/librrd/}.
It implements an order over sequence wraps (local wrapping) that we have found to work well in practice, which first minimizes $\max(0, \variable{w}.\literal{max-content} - \variable{target-width})^2 + 10 \cdot \plty{w}(\variable{w})$ with wrap-length penalty $\plty{\ell}(\ell) = \ell$ and depth penalty $\plty{d}(d) = 2^{2d}$, and then minimizes \literal{max-content}.

\section{Evaluation}
\label{sec:eval}

We started \autoref{sec:formal} by defining a language to express railroad diagrams, and by the end of~\autoref{sec:threesteps}, we had a principled algorithm to produce layouts ready for rendering.
In this section, we evaluate whether both ends of this pipeline are practical.
For the front end, we show that our diagram language is well-suited for common applications, viz.~illustrating formal grammars, by describing how regular expressions and Backus-Naur form can be compiled to it~(\autoref{sec:frontends}).
For the back end, we argue that our compiler is practical by comparing its output to diagrams laid out by hand~(\autoref{sec:evalmanual}) and by other tools~(\autoref{sec:evalauto}).
We also show that our compiler is performant enough for interactive use~(\autoref{sec:performance}).

\begin{figure}[t]
\begin{subfigure}[b]{0.58\linewidth}\centering\begin{tabular}{cc>{\ttfamily}c>{\hspace{-2mm}}c}
\textbf{regex} & \textbf{BNF rule} & \textbf{\textsf{diagram}} & \textbf{(a layout)} \\[1mm]
$\varepsilon$ & (empty) & () & \centeralignimage{\includegraphics[scale=0.9]{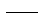}} \\[1mm]
\literal{a} & \literal{a} & "a" & \centeralignimage{\includegraphics[scale=0.9]{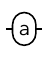}} \\[4mm]
(n/a) & <\literal{a}> & [a] & \centeralignimage{\includegraphics[scale=0.9]{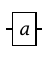}} \\[4mm]
$r \cdot s$ & $r$ $s$ & ((rrd $r$) (rrd $s$)) & \centeralignimage{\includegraphics[scale=0.9]{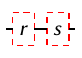}} \\[4mm]
$r\:|\:s$ & $r\:|\:s$ & \parbox[m]{6em}{(+ (rrd $r$) \\\hspace*{1em} (rrd $s$))} & \centeralignimage{\includegraphics[scale=0.9]{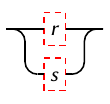}} \\[8mm]
$r^*$ & (n/a) & (- () (rrd $r$)) & \centeralignimage{\includegraphics[scale=0.9]{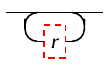}}
\end{tabular}
\medskip
\caption{Translating nonempty regular expressions and BNF rules to railroad diagrams. $r, s$ are nonempty regular expressions or rules, and \literal{a} is any literal symbol or rule name. Dashed red boxes denote recursive translations.}
\label{tab:translate-rrd}
\end{subfigure}\hfill\begin{subfigure}[b]{0.38\linewidth}
\centering
\begin{tabular}{p{14em}}
\centering
$\literal{[} \cdot \big(\varepsilon \:|\: \variable{item} \cdot (\literal{,} \cdot \variable{item})^*\big) \cdot \literal{]}$
\\[-1mm]
\topalignimage{\includegraphics[scale=0.83]{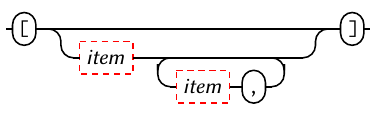}} \\[5mm]
{\ttfamily\raggedright\fontsize{10}{10}\selectfont\verb'  list  := [ <items>? ]'\par
\verb'  items := <item>'\par
\verb'         | <item> , <items>'\par
\verb' '} \\[-1mm]
\hspace{-7.8em}\variable{list}\par\vspace{-2.5mm}
\topalignimage{\includegraphics[scale=0.83]{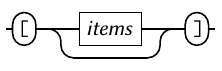}} \\[2mm]
\hspace{-8em}\variable{items}\par\vspace{-4mm}
\topalignimage{\includegraphics[scale=0.83]{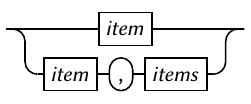}}
\end{tabular}
\smallskip
\caption{Syntax of a JSON list, as a regular expression and in BNF, and the corresponding diagrams. Parentheses are only to indicate order of operations.}
\label{tab:json-list}
\end{subfigure}
\caption{Translating regular expressions and rules in Backus-Naur form to railroad diagrams.}
\end{figure}

\subsection{Regular expression and Backus-Naur frontends}
\label{sec:frontends}

A recursive translation from nonempty regular expressions to our diagram language is given in \autoref{tab:translate-rrd}. (It is identical to the construction in~\cite{HinzeSelfcertifyingRailroadDiagrams2019}.)
An empty regular expression (i.e.~0, the empty language) cannot be drawn;
to draw an arbitrary regular expression, 0s must first be eliminated using standard equivalences like $r \cdot 0 = 0$.

For a grammar specified in Backus-Naur form (BNF), the right-hand side of each rule is translated to a railroad diagram.
A rule can refer to another of name \literal{a} with the syntax <\literal{a}>, represented as a nonterminal token \literal{[a]}.
The rest of the translation is also in \autoref{tab:translate-rrd}.

Although technically sufficient, the treatment of iteration/recursion in both translations above is a little unsatisfying.
Consider the definition of JSON list syntax as a regular expression and in BNF in \autoref{tab:json-list}, and the translation to railroad diagrams, assuming a list item is separately defined.
In the regular expression diagram, we may prefer \variable{item} to appear only once, as the top of the negative stack;
similarly, we may prefer to make the recursion explicit in the BNF diagram as a negative stack.
But we may prefer this transformation \emph{not} to happen if there are several kinds of lists, and it just happens to be the case in the one depicted above that the first item is of the same kind as the rest.
We cannot accommodate such preferences merely by adding rules to our translation table.
Rather, seen as a notation for syntax diagrams, both regular expressions and BNF have a conflict between \emph{canonicity} -- objects are equivalent iff their representations are equal -- and \emph{idiomaticity} -- the ability to conveniently express common representation idioms, like negative stacks for recursion.
(This terminology was introduced by~\cite{ChiplunkarPit-ClaudelDiagrammaticNotationsInteractive2023}.)

Our diagram language, together with our approach to its compilation, is idiomatic, as it is specifically designed to express railroad diagrams under common constraints.
However, it is not canonical: nested positive stacks can be reassociated without affecting the rendering, and the diagram \mbox{\literal{(- (- \variable{D\tss{1}} \variable{D\tss{2}}) \variable{D\tss{3}})}} can seemingly be rewritten to \mbox{\literal{(- \variable{D\tss{1}} (+ \variable{D\tss{2}} \variable{D\tss{3}}))}}.
The deeper problem is that railroad diagrams do not yet enjoy the solid theoretical foundations of other syntax notations, from either the theory of computation or graph automata, that would let us reason more precisely about rewriting, canonicity, and semantics in general.

\subsection{Manual layout in the wild}
\label{sec:evalmanual}
\begin{figure}[p]
\centering\vspace{-3mm}
\begin{minipage}[c]{0.59\linewidth}
\begin{subfigure}{\linewidth}
\centering
\topalignimage{\includegraphics[scale=0.65]{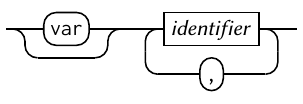}}
\caption{\autoref{fig:counterexamples:ill-nested}, but as a well-nested diagram.}
\end{subfigure}

\bigskip\medskip
\begin{subfigure}{\linewidth}
\addtocounter{subfigure}{1}
\raggedright
\topalignimage{\includegraphics[width=0.9\linewidth]{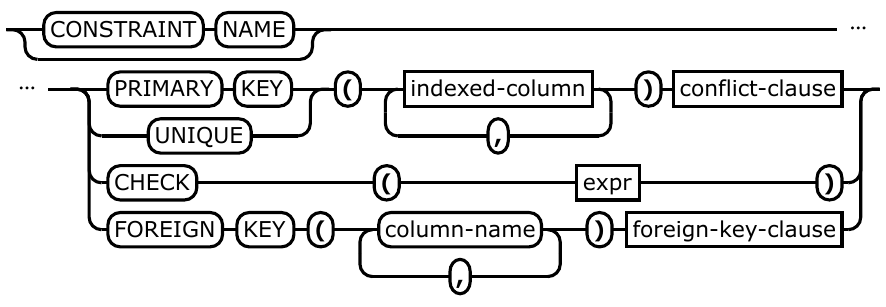}}
\caption{\autoref{fig:counterexamples:semantic-vertical}, but with a single semantic axis.}
\label{fig:approx-counter:sqlite}
\end{subfigure}
\end{minipage}\begin{minipage}[c]{0.41\linewidth}
\begin{subfigure}{\linewidth}
\addtocounter{subfigure}{-2}
\centering
\topalignimage{\includegraphics[scale=0.42]{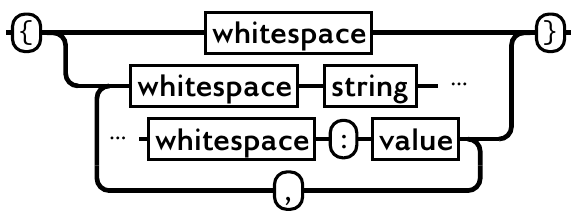}}
\caption{\autoref{fig:counterexamples:semantic-nesting}, but as a well-nested layout.}
\label{fig:approx-counter:json}
\end{subfigure}

\bigskip\medskip
\begin{subfigure}{\linewidth}
\addtocounter{subfigure}{1}
\centering
\topalignimage{\includegraphics[scale=0.65]{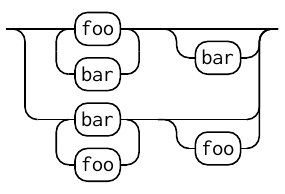}}
\caption{\autoref{fig:counterexamples:unseen}, but using common constructs.}
\end{subfigure}
\end{minipage}
\caption{Our approximations of the diagrams excluded from our scope in \autoref{fig:counterexamples}.}
\label{fig:approx-counter}
\end{figure}

\enlargethispage{-.6\baselineskip}
\autoref{fig:counterexamples} showed a few examples of diagrams outside our scope, chosen due to their broad variety of complex layout features, from corpuses that are often considered a reference by other illustrators to emulate.
In \autoref{fig:approx-counter}, we show our best approximation of each.
It would not make sense to directly compare rendering properties like width and height, because they depend as much on stylistic choices like fonts and spacing as on layout.
Neither would it make sense to compare wrap penalties, because we cannot infer the penalty functions implicitly expressed in those diagrams.
We can only argue that our layouts are comparably compact and readable with the additional merit of being automatically laid out.
Meanwhile, it is worth noting that the diagrams in \autoref{fig:counterexamples} (except~\ref{fig:counterexamples:unseen}) were laid out by hand to document very widely used syntaxes: Apple Pascal, JSON, and SQLite.
High-impact settings may motivate creators to invest in custom manual layout even when automated tools are available.

\begin{figure}[p]
\vspace*{3mm}
\centering
\topalignimage{\includegraphics[scale=0.15]{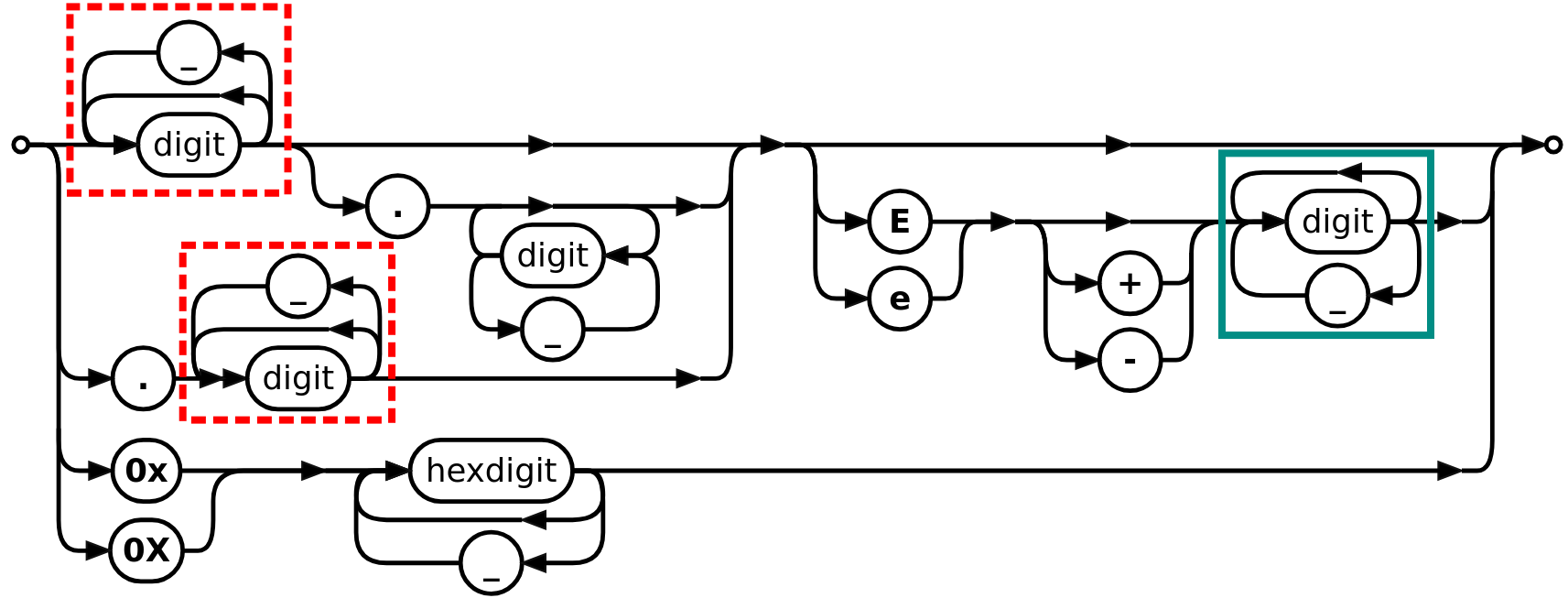}}
\caption{The two dashed red rectangles indicate identical layouts, but the bottom one has an extra arrowhead.
The solid blue rectangle indicates the same diagram as those two, but laid out differently, even when
laying them all out the same way would not have changed the overall height of the diagram.
(SQLite \literal{numeric-literal} diagram from~\cite{SQLitecontributorsSyntaxDiagramsSQLite2024}.)}
\label{fig:sqlite-inconsistency:both}
\end{figure}

\begin{figure}[p]
\vspace*{3mm}
\begin{subfigure}[b]{0.45\linewidth}
\centering
\includegraphics[scale=0.15]{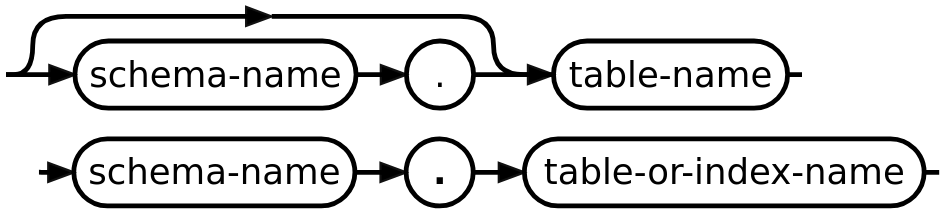}
\smallskip
\caption{An example of inconsistent rendering.
Punctuation terminals are generally in bold, but the one on top isn't.
(From \literal{alter-table-stmt}.)}
\label{fig:sqlite-inconsistency:stylistic}
\end{subfigure}
\hspace{3em}
\begin{subfigure}[b]{0.45\linewidth}
\centering
\includegraphics[scale=0.15]{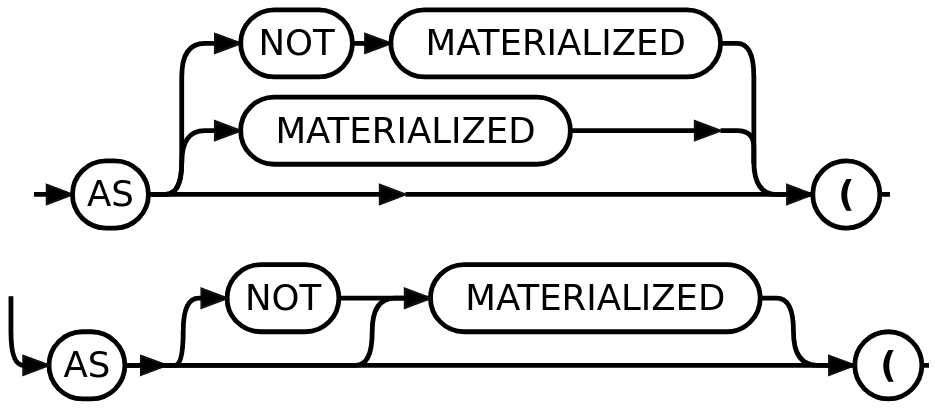}
\caption{An example of inconsistent layout of the same subdiagram in the same context.
(From \literal{common-table-expression} and \literal{with-clause}.)}
\label{fig:sqlite-inconsistency:layout1}
\end{subfigure}

\bigskip
\begin{subfigure}[t]{0.45\linewidth}
\centering
\includegraphics[scale=0.15]{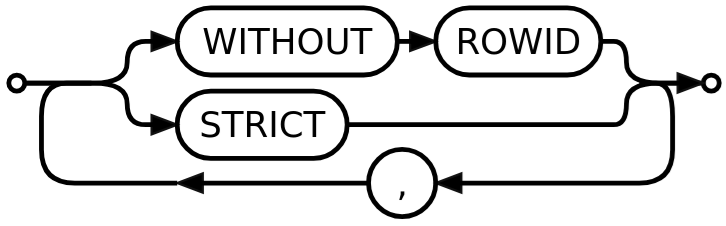}
\caption{\literal{table-options} is the only diagram of~71 that (vertically) center-aligns a component such that the tips are not aligned with a logical row.}
\label{fig:sqlite-inconsistency:layout2}
\end{subfigure}
\hspace{3em}
\begin{subfigure}[t]{0.45\linewidth}
\centering
\includegraphics[scale=0.13]{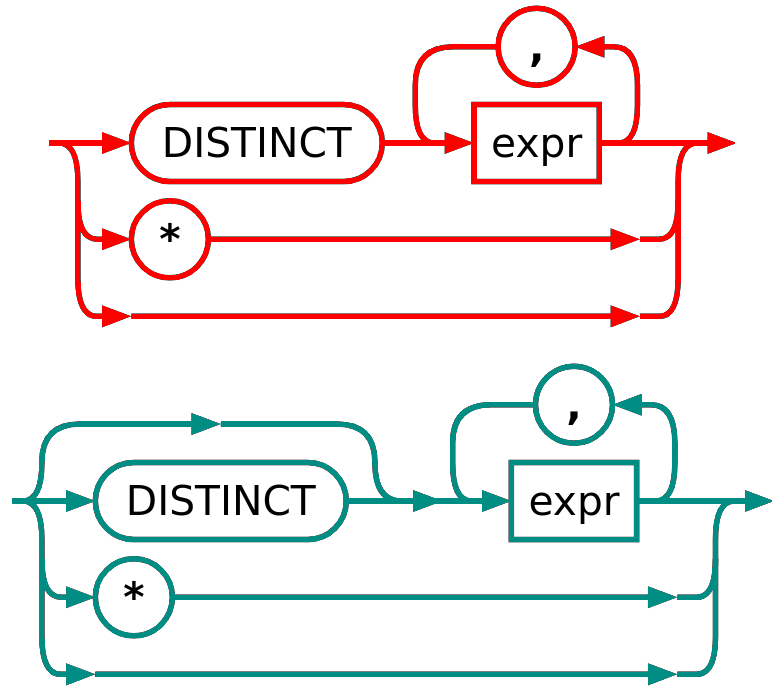}
\caption{Diff of SQLite \literal{expr} diagram in check-in \literal{12118bdbb8}: ``Fix the expression syntax diagram. DISTINCT is not required […]''  In this case, the disconnection between the source grammar and the diagram leads to a semantic bug.}
\label{fig:sqlite-error}
\end{subfigure}
\caption{Layout and rendering inconsistencies in SQLite syntax diagrams~\cite{SQLitecontributorsSyntaxDiagramsSQLite2024}.}
\label{fig:sqlite-inconsistency}
\end{figure}

SQLite has made a particular effort to maintain high-quality railroad diagrams documenting its complete input syntax over many years, which makes for an instructive case study.
A brief history (from check-ins \href{https://www.sqlite.org/docsrc/info/289df32643f94b1c}{\texttt{289df32643}} and \href{https://www.sqlite.org/docsrc/info/9f5383c82461ffe9}{\texttt{9f5383c824}}):
the diagrams were first introduced in~2008,
using Tcl~scripts to render from a railroad layout DSL.
In~2020,
striving for ``additional flexibility in the formatting of syntax diagrams [to make them] easier to read and understand [and] maintain'',
the project moved to directly specifying \emph{renderings} in Pikchr, a modernized implementation of the PIC graphics language~\cite{PikchrcontributorsHowPikchrGenerates2024,KernighanPICGraphicsLanguage1991}.
The manual renderings indeed have many features the automatic ones did not: custom alignment and justification, more wrapping and collapsing, etc.
But, being hand-written, they sometimes introduced stylistic inconsistencies (e.g.~\autoref{fig:sqlite-inconsistency:both}, \autoref{fig:sqlite-inconsistency:stylistic}).
Nevertheless, in this tricky tradeoff between manual layout and manual rendering, the creators evidently chose the latter.

Our approach is an appealing third option in such a tradeoff space.
Not only would it recover the benefits of automatic rendering for stylistic consistency, but it would also guarantee \emph{layout consistency}, going beyond both manual layout and manual rendering.
For instance, our notion of alignment would avoid the inconsistencies in \autoref{fig:sqlite-inconsistency:both}, \autoref{fig:sqlite-inconsistency:layout1}, and \autoref{fig:sqlite-inconsistency:layout2} in current SQLite documentation.
(We have reported the issues and submitted a patch fixing them.)
Moreover, the translations presented in~\autoref{sec:frontends} open up the possibility of generating documentation diagrams for a language from the same standard syntax notation its parser uses, avoiding issues like in \autoref{fig:sqlite-error}.
These benefits would come at only a small cost: out of SQLite's 71~syntax diagrams,
\begin{itemize}
\item 46~can be expressed in our layout language, of which 41~perfectly and 5~differing only in permitted collapsing;
\item 20~are ill-nested layouts that can still be expressed in our diagram language and compiled to comparable layouts; and
\item 5~are ill-nested diagrams that would have to be rewritten to equivalent well-nested ones.
\end{itemize}
\smallskip
One of the 20~from the second category above is shown in \autoref{tab:sqlite-evolution}, along with its specification in the pre-2020 SQLite DSL, our diagram language, and Pikchr.
(Its pre-2020 layout was perfectly expressible in our layout language but its modern one is not, due to a vertical semantic~ε.)
On one hand, the rendering from our prototype compiler is not significantly less readable or compact, and it corresponds to just one of many possible configurations for wrapping, alignment, and justification.
On the other, our specification is the same size as the pre-2020 manual layout, and simpler, as it does not use layout-specific constructors like \literal{stack} (for wrapping) or \literal{optx} (for alignment), instead leaving those decisions to the compiler.
To be clear, we do not aim to cast our prototype as a surefire replacement for SQLite's current diagrams, but rather as a reasonable alternative in contexts with similar demands for consistency, flexibility, and maintainability.

\begin{table}[p]
\centering
\caption{SQLite \literal{create-table-stmt} diagram in the pre-2020 DSL (\href{https://www.sqlite.org/docsrc/file?name=art/syntax/bubble-generator-data.tcl\&ci=bd9cdee968a7bcc0}{check-in \literal{bd9cdee968}}), our diagram language, and current Pikchr syntax (\href{https://www.sqlite.org/docsrc/file?name=art/syntax/create-table-stmt.pikchr\&ci=e668c1ce28f03f19}{check-in \literal{e668c1ce28}}).
Note that wrapping is specified explicitly with \literal{stack} in the old DSL, but done automatically in our tool, and our \literal{+} constructor is variously expressed as \literal{or}, \literal{opt}, or \literal{optx} in the old DSL, depending on the desired layout.
(The pre-2020 \literal{"WITHOUT"} \literal{"ROWID"} was replaced by \literal{[table-options]} as the language and documentation evolved over time.)}
\setlength{\tabcolsep}{0pt}
\begin{tabular}{>{\fontsize{8pt}{10pt}\selectfont}m{0.5\linewidth}c}
\begin{verbatim}
{stack
  {line CREATE {or {} TEMP TEMPORARY} TABLE
    {opt IF NOT EXISTS}}
  {line {optx /schema-name .} /table-name}
  {or {line ( {loop column-def ,}
        {loop {} {, table-constraint}} )
        {opt WITHOUT ROWID}}
    {line AS select-stmt}}}
\end{verbatim} & \centeralignimage{\includegraphics[width=0.5\linewidth]{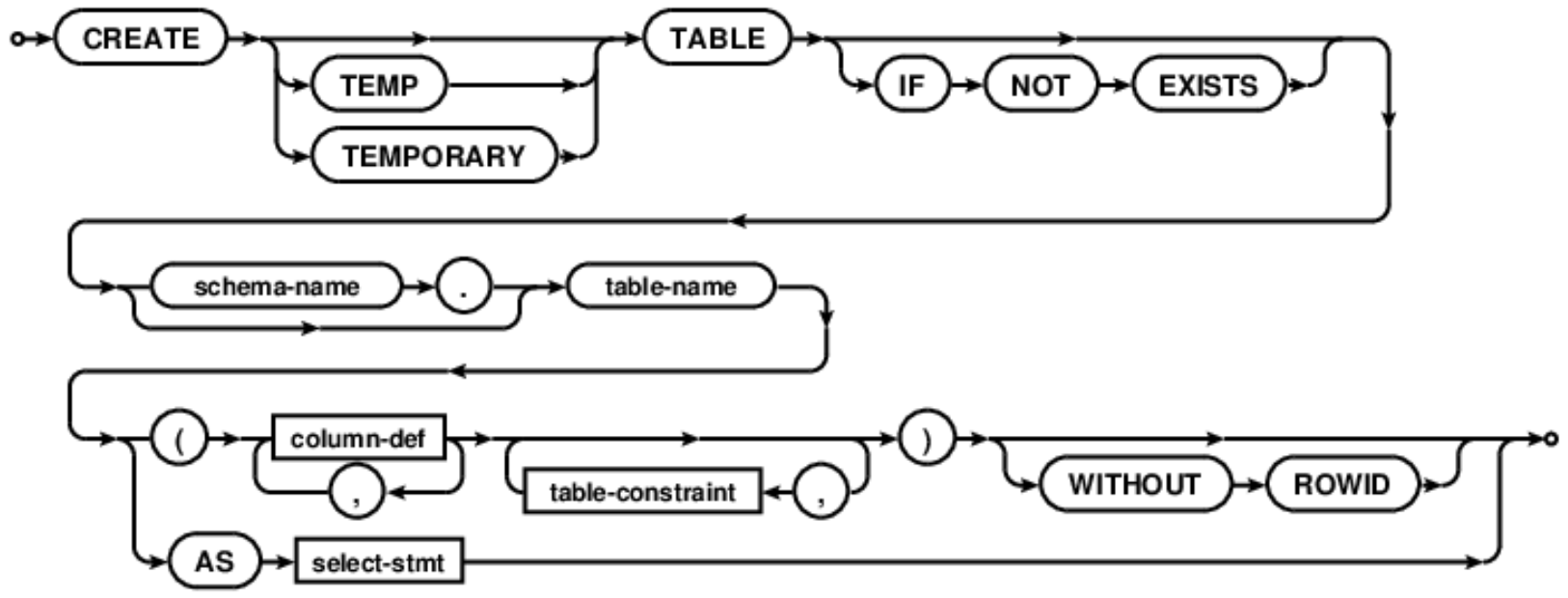}} \\[2mm]
\begin{verbatim}
("CREATE"
 (+ (+ () "TEMP") "TEMPORARY") "TABLE"
 (+ ("IF" "NOT" "EXISTS") ())
 (+ ("schema-name" ".") ()) "table-name"
 (+ ("AS" [select-stmt])
    ("(" (- [column-def] ",")
     (- () ("," [table-constraint])) ")"
     (+ () [table-options]))))
\end{verbatim} & \centeralignimage{\includegraphics[width=0.5\linewidth]{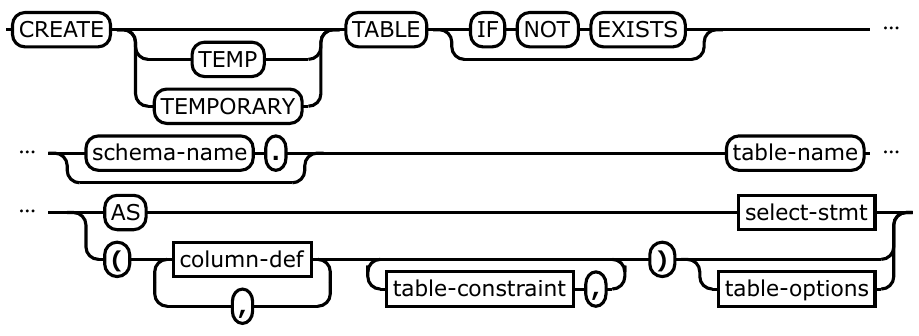}} \\[2mm]
{\fontsize{4pt}{4pt}\selectfont\begin{verbatim}
     linerad = 10px
     linewid *= 0.5
     $h = 0.21

     circle radius 10%
A0:  arrow 2*arrowht
CR:  oval "CREATE" fit
T1:  oval "TEMP" fit with .w at (linewid right of CR.e,.8*$h below CR)
T2:  oval "TEMPORARY" fit with .w at 1.25*$h below T1.w
TBL: oval "TABLE" fit with .w at (linewid right of T2.e,CR)
     arrow from CR.e right even with T2; arrow to TBL.w
     arrow from CR.e right linerad then down even with T1 then to T1.w
     arrow from CR.e right linerad then down even with T2 then to T2.w
     line from T2.e right linerad then up even with TBL \
        then to arrowht left of TBL.w
     line from T1.e right even with linerad right of T2.e then up linerad
     arrow from TBL.e right
     oval "IF" fit
     arrow right 2*arrowht
     oval "NOT" fit
     arrow 2*arrowht
ETS: oval "EXISTS" fit

     # IF NOT EXISTS bypass
Y1:  .5*$h below T2.s  # vertical position of back-arrow
     arrow from TBL.e right linerad then down even with Y1 then left even with T2
     arrow from ETS.e right linerad then down even with Y1 \
        then left even with ETS.w
     line left even with TBL.w
\end{verbatim}}\par
\textbf{\fontsize{10pt}{12pt}\selectfont\qquad (62 more hand-written lines…)} & \centeralignimage{\includegraphics[width=0.5\linewidth]{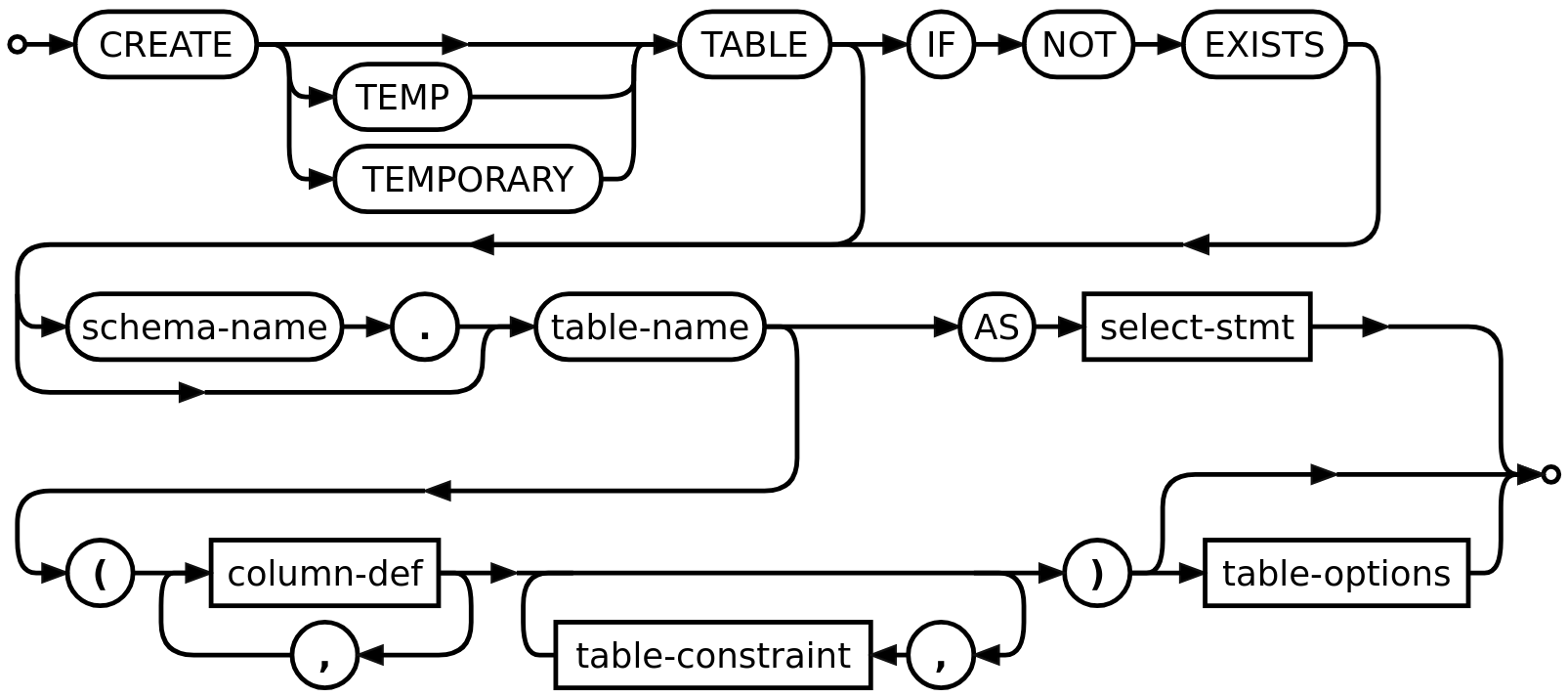}}
\end{tabular}
\medskip
\label{tab:sqlite-evolution}
\end{table}

\begin{table}[p]
\centering
\caption{Layout time statistics for a few sample diagrams from existing documentation~\cite{KrishnamurthySQLLanguageReference2025,SQLitecontributorsSyntaxDiagramsSQLite2024,CrockfordJSONObjectSyntax2001}.
The last four are among the largest we have found.}
\begin{tabular}{lr<{\hspace{2.3em}}r<{\hspace{1.85em}}r<{\hspace{2.7em}}}
\textbf{Diagram} & \parbox[b]{6em}{\centering\textbf{\# of \\ constructors}}\hspace{-2.3em} & \parbox[b]{6.5em}{\centering\textbf{Mean layout \\ time (ms)}}\hspace{-1.85em} & \textbf{Std.\ dev.\ (ms)}\hspace{-2.7em} \\\hline
JSON list (\autoref{fig:approx-counter:json}) & 13 & 46.9 & 2.2 \\
SQLite \literal{table-constraint} (\autoref{fig:approx-counter:sqlite}) & 34 & 54.4 & 1.8 \\
SQLite \literal{create-table-stmt} (\autoref{tab:sqlite-evolution}) & 38 & 54.1 & 1.0 \\
SQLite \literal{insert-stmt} (\autoref{fig:sqlite-insert}) & 74 & 67.6 & 2.8 \\
Oracle SQL \literal{CREATE TABLE} & 94 & 133.5 & 8.7 \\
SQLite \literal{select} & 96 & 75.0 & 1.8 \\
Oracle SQL \literal{ALTER INDEX} & 101 & 72.2 & 1.9 \\
SQLite \literal{expr} & 180 & 85.2 & 10.5
\end{tabular}
\label{tab:perf}
\end{table}

\subsection{Automatic layout in the wild}
\label{sec:evalauto}

\enlargethispage{1.9\baselineskip} 
Existing railroad diagram tools fall into two broad categories: \emph{rendering tools} that render layout descriptions, and \emph{grammar translation tools} that turn grammar notations into renderings.
We describe the capabilities and limitations of a few representatives of each.
In comparison, we argue that our system fills a niche in between, and can expand the capabilities of both.

Rendering tools define DSLs for specifying railroad layouts and compile these specifications to various backends.
\cite{ThibedeauSyntrax2019} cites the pre-2020 SQLite DSL as inspiration, and offers similar primitives for manual wrapping, manual alignment, etc.
The design of the following tools is largely similar: \cite{LuegRailroad2024,atp-miptJSyntrax2024,GroenUtfrailroad2020,DaenenRailroad2024,TakadaRailroadDiagrams2025,HollemansRailroadDiagramsSwift2019}.
One that stands out is~\cite{AtkinsRailroaddiagramGenerator2024}, for two reasons:
(i)~its layout DSL is considerably larger, with rare and complex constructs such as \autoref{fig:counterexamples:unseen}; and
(ii)~it enables a greater range of custom styling by letting users tag layout components to style their output SVG counterparts with CSS.
These features, along with the quality of its renderings and its parallel implementations in two popular languages (Python and JavaScript), may explain why several other tools build upon it (e.g.~\cite{LopezPyRailroad2024,KraussRRDANTLR42021,LischkeVSCodeANTLR42024,DundalekGrammKit2023,chevrotaincontributorsChevrotain2026}).
\cite{AbrechtRailroadCSS2023}, an in-progress work, tries to use CSS not just for styling but also for rendering, with an XML-like layout language directly embedded in a webpage.
Lastly, \cite{KlockerSyngen1996,PieperEtAlSyntaxdi2020,BarthelmannRooijakkersRail1998,WoodingSyntaxmdw1996} have a more limited range of output but work within the constraints of the \LaTeX\ and TikZ/PGF environments.

\begin{figure}[t]
\centering
\begin{subfigure}[t]{0.45\linewidth}
\includegraphics[width=\textwidth]{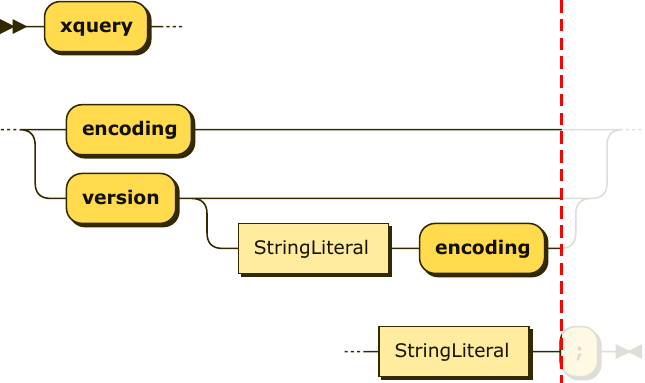}
\end{subfigure}
\hspace{3em}
\begin{subfigure}[t]{0.38\linewidth}
\includegraphics[width=\textwidth]{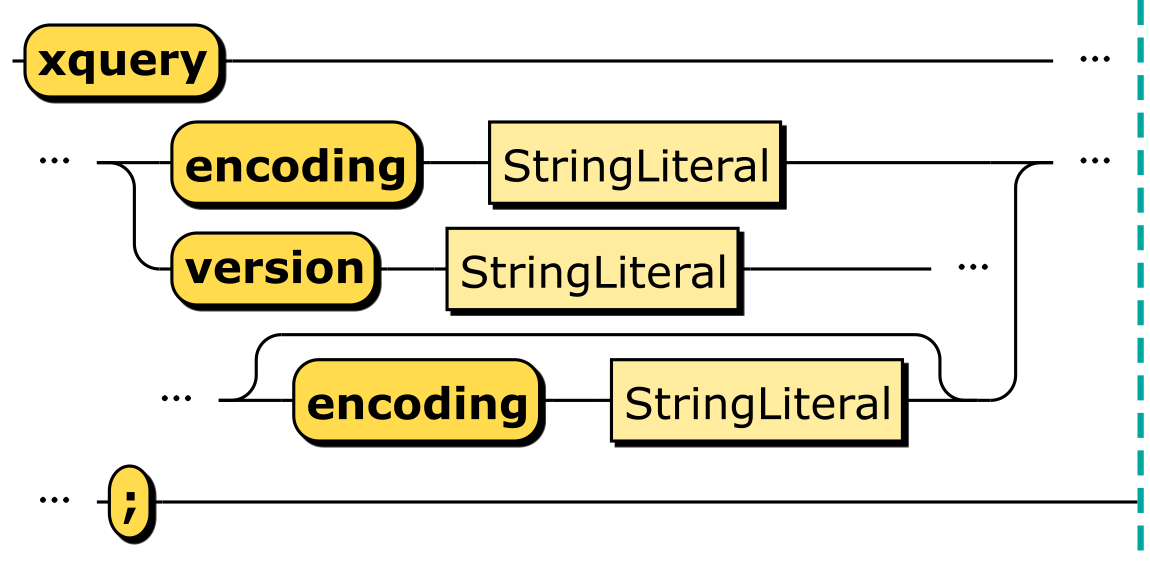}
\end{subfigure}
\caption{A diagram laid out with a state-of-the-art grammar translation tool~\cite{RademacherRr2025} (left) and our tool (right), both with target width 360~px.
\cite{RademacherRr2025} cannot wrap internally and hence overflows.
Note also that~\cite{RademacherRr2025} has a fixed layout style, while ours shows just one of many possible choices for justification, fonts, etc.
Lastly, we remark that~\cite{RademacherRr2025} automatically factors out the rightmost \literal{StringLiteral}.
This might improve readability for some rules in a grammar, but the tool offers no finer control than ``always'' or ``never'', illustrating the canonicity-idiomaticity conflict (\autoref{sec:frontends}).}
\label{fig:autoeval-rademacher}
\end{figure}

Grammar translation tools take descriptions of \emph{grammars} instead of layouts as input.
\cite{AvalloneRegexper2020}~renders from JavaScript-style regular expressions,
\cite{RademacherRr2025}~from EBNF,
and~\cite{AndersonPickettSvgsyntaxdiagrams2019} (e.g.~\autoref{fig:wild:ibm}) from a syntax representation fragment of the XML-based DITA~markup language~\cite{EberleinAndersonDarwinInformationTyping2018}.
In all grammar translators, the conflict between canonicity and idiomaticity that we described in~\autoref{sec:frontends} is evident.
For example, \cite{RademacherRr2025}~implements additional translation rules to better handle recursion, as well as some inlining of nonterminals and factoring of common subexpressions, but these are coarse-grained settings that affect all rules in a grammar (see \autoref{fig:autoeval-rademacher}).
\cite{AvalloneRegexper2020}~supports a variety of regex constructs, like ranged quantifiers, but does not produce nonempty loops at all.
Many other tools run into the same conflict -- \cite{DeckersRRDiagram2024,KirpichenkoVirtDiagramsDSL2022,HydeClapham2009,DoplerSchorgenhumerEBNFVisualizer2005,ThiemannEbnf2psPeterSyntax2010,PyeattEbnf2tikz2023,VoglspergerEBNFRailroadVisualizer2024,FlavelKatesGrammarTool2025,JacquesDrawGrammar2018,BrissonMethodsSystemConverting1997a,SnyderSpecifyingTextualGraphical1991} -- and some choose to introduce layout-specific annotations (e.g., to explicitly notate loops) into their otherwise grammar-focused notations, while others have a fixed translation outside the user's control.
\cite{VoglspergerEBNFRailroadVisualizer2024}~is unique in letting users click to expand and collapse nonterminals inline, making it the only tool we have seen that uses a dynamic rendering format.

We found only four existing tools that perform automatic wrapping of any kind.
\cite{AndersonPickettSvgsyntaxdiagrams2019} and~\cite{RademacherRr2025} use a greedy algorithm, and~\cite{WoodingSyntaxmdw1996}~co\"opts \TeX{}'s line-breaking;
none perform internal or nested wrapping or have any user-facing parameters besides the target width, as illustrated in \autoref{fig:autoeval-rademacher}.
\cite{BrissonMethodsSystemConverting1997a}~takes a different approach, with an algorithm for ``compressing'' diagrams translated from BNF by inlining or abstracting grammar rules in a diagram according to the available line width.

Three previous attempts at formalizing railroad layout have had varying results.
The first, \cite{HerriotStructuredSyntaxDiagrams1976}, proposed to address perceived flaws in conventional diagrams with an alternative ``structured'' diagrammatic notation, which seems to not have caught on.
The second, \cite{SnyderSpecifyingTextualGraphical1991}, framed the layout problem as compilation from EBNF to a custom rendering language, all declaratively specified in Prolog.
Despite the limitations of graphical environments of its time, the formal treatment quickly surfaces many finer properties of railroad layout such as nested wrapping and alignment, which we study in the present work.
More recently, \cite{BannisterEtAlConfluentOrthogonalDrawings2015} introduced a graph-based formalism for railroad diagrams and a corresponding layout algorithm that combines layered layout with railroad-specific heuristics.
However, the resulting layouts do not always look conventional, due to edge crossings, unnatural edge routing, etc.

Our approach starts from a specification of neither a layout nor a grammar, but a diagram, and compiles it to a layout independent of rendering.
Grammar notations are intended for describing formal languages, not their visual representations, while previous layout DSLs are tightly coupled with rendering and have no definite notions of diagrammatic equivalence;
our diagram language is the missing intermediate representation.
The way we characterize the compilation problem lets us implement at least four features that no other tool has:
\begin{itemize}
\item parametric wrapping, either globally or locally;
\item independent alignment of the two sides of a sublayout, e.g.~to collapse one or both sides of a stack automatically in valid contexts, such as happens several times in \autoref{fig:approx-counter:sqlite};
\item direction-aware justification; and
\item wrapping and justification that account for the possible internal wrapping of sublayouts.
\end{itemize}
We posit that our layouts can easily be rendered with additional backends or features from the rendering tools above, and that our diagram language can be an idiomatic target for grammar translation tools that would let them address canonicity independently of layout.

\subsection{Performance}
\label{sec:performance}

We measured how long our prototype compiler took to lay out a few sample diagrams with the same parameters that produced the layout of each shown in this paper.
We also measured a few of the largest diagrams we found in the SQLite and Oracle SQL documentation~\cite{KrishnamurthySQLLanguageReference2025,SQLitecontributorsSyntaxDiagramsSQLite2024}.
For each diagram, we computed the mean and standard deviation over 10~runs, in milliseconds.
We measured layout time on a laptop with an i7-1265U processor at 2.7~GHz and 32~GB of memory, running the Scala.js~1.19.0 compiler and the Node.js~20.10.0 runtime under Ubuntu~22.04.5.
\autoref{tab:perf} shows the results.

As a point of comparison, we cite Penrose~\cite{YeEtAlPenroseMathematicalNotation2020}, an optimization-based mathematical diagramming tool intended for iterative, exploratory use.
Our tool, despite being written with no explicit attention to performance, is well under the 500~ms limit Penrose considers the target for ``data visualization, live programming, and other exploratory creative tools''.

\begin{figure}
\centering
\includegraphics[width=0.8\linewidth]{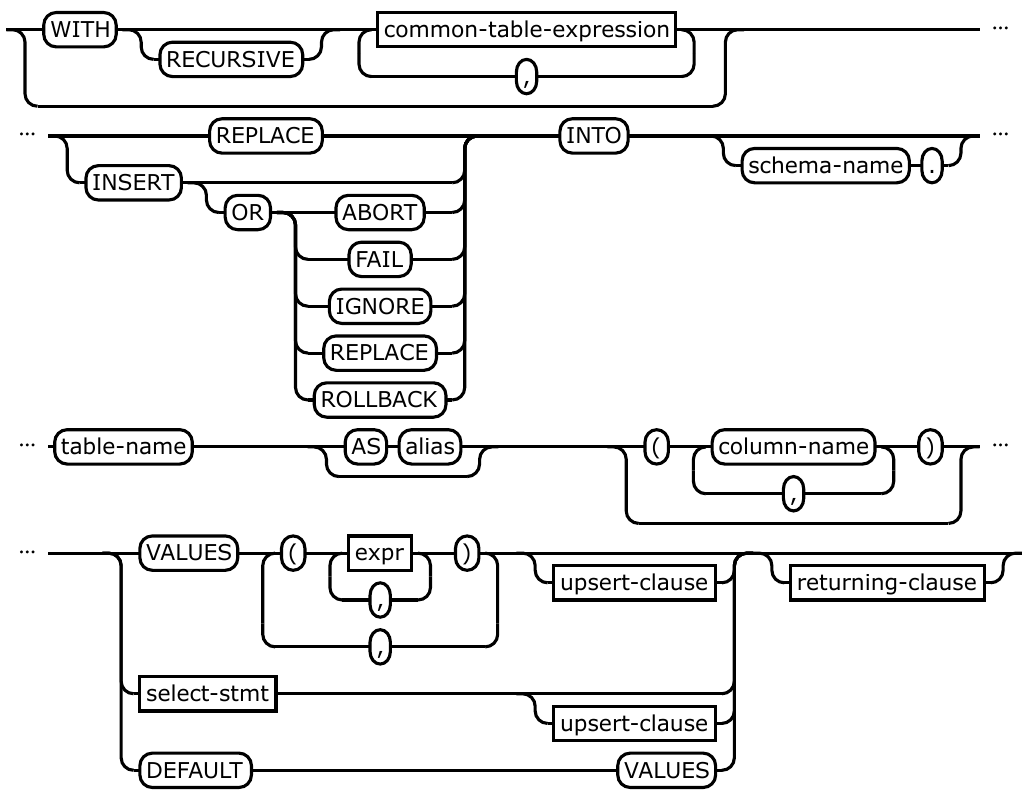}
\caption{Our reproduction of a large SQLite diagram, for \literal{insert-stmt}~\cite{SQLitecontributorsSyntaxDiagramsSQLite2024}.}
\label{fig:sqlite-insert}
\end{figure}

\section{Related work}
\label{sec:related}

In~\autoref{sec:introduction}, we described how railroad layout is a \emph{1.5-dimensional} problem falling outside the scope of conventional 1- and 2-dimensional approaches to layout.
In this section, we briefly survey those approaches to expand on that idea.

\enlargethispage{1.9\baselineskip} 
A classic 1-dimensional layout problem is pretty-printing.
Starting from~\cite{OppenPrettyprinting1980}, a steady stream of research -- including~\cite{BernardyPrettyNotGreedy2017,HughesDesignPrettyprintingLibrary1995,KiselyovEtAlLazyYieldIncremental2012,PughSinofskyNewLanguageindependentPrettyprinting1987,SwierstraChitilLinearBoundedFunctional2009,WadlerPrettierPrinter2003,BonichonWeisFormatUnraveled2017}, and most recently~\cite{PorncharoenwaseEtAlPrettyExpressivePrinter2023} -- has explored the space of document specification languages, optimality criteria, algorithmic efficiency, correctness, and programming techniques.
Like 1-dimensional pretty-printed code, components of a railroad diagram follow the reading order, wrap, and reflect the nested structure of the source term, but pretty-printers crucially cannot express the branching and remerging of rows that independently follow the reading order, nor the reversed layout in loops.

Some 1-dimensional layouts work with flexible boxes instead of rigid characters and spaces.
\cite{KnuthPlassBreakingParagraphsLines1981}, designed for \TeX, uses boxes to line-wrap, justify, and hyphenate a paragraph of text, with a practical heuristic for a global optimization problem;
however, it gives no special consideration to nested structure beyond lines, words, and characters.
CSS flexbox~\cite{W3CCSSFlexibleBox2018} solves the same problem for arbitrary boxes, possibly with nested layouts, but (unlike our algorithm) does not account for the differing stretchability of boxes during wrapping or justification, and uses a simple greedy wrapping algorithm.
Neither \TeX\ nor flexbox supports independent left and right baselines for boxes as is needed for railroad layout.

Graph layout is perhaps the best-studied 2-dimensional layout problem.
\cite{TamassiaHandbookGraphDrawing2013} surveys the research behind powerful widely-used tools like Graphviz~\cite{GansnerNorthOpenGraphVisualization2000} and the algorithms in D3~\cite{BostockEtAlD3DatadrivenDocuments2011}.
Railroad diagrams could be seen as labeled directed graphs with stylized 2-dimensional layouts, but are much more rigid and restricted than the conventional presentations.
Arbitrary mathematical graphs do not have an inherent reading order like code or text;
many layout algorithms, like force-directed layout, are accordingly rotationally symmetric.
Some algorithms designed for graphs with additional structure, like trees and DAGs, favor one direction over the other.
Graphs that come closest to railroad diagrams are \emph{two-terminal series-parallel graphs} or \emph{SP graphs} -- graphs that are either just two vertices with an edge, or a series or parallel composition of other series-parallel graphs~\cite{BrandstadtEtAlAlgebraicCompositionsRecursive1999} -- and the yWorks graph drawing tool~\cite{yWorksYFilesHTML2025} offers a series-parallel layout visually similar to railroad layout, but performs no wrapping.
A more fundamental difference, however, is that series-parallel graphs are defined to be either undirected or directed but acyclic~\cite{AlurEtAlRobustTheorySeries2023}, whereas railroad layouts specifically have loops, and hence more constraints on collapsing.

\enlargethispage{\baselineskip}
Many other 2-dimensional layout problems and tools are further afield from railroad diagrams.
A large category is statistical graphs and data-driven visualizations, starting with~\cite{WilkinsonGrammarGraphics2005}'s seminal work on statistical graphs, inspiring libraries like ggplot2~\cite{WickhamGgplot2ElegantGraphics2016}, Vega and Vega-Lite~\cite{HeerBostockDeclarativeLanguageDesign2010,SatyanarayanEtAlVegaLiteGrammarInteractive2017}, D3~\cite{BostockEtAlD3DatadrivenDocuments2011}, and Matplotlib~\cite{Hunter:2007}.
Another category is graphics programming, i.e.~languages and tools to specify individual pictures with low-level constructs, like PIC~\cite{KernighanPICGraphicsLanguage1991}, SVG~\cite{W3CScalableVectorGraphics2011}, TikZ~\cite{TantauTikZPGFPackages2023}, and Bluefish~\cite{PollockEtAlBluefishComposingDiagrams2024}.
Some domain-specific visualization tools provide a higher level of abstraction for narrower domains, like Penrose~\cite{YeEtAlPenroseMathematicalNotation2020}, Mermaid~\cite{MermaidcontributorsMermaidDiagrammingCharting}, TikZ libraries and frontends like TikZiT~\cite{KissingerTikZiT2020}, and program state visualizations like PythonTutor~\cite{GuoOnlinePythonTutor2013} and DDD~\cite{ZellerLutkehausDDDFreeGraphical1996}.
Lastly, frameworks like Lean Widgets~\cite{NawrockiEtAlExtensibleUserInterface2023} and Alectryon~\cite{Pit-ClaudelUntanglingMechanizedProofs2020} enable the use of existing visualization tools in the context of computerized proofs.

Our definition of ``1.5-dimensional'' layout problems not only reflects their intermediate nature between the 1- and 2-dimensional ones described above, but also aligns with previous uses of the term.
\cite{BurchEtAlIndentedPixelTree2010} lays out sibling nodes in a tree following a left-to-right reading order, while a child node is drawn under its parent.
For~\cite{ShiEtAl15DEgocentricDynamic2015}, there is only one top-to-bottom sublayout, while the rest of the graph is laid out freely in 2 dimensions.
Neither considers the possibility of wrapping, unlike~\cite{RueggvonHanxledenWrappingLayeredGraphs2018}, which treats layered layouts as having a reading order, albeit without nesting.
Lastly, the term ``1.5-dimensional'' is also used in cutting and packing (C\&P) problems, but they are fundamentally different from the other layout problems outlined above because items have no structure or relationships that their layout must respect~\cite{DyckhoffTypologyCuttingPacking1990}.
C\&P use of the term is thus most properly seen as coincidental.

\section{Conclusion}

Railroad diagrams are a common visualization of grammars, but limited tooling and a lack of formal attention to their layout has mostly confined them to hand-drawn documentation.
In this paper, we presented the first formal treatment of railroad layout, along with a principled implementation that performs \emph{line wrapping} to meet a target width, as well as vertical \emph{alignment} and horizontal \emph{justification} per user-specified policies.
We presented a practical heuristic for the optimization problem of nested line wrapping.

We then showed how our approach to automatic railroad layout has practical value for the programming languages, formal methods, and software engineering community.
It can replace existing uses of (manually laid-out) railroad diagrams and enable new ones, including in interactive settings like proof assistants.
Both our layout language and our approach to compilation can be extended to accommodate further variation in hand-drawn layouts.

The unique nature of railroad layout as a 1.5-dimensional layout problem suggests a number of directions for future work.
On one hand, future work could explore whether railroad layout can subsume pretty-printing, and whether the elegant successes of techniques like program calculation for pretty-printers replicate.
On the other, a generalization of series-parallel graphs with reversed parallel composition could model railroad diagrams, and serve as a semantic foundation similar to automata for other grammar notations.

\bibliography{rrd}

@inproceedings{GuoOnlinePythonTutor2013,
  title = {Online {{Python Tutor}}: Embeddable Web-Based Program Visualization for {{CS}} Education},
  shorttitle = {Online Python Tutor},
  booktitle = {Proceedings of the 44th {{ACM Technical Symposium}} on {{Computer Science Education}}},
  author = {Guo, Philip J.},
  year = 2013,
  month = mar,
  series = {{{SIGCSE}} 2013},
  pages = {579--584},
  publisher = {Association for Computing Machinery},
  address = {New York, NY, USA},
  doi = {10.1145/2445196.2445368},
  urldate = {2023-03-31},
  isbn = {978-1-4503-1868-6}
}

@article{PanchekhaEtAlModularVerificationWeb2019,
  title = {Modular Verification of Web Page Layout},
  author = {Panchekha, Pavel and Ernst, Michael D. and Tatlock, Zachary and Kamil, Shoaib},
  year = 2019,
  month = oct,
  journal = {Proceedings of the ACM on Programming Languages},
  volume = {3},
  number = {OOPSLA},
  pages = {151:1--151:26},
  doi = {10.1145/3360577},
  urldate = {2023-04-09}
}

@article{OppenPrettyprinting1980,
  title = {Prettyprinting},
  author = {Oppen, Derek C.},
  year = 1980,
  month = oct,
  journal = {ACM Transactions on Programming Languages and Systems},
  volume = {2},
  number = {4},
  pages = {465--483},
  issn = {0164-0925},
  doi = {10.1145/357114.357115},
  urldate = {2023-04-10}
}

@article{BernardyPrettyNotGreedy2017,
  title = {A Pretty but Not Greedy Printer (Functional Pearl)},
  author = {Bernardy, Jean-Philippe},
  year = 2017,
  month = aug,
  journal = {Proceedings of the ACM on Programming Languages},
  volume = {1},
  number = {ICFP},
  pages = {6:1--6:21},
  doi = {10.1145/3110250},
  urldate = {2023-04-10}
}

@article{KnuthPlassBreakingParagraphsLines1981,
  title = {Breaking Paragraphs into Lines},
  author = {Knuth, Donald E. and Plass, Michael F.},
  year = 1981,
  month = nov,
  journal = {Software: Practice and Experience},
  volume = {11},
  number = {11},
  pages = {1119--1184},
  issn = {00380644, 1097024X},
  doi = {10.1002/spe.4380111102},
  urldate = {2023-04-10},
  langid = {english}
}

@incollection{WadlerPrettierPrinter2003,
  title = {A Prettier Printer},
  booktitle = {The {{Fun}} of {{Programming}}},
  author = {Wadler, Philip},
  editor = {Gibbons, Jeremy and {de Moor}, Oege},
  year = 2003,
  pages = {223--243},
  publisher = {Macmillan Education UK},
  address = {London},
  doi = {10.1007/978-1-349-91518-7_11},
  urldate = {2023-04-10},
  isbn = {978-0-333-99285-2 978-1-349-91518-7},
  langid = {english}
}

@inproceedings{HughesDesignPrettyprintingLibrary1995,
  title = {The Design of a Pretty-Printing Library},
  booktitle = {Advanced {{Functional Programming}}},
  author = {Hughes, John},
  editor = {Jeuring, Johan and Meijer, Erik},
  year = 1995,
  month = may,
  series = {Lecture {{Notes}} in {{Computer Science}}},
  volume = {925},
  pages = {53--96},
  publisher = {Springer-Verlag},
  address = {B\aa stad, Sweden},
  doi = {10.1007/3-540-59451-5_3},
  urldate = {2023-04-10},
  isbn = {978-3-540-49270-2}
}

@inproceedings{KiselyovEtAlLazyYieldIncremental2012,
  title = {Lazy v. {{Yield}}: Incremental, Linear Pretty-Printing},
  shorttitle = {Lazy v. {{Yield}}},
  booktitle = {Programming {{Languages}} and {{Systems}}},
  author = {Kiselyov, Oleg and {Peyton-Jones}, Simon and Sabry, Amr},
  editor = {Jhala, Ranjit and Igarashi, Atsushi},
  year = 2012,
  series = {Lecture {{Notes}} in {{Computer Science}}},
  pages = {190--206},
  publisher = {Springer},
  address = {Berlin, Heidelberg},
  doi = {10.1007/978-3-642-35182-2_14},
  isbn = {978-3-642-35182-2},
  langid = {english}
}

@article{YeEtAlPenroseMathematicalNotation2020,
  title = {Penrose: From Mathematical Notation to Beautiful Diagrams},
  shorttitle = {Penrose},
  author = {Ye, Katherine and Ni, Wode and Krieger, Max and Ma'ayan, Dor and Wise, Jenna and Aldrich, Jonathan and Sunshine, Joshua and Crane, Keenan},
  year = 2020,
  month = aug,
  journal = {ACM Transactions on Graphics},
  volume = {39},
  number = {4},
  issn = {0730-0301, 1557-7368},
  doi = {10.1145/3386569.3392375},
  urldate = {2023-04-10},
  langid = {english}
}

@techreport{PughSinofskyNewLanguageindependentPrettyprinting1987,
  title = {A New Language-Independent Prettyprinting Algorithm},
  author = {Pugh, William W. and Sinofsky, Steven J.},
  year = 1987,
  month = jan,
  number = {TR 87-808},
  institution = {Cornell University},
  url = {https://www.cs.tufts.edu/~nr/cs257/archive/steven-sinofsky/prettyprinting-87-808.pdf},
  urldate = {2023-04-10}
}

@article{SwierstraChitilLinearBoundedFunctional2009,
  title = {Linear, Bounded, Functional Pretty-Printing},
  author = {Swierstra, S. Doaitse and Chitil, Olaf},
  year = 2009,
  month = jan,
  journal = {Journal of Functional Programming},
  volume = {19},
  number = {1},
  pages = {1--16},
  issn = {0956-7968, 1469-7653},
  doi = {10.1017/S0956796808006990},
  urldate = {2023-04-10},
  langid = {english}
}

@misc{TantauTikZPGFPackages2023,
  title = {The {{TikZ}} and {{PGF Packages}}},
  author = {Tantau, Till},
  year = 2023,
  month = may,
  url = {www.ctan.org/pkg/pgf}
}

@article{GansnerNorthOpenGraphVisualization2000,
  title = {An Open Graph Visualization System and Its Applications to Software Engineering},
  author = {Gansner, Emden R. and North, Stephen C.},
  year = 2000,
  month = aug,
  journal = {Software: Practice and Experience},
  volume = {30},
  number = {11},
  pages = {1203--1233},
  issn = {1097-024X},
  doi = {10.1002/1097-024X(200009)30:11<1203::AID-SPE338>3.0.CO;2-N},
  url = {https://graphviz.org/documentation/GN99.pdf},
  urldate = {2023-06-20},
  copyright = {Copyright \copyright{} 2000 John Wiley \& Sons, Ltd.},
  langid = {english}
}

@inproceedings{PanchekhaTorlakAutomatedReasoningWeb2016,
  title = {Automated Reasoning for Web Page Layout},
  booktitle = {Proceedings of the 2016 {{ACM SIGPLAN International Conference}} on {{Object-Oriented Programming}}, {{Systems}}, {{Languages}}, and {{Applications}}},
  author = {Panchekha, Pavel and Torlak, Emina},
  year = 2016,
  month = oct,
  series = {{{OOPSLA}} 2016},
  pages = {181--194},
  publisher = {ACM},
  address = {Amsterdam, Netherlands},
  doi = {10.1145/2983990.2984010},
  urldate = {2023-06-20},
  isbn = {978-1-4503-4444-9}
}

@inproceedings{Pit-ClaudelUntanglingMechanizedProofs2020,
  title = {Untangling Mechanized Proofs},
  booktitle = {Proceedings of the 13th {{ACM SIGPLAN International Conference}} on {{Software Language Engineering}}},
  author = {{Pit-Claudel}, Cl{\'e}ment},
  year = 2020,
  month = nov,
  series = {{{SLE}} 2020},
  pages = {155--174},
  publisher = {ACM},
  address = {Virtual},
  doi = {10.1145/3426425.3426940},
  urldate = {2023-07-13},
  isbn = {978-1-4503-8176-5},
  langid = {english}
}

@article{BostockEtAlD3DatadrivenDocuments2011,
  title = {D3: Data-Driven Documents},
  author = {Bostock, Michael and Ogievetsky, Vadim and Heer, Jeffrey},
  year = 2011,
  month = nov,
  journal = {IEEE Transactions on Visualization \& Computer Graphics},
  volume = {17},
  number = {12},
  pages = {2301--2309},
  issn = {1941-0506},
  doi = {10.1109/TVCG.2011.185},
  url = {http://vis.stanford.edu/papers/d3}
}

@misc{MermaidcontributorsMermaidDiagrammingCharting,
  title = {Mermaid \textbar{} {{Diagramming}} and Charting Tool},
  author = {{Mermaid contributors}},
  url = {https://mermaid.js.org/},
  urldate = {2023-07-13},
  lastaccessed = {2023-07-13}
}

@misc{KissingerTikZiT2020,
  title = {{{TikZiT}}},
  author = {Kissinger, Aleks},
  year = 2020,
  month = aug,
  url = {https://tikzit.github.io/},
  urldate = {2023-07-13}
}

@inproceedings{NawrockiEtAlExtensibleUserInterface2023,
  title = {An Extensible User Interface for {{Lean}} 4},
  booktitle = {14th {{International Conference}} on {{Interactive Theorem Proving}} ({{ITP}} 2023)},
  author = {Nawrocki, Wojciech and Ayers, Edward W and Ebner, Gabriel},
  year = 2023,
  series = {Leibniz {{International Proceedings}} in {{Informatics}} ({{LIPIcs}})},
  pages = {24:1--24:20},
  publisher = {Schloss Dagstuhl -- Leibniz-Zentrum f\"ur Informatik},
  address = {Dagstuhl, Germany},
  doi = {10.4230/LIPIcs.ITP.2023.24},
  langid = {english}
}

@book{TamassiaHandbookGraphDrawing2013,
  title = {Handbook of Graph Drawing and Visualization},
  editor = {Tamassia, Roberto},
  year = 2013,
  edition = {1st},
  publisher = {Chapman \& Hall/CRC},
  address = {New York, NY, USA},
  url = {https://cs.brown.edu/people/rtamassi/gdhandbook/},
  isbn = {978-1-138-03424-2}
}

@techreport{KernighanPICGraphicsLanguage1991,
  title = {{{PIC}} --- a Graphics Language for Typesetting User Manual},
  author = {Kernighan, Brian W},
  year = 1991,
  month = may,
  number = {116},
  address = {Murray Hill, NJ 07974},
  institution = {AT\&T Bell Laboratories},
  url = {https://pikchr.org/home/uv/pic.pdf},
  langid = {english}
}

@book{WilkinsonGrammarGraphics2005,
  title = {The {{Grammar}} of {{Graphics}}},
  author = {Wilkinson, Leland},
  year = 2005,
  series = {Statistics and {{Computing}}},
  edition = {2nd},
  publisher = {Springer-Verlag},
  address = {New York, NY, USA},
  doi = {10.1007/0-387-28695-0},
  urldate = {2023-10-26},
  isbn = {978-0-387-24544-7},
  langid = {english}
}

@inproceedings{ChiplunkarPit-ClaudelDiagrammaticNotationsInteractive2023,
  title = {Diagrammatic Notations for Interactive Theorem Proving},
  booktitle = {4th {{International Workshop}} on {{Human Aspects}} of {{Types}} and {{Reasoning Assistants}} ({{HATRA}} 2023)},
  author = {Chiplunkar, Shardul and {Pit-Claudel}, Cl{\'e}ment},
  year = 2023,
  month = oct,
  doi = {10.5075/epfl-SYSTEMF-305144},
  url = {https://infoscience.epfl.ch/record/305144}
}

@inproceedings{HinzeSelfcertifyingRailroadDiagrams2019,
  title = {Self-Certifying Railroad Diagrams},
  booktitle = {Mathematics of {{Program Construction}}},
  author = {Hinze, Ralf},
  editor = {Hutton, Graham},
  year = 2019,
  series = {Lecture {{Notes}} in {{Computer Science}}},
  volume = {11825},
  pages = {103--137},
  publisher = {Springer International Publishing},
  address = {Cham},
  doi = {10.1007/978-3-030-33636-3_5},
  isbn = {978-3-030-33636-3},
  langid = {english}
}

@article{PiedeleuZanasiFiniteAxiomatisationFinitestate2023,
  title = {A Finite Axiomatisation of Finite-State Automata Using String Diagrams},
  author = {Piedeleu, Robin and Zanasi, Fabio},
  year = 2023,
  month = feb,
  journal = {Logical Methods in Computer Science},
  volume = {19},
  number = {1},
  pages = {13:1--13:38},
  issn = {1860-5974},
  doi = {10.46298/lmcs-19(1:13)2023},
  url = {https://lmcs.episciences.org/10963},
  urldate = {2024-06-26}
}

@techreport{WirthProgrammingLanguagePascal1970,
  type = {Report},
  title = {The Programming Language {{Pascal}}},
  author = {Wirth, Niklaus},
  year = 1970,
  eprint = {20.500.11850/68712},
  eprinttype = {hdl},
  address = {Z\"urich, Switzerland},
  institution = {ETH Z\"urich},
  doi = {10.3929/ethz-a-004150844},
  url = {http://hdl.handle.net/20.500.11850/68712},
  urldate = {2025-03-04},
  copyright = {http://rightsstatements.org/page/InC-NC/1.0/},
  langid = {english}
}

@misc{SteeleItsTimeNew2017,
  title = {It's Time for a New Old Language},
  author = {Steele, Jr., Guy Lewis},
  year = 2017,
  month = apr,
  address = {MIT, Cambridge, MA, USA},
  url = {https://groups.csail.mit.edu/mac/users/gjs/6.945/readings/Steele-MIT-April-2017.pdf},
  urldate = {2025-03-04},
  langid = {english},
  lastaccessed = {2025-03-04}
}

@techreport{W3CCSSFlexibleBox2018,
  type = {Candidate {{Recommendation}}},
  title = {{{CSS Flexible Box Layout Module Level}} 1},
  author = {{W3C}},
  editor = {Atkins, Jr., Tab and Etemad, Elika J. "fantasai" and Atanassov, Rossen},
  year = 2018,
  month = nov,
  institution = {W3C},
  url = {https://www.w3.org/TR/css-flexbox-1/},
  urldate = {2025-03-06}
}

@article{KozenKleeneAlgebraTests1997,
  title = {Kleene Algebra with Tests},
  author = {Kozen, Dexter},
  year = 1997,
  month = may,
  journal = {ACM Transactions on Programming Languages and Systems},
  volume = {19},
  number = {3},
  pages = {427--443},
  issn = {0164-0925},
  doi = {10.1145/256167.256195},
  urldate = {2025-03-06}
}

@article{PorncharoenwaseEtAlPrettyExpressivePrinter2023,
  title = {A Pretty Expressive Printer},
  author = {Porncharoenwase, Sorawee and Pombrio, Justin and Torlak, Emina},
  year = 2023,
  month = oct,
  journal = {Proceedings of the ACM on Programming Languages},
  volume = {7},
  number = {OOPSLA2},
  pages = {261:1122--261:1149},
  doi = {10.1145/3622837},
  urldate = {2025-03-12}
}

@misc{KamifujiRaskinApplePascalSyntax1979,
  title = {Apple {{Pascal}} Syntax Chart},
  author = {Kamifuji, Tom and Raskin, Jef},
  year = 1979,
  url = {http://www.danamania.com/print/Apple%20Pascal%20Poster/PascalPosterV3%20A1.pdf},
  urldate = {2025-03-12},
  howpublished = {Courtesy Dana Sibera ``NanoRaptor''}
}

@misc{CrockfordJSONObjectSyntax2001,
  title = {{{JSON}} Object Syntax},
  author = {Crockford, Douglas},
  year = 2001,
  url = {https://www.json.org/json-en.html}
}

@misc{SQLitecontributorsSyntaxDiagramsSQLite2024,
  title = {Syntax Diagrams for {{SQLite}}},
  author = {{SQLite contributors}},
  year = 2024,
  url = {https://www.sqlite.org/syntaxdiagrams.html}
}

@misc{AtkinsRailroaddiagramGenerator2024,
  title = {Railroad-Diagram Generator},
  author = {Atkins, Jr., Tab},
  year = 2024,
  month = jan,
  url = {https://github.com/tabatkins/railroad-diagrams/}
}

@incollection{BrandstadtEtAlAlgebraicCompositionsRecursive1999,
  title = {Algebraic Compositions and Recursive Definitions},
  booktitle = {Graph {{Classes}}: {{A Survey}}},
  author = {Brandst{\"a}dt, Andreas and Le, Van Bang and Spinrad, Jeremy P.},
  year = 1999,
  series = {Discrete {{Mathematics}} and {{Applications}}},
  pages = {167--185},
  publisher = {{Society for Industrial and Applied Mathematics}},
  address = {Philadelphia, PA, USA},
  doi = {10.1137/1.9780898719796.ch11},
  urldate = {2025-03-13},
  isbn = {978-0-89871-432-6},
  langid = {english}
}

@article{AlurEtAlRobustTheorySeries2023,
  title = {A {{Robust Theory}} of {{Series Parallel Graphs}}},
  author = {Alur, Rajeev and Stanford, Caleb and Watson, Christopher},
  year = 2023,
  month = jan,
  journal = {Proceedings of the ACM on Programming Languages},
  volume = {7},
  number = {POPL},
  pages = {1058--1088},
  issn = {2475-1421},
  doi = {10.1145/3571230},
  urldate = {2025-03-13},
  langid = {english}
}

@misc{AvalloneRegexper2020,
  title = {Regexper},
  author = {Avallone, Jeff},
  year = 2020,
  month = sep,
  url = {https://regexper.com}
}

@misc{RademacherRr2025,
  title = {{{rr}}},
  author = {Rademacher, Gunther},
  year = 2025,
  month = feb,
  url = {https://github.com/GuntherRademacher/rr/}
}

@misc{AndersonPickettSvgsyntaxdiagrams2019,
  title = {{{svg-syntaxdiagrams}}},
  author = {Anderson, Robert D. and Pickett, Deborah},
  year = 2019,
  month = oct,
  url = {https://github.com/robander/svg-syntaxdiagrams},
  howpublished = {IBM}
}

@techreport{IBMCorporationIBMMQReference2025,
  title = {{{IBM MQ}} 9.4 {{Reference}}},
  author = {{IBM Corporation}},
  year = 2025,
  month = jan,
  institution = {IBM Corporation},
  url = {https://www.ibm.com/docs/en/ibm-mq/9.4?topic=mq-reference}
}

@misc{EcmaJSONDataInterchange2017,
  title = {The {{JSON Data Interchange Syntax}}},
  author = {{Ecma}},
  year = 2017,
  number = {ECMA-404},
  publisher = {Ecma International},
  address = {Geneva, Switzerland},
  url = {https://ecma-international.org/publications-and-standards/standards/ecma-404/},
  langid = {english}
}

@techreport{W3CCSSBoxAlignment2025,
  type = {Working {{Draft}}},
  title = {{{CSS Box Alignment Module Level}} 3},
  author = {{W3C}},
  editor = {Etemad, Elika J. "fantasai" and Atkins, Jr., Tab},
  year = 2025,
  month = mar,
  institution = {W3C},
  url = {https://www.w3.org/TR/css-align-3/},
  urldate = {2025-03-19}
}

@misc{ChengsuXizhiLantingjiXuScroll650,
  title = {Lantingji {{Xu}} Scroll ({{Shenlong}} Version)},
  author = {Feng, Chengsu and Wang, Xizhi},
  year = 650,
  url = {https://commons.wikimedia.org/wiki/File:%E7%A5%9E%E9%BE%8D%E8%98%AD%E4%BA%AD%E5%BA%8F%E5%85%A8.JPG},
  urldate = {2025-03-20},
  copyright = {Public domain},
  howpublished = {Courtesy user Aoser on Wikimedia Commons}
}

@misc{yWorksYFilesHTML2025,
  title = {{{yFiles}} for {{HTML}}},
  author = {{yWorks}},
  year = 2025,
  month = feb,
  url = {https://www.yworks.com/yfiles-overview},
  urldate = {2025-03-23},
  howpublished = {yWorks}
}

@misc{HydeClapham2009,
  title = {Clapham},
  author = {Hyde, Julian},
  year = 2009,
  month = may,
  url = {http://clapham.hydromatic.net/},
  urldate = {2025-03-23}
}

@misc{DoplerSchorgenhumerEBNFVisualizer2005,
  title = {{{EBNF}} Visualizer},
  author = {Dopler, Markus and Sch{\"o}rgenhumer, Stefan},
  year = 2005,
  url = {http://dotnet.jku.at/applications/Visualizer/}
}

@misc{ThiemannEbnf2psPeterSyntax2010,
  title = {Ebnf2ps: {{Peter}}'s Syntax Diagram Drawing Tool},
  author = {Thiemann, Peter},
  year = 2010,
  month = jun,
  url = {http://www2.informatik.uni-freiburg.de/~thiemann/haskell/ebnf2ps/},
  urldate = {2025-03-23}
}

@misc{VoglspergerEBNFRailroadVisualizer2024,
  title = {{{EBNF}} Railroad Visualizer},
  author = {Voglsperger, Alexander},
  year = 2024,
  month = dec,
  url = {https://github.com/MrMinemeet/ebnf_railroad_visualizer},
  urldate = {2025-03-23},
  copyright = {CC-BY-4.0}
}

@misc{AbrechtRailroadCSS2023,
  title = {Railroad {{CSS}}},
  author = {Abrecht, Daniel},
  year = 2023,
  month = feb,
  url = {https://daniel-abrecht.github.io/railroad-css/},
  urldate = {2025-03-23}
}

@misc{PyeattEbnf2tikz2023,
  title = {Ebnf2tikz},
  author = {Pyeatt, Larry D.},
  year = 2023,
  month = dec,
  url = {https://github.com/pyeatt/ebnf2tikz/},
  urldate = {2025-03-23}
}

@misc{LuegRailroad2024,
  title = {Railroad},
  author = {Lueg, Lukas},
  year = 2024,
  month = jul,
  url = {https://github.com/lukaslueg/railroad}
}

@misc{ThibedeauSyntrax2019,
  title = {Syntrax},
  author = {Thibedeau, Kevin},
  year = 2019,
  month = feb,
  url = {https://github.com/kevinpt/syntrax},
  urldate = {2025-03-23},
  copyright = {MIT}
}

@misc{atp-miptJSyntrax2024,
  title = {{{JSyntrax}}},
  author = {{atp-mipt}},
  year = 2024,
  month = may,
  url = {https://github.com/atp-mipt/jsyntrax},
  urldate = {2025-03-23},
  copyright = {MIT}
}

@misc{GroenUtfrailroad2020,
  title = {Utf-Railroad},
  author = {Groen, Matthijs},
  year = 2020,
  month = oct,
  url = {https://github.com/matthijsgroen/utf-railroad},
  urldate = {2025-03-23}
}

@misc{DaenenRailroad2024,
  title = {Railroad},
  author = {Daenen, Quint},
  year = 2024,
  month = apr,
  url = {https://github.com/0x51-dev/railroad},
  urldate = {2025-03-23},
  copyright = {Apache-2.0}
}

@misc{TakadaRailroadDiagrams2025,
  title = {Railroad Diagrams},
  author = {Takada, Yudai},
  year = 2025,
  month = feb,
  url = {https://github.com/ydah/railroad_diagrams},
  urldate = {2025-03-23}
}

@misc{KirpichenkoVirtDiagramsDSL2022,
  title = {{{VirtDiagramsDSL}}},
  author = {Kirpichenko, Stanislav},
  year = 2022,
  url = {https://github.com/Stasychbr/VirtDiagramsDSL},
  urldate = {2025-03-23},
  copyright = {MIT}
}

@misc{LopezPyRailroad2024,
  title = {{{PyRailroad}}},
  author = {Lopez, Rafael},
  year = 2024,
  month = oct,
  url = {https://github.com/Epithumia/pyrailroad},
  urldate = {2025-03-23},
  copyright = {MIT}
}

@misc{DeckersRRDiagram2024,
  title = {{{RRDiagram}}},
  author = {Deckers, Christopher},
  year = 2024,
  month = jan,
  url = {https://github.com/Chrriis/RRDiagram},
  urldate = {2025-03-23},
  copyright = {Apache-2.0}
}

@misc{KraussRRDANTLR42021,
  title = {{{RRD-ANTLR4}}},
  author = {Krauss, Philip},
  year = 2021,
  month = jun,
  url = {https://github.com/flashpixx/RRD-ANTLR4}
}

@misc{HollemansRailroadDiagramsSwift2019,
  title = {Railroad-{{Diagrams-Swift}}},
  author = {Hollemans, Matthijs},
  year = 2019,
  month = may,
  url = {https://github.com/hollance/Railroad-Diagrams-Swift},
  urldate = {2025-03-23}
}

@misc{PikchrcontributorsHowPikchrGenerates2024,
  title = {How {{Pikchr}} Generates the {{SQLite}} Syntax Diagrams},
  author = {{Pikchr contributors}},
  year = 2024,
  month = feb,
  journal = {Pikchr},
  url = {https://pikchr.org/home/doc/trunk/doc/sqlitesyntax.md},
  urldate = {2025-03-24},
  lastaccessed = {2025-03-24}
}

@misc{LischkeVSCodeANTLR42024,
  title = {{{VSCode-ANTLR4}}},
  author = {Lischke, Mike},
  year = 2024,
  month = aug,
  url = {https://github.com/mike-lischke/vscode-antlr4},
  urldate = {2025-03-25}
}

@misc{EberleinAndersonDarwinInformationTyping2018,
  title = {Darwin {{Information Typing Architecture}} ({{DITA}})},
  author = {Eberlein, Kristen James and Anderson, Robert D.},
  year = 2018,
  month = jun,
  publisher = {OASIS},
  url = {http://docs.oasis-open.org/dita/dita/v1.3/dita-v1.3-part3-all-inclusive.html},
  urldate = {2025-03-25}
}

@inproceedings{BonichonWeisFormatUnraveled2017,
  title = {Format Unraveled},
  booktitle = {Journ\'ees {{Francophones}} Des {{Langages Applicatifs}}: {{JFLA}} 2017},
  author = {Bonichon, Richard and Weis, Pierre},
  year = 2017,
  month = jan,
  address = {Gourette, France},
  url = {https://hal.science/hal-01503081},
  urldate = {2025-03-26}
}

@article{Hunter:2007,
  title = {Matplotlib: {{A 2D}} Graphics Environment},
  author = {Hunter, John D.},
  year = 2007,
  journal = {Computing in Science \& Engineering},
  volume = {9},
  number = {3},
  pages = {90--95},
  publisher = {IEEE},
  issn = {1521-9615},
  doi = {10.1109/MCSE.2007.55}
}

@article{HeerBostockDeclarativeLanguageDesign2010,
  title = {Declarative Language Design for Interactive Visualization},
  author = {Heer, Jeffrey and Bostock, Michael},
  year = 2010,
  month = nov,
  journal = {IEEE Transactions on Visualization and Computer Graphics},
  volume = {16},
  number = {6},
  pages = {1149--1156},
  issn = {1941-0506},
  doi = {10.1109/TVCG.2010.144},
  url = {https://idl.uw.edu/papers/protovis-design},
  urldate = {2025-03-26},
  langid = {english}
}

@article{SatyanarayanEtAlVegaLiteGrammarInteractive2017,
  title = {Vega-{{Lite}}: A Grammar of Interactive Graphics},
  shorttitle = {Vega-Lite},
  author = {Satyanarayan, Arvind and Moritz, Dominik and Wongsuphasawat, Kanit and Heer, Jeffrey},
  year = 2017,
  month = jan,
  journal = {IEEE Transactions on Visualization and Computer Graphics},
  volume = {23},
  number = {1},
  pages = {341--350},
  issn = {1941-0506},
  doi = {10.1109/TVCG.2016.2599030},
  url = {https://ieeexplore.ieee.org/document/7539624},
  urldate = {2025-03-26}
}

@book{WickhamGgplot2ElegantGraphics2016,
  title = {{{ggplot2}}: Elegant Graphics for Data Analysis},
  author = {Wickham, Hadley},
  year = 2016,
  publisher = {Springer-Verlag},
  address = {New York, NY, USA},
  url = {https://ggplot2.tidyverse.org},
  isbn = {978-3-319-24277-4}
}

@misc{W3CScalableVectorGraphics2011,
  type = {Recommendation},
  title = {Scalable {{Vector Graphics}} ({{SVG}}) 1.1 ({{Second Edition}})},
  author = {{W3C}},
  year = 2011,
  month = aug,
  url = {https://www.w3.org/TR/SVG11/}
}

@article{ZellerLutkehausDDDFreeGraphical1996,
  title = {{{DDD}}---a Free Graphical Front-End for {{UNIX}} Debuggers},
  author = {Zeller, Andreas and L{\"u}tkehaus, Dorothea},
  year = 1996,
  month = jan,
  journal = {ACM SIGPLAN Notices},
  volume = {31},
  number = {1},
  pages = {22--27},
  issn = {0362-1340, 1558-1160},
  doi = {10.1145/249094.249108},
  urldate = {2025-03-26},
  langid = {english}
}

@techreport{KrishnamurthySQLLanguageReference2025,
  title = {{{SQL Language Reference}} 23ai},
  author = {Krishnamurthy, Usha},
  year = 2025,
  month = feb,
  number = {F47038-21},
  institution = {Oracle Corporation},
  url = {https://docs.oracle.com/en/database/oracle/oracle-database/23/sqlrf/index.html},
  urldate = {2025-04-15},
  langid = {american}
}

@article{PeelsEtAlDocumentArchitectureText1985,
  title = {Document Architecture and Text Formatting},
  author = {Peels, Arno J. H. and Janssen, Norbert J. M. and Nawijn, Wop},
  year = 1985,
  month = oct,
  journal = {ACM Transactions on Information Systems},
  volume = {3},
  number = {4},
  pages = {347--369},
  issn = {1046-8188, 1558-2868},
  doi = {10.1145/4656.4657},
  urldate = {2025-06-20},
  langid = {english}
}

@article{deMoorGibbonsBridgingAlgorithmGap1999,
  title = {Bridging the Algorithm Gap: A Linear-Time Functional Program for Paragraph Formatting},
  shorttitle = {Bridging the Algorithm Gap},
  author = {{de Moor}, Oege and Gibbons, Jeremy},
  year = 1999,
  month = sep,
  journal = {Science of Computer Programming},
  volume = {35},
  number = {1},
  pages = {3--27},
  issn = {0167-6423},
  doi = {10.1016/S0167-6423(99)00005-2},
  urldate = {2025-06-20}
}

@article{BirdTransformationalProgrammingParagraph1986,
  title = {Transformational Programming and the Paragraph Problem},
  author = {Bird, Richard S.},
  year = 1986,
  month = jan,
  journal = {Science of Computer Programming},
  volume = {6},
  pages = {159--189},
  issn = {0167-6423},
  doi = {10.1016/0167-6423(86)90023-7},
  urldate = {2025-06-20}
}

@article{JohariEtAlAutomaticYellowPagesPagination1997,
  title = {Automatic {{Yellow-Pages}} Pagination and Layout},
  author = {Johari, Ramesh and Marks, Joe and Partovi, Ali and Shieber, Stuart},
  year = 1997,
  month = mar,
  journal = {Journal of Heuristics},
  volume = {2},
  number = {4},
  pages = {321--342},
  issn = {1572-9397},
  doi = {10.1007/BF00132503},
  urldate = {2025-06-20},
  langid = {english}
}

@article{JacobsEtAlAdaptiveGridbasedDocument2003,
  title = {Adaptive Grid-Based Document Layout},
  author = {Jacobs, Charles and Li, Wilmot and Schrier, Evan and Bargeron, David and Salesin, David},
  year = 2003,
  month = jul,
  journal = {ACM Transactions on Graphics},
  volume = {22},
  number = {3},
  pages = {838--847},
  issn = {0730-0301},
  doi = {10.1145/882262.882353},
  urldate = {2025-06-20}
}

@article{CiancariniEtAlHighqualityPaginationPublishing2011,
  title = {High-Quality Pagination for Publishing},
  author = {Ciancarini, Paolo and Iorio, Angelo Di and Furini, Luca and Vitali, Fabio},
  year = 2011,
  month = jul,
  journal = {Software: Practice and Experience},
  volume = {42},
  number = {6},
  pages = {733--751},
  publisher = {John Wiley \& Sons, Ltd},
  issn = {0038-0644},
  doi = {10.1002/spe.1096},
  urldate = {2025-06-20}
}

@inproceedings{PollockEtAlBluefishComposingDiagrams2024,
  title = {Bluefish: {{Composing Diagrams}} with {{Declarative Relations}}},
  shorttitle = {Bluefish},
  booktitle = {Proceedings of the 37th {{Annual ACM Symposium}} on {{User Interface Software}} and {{Technology}}},
  author = {Pollock, Josh and Mei, Catherine and Huang, Grace and Evans, Elliot and Jackson, Daniel and Satyanarayan, Arvind},
  year = 2024,
  month = oct,
  pages = {1--21},
  publisher = {ACM},
  address = {Pittsburgh, PA, USA},
  doi = {10.1145/3654777.3676465},
  url = {https://dl.acm.org/doi/10.1145/3654777.3676465},
  urldate = {2025-07-30},
  isbn = {979-8-4007-0628-8},
  langid = {english}
}

@inproceedings{RueggvonHanxledenWrappingLayeredGraphs2018,
  title = {Wrapping Layered Graphs},
  booktitle = {Diagrammatic {{Representation}} and {{Inference}}},
  author = {R{\"u}egg, Ulf and {von Hanxleden}, Reinhard},
  editor = {Chapman, Peter and Stapleton, Gem and Moktefi, Amirouche and {Perez-Kriz}, Sarah and Bellucci, Francesco},
  year = 2018,
  month = jun,
  series = {Lecture {{Notes}} in {{Computer Science}}},
  volume = {10871},
  pages = {743--747},
  publisher = {Springer International Publishing},
  address = {Cham},
  doi = {10.1007/978-3-319-91376-6_72},
  urldate = {2025-07-31},
  isbn = {978-3-319-91376-6},
  langid = {english}
}

@inproceedings{BannisterEtAlConfluentOrthogonalDrawings2015,
  title = {Confluent {{Orthogonal Drawings}} of {{Syntax Diagrams}}},
  booktitle = {Graph {{Drawing}} and {{Network Visualization}}},
  author = {Bannister, Michael J. and Brown, David A. and Eppstein, David},
  editor = {Di Giacomo, Emilio and Lubiw, Anna},
  year = 2015,
  series = {Lecture {{Notes}} in {{Computer Science}}},
  volume = {9411},
  pages = {260--271},
  publisher = {Springer International Publishing},
  address = {Cham},
  doi = {10.1007/978-3-319-27261-0_22},
  isbn = {978-3-319-27261-0},
  langid = {english}
}

@article{BellGilbertLearningRecursionSyntax1974,
  title = {Learning Recursion with Syntax Diagrams},
  author = {Bell, Stoughton and Gilbert, Edgar J.},
  year = 1974,
  month = sep,
  journal = {SIGCSE Bulletin},
  volume = {6},
  number = {3},
  pages = {44--45},
  issn = {0097-8418},
  doi = {10.1145/988881.988890},
  urldate = {2026-01-08}
}

@article{BrazVisualSyntaxDiagrams1990,
  title = {Visual Syntax Diagrams for Programming Language Statements},
  author = {Braz, Lisa M.},
  year = 1990,
  month = sep,
  journal = {SIGDOC Asterisk Journal of Computer Documentation},
  volume = {14},
  number = {4},
  pages = {23--27},
  issn = {0731-1001},
  doi = {10.1145/97435.97987},
  urldate = {2026-01-08}
}

@article{HerriotStructuredSyntaxDiagrams1976,
  title = {Structured Syntax Diagrams},
  author = {Herriot, Robert G.},
  year = 1976,
  month = jan,
  journal = {Computer Languages},
  volume = {2},
  number = {1},
  pages = {9--19},
  issn = {0096-0551},
  doi = {10.1016/0096-0551(76)90009-6},
  urldate = {2026-01-08}
}

@article{ShiEtAl15DEgocentricDynamic2015,
  title = {1.{{5D}} Egocentric Dynamic Network Visualization},
  author = {Shi, Lei and Wang, Chen and Wen, Zhen and Qu, Huamin and Lin, Chuang and Liao, Qi},
  year = 2015,
  month = may,
  journal = {IEEE Transactions on Visualization and Computer Graphics},
  volume = {21},
  number = {5},
  pages = {624--637},
  issn = {1941-0506},
  doi = {10.1109/TVCG.2014.2383380},
  urldate = {2026-01-12}
}

@inproceedings{BurchEtAlIndentedPixelTree2010,
  title = {Indented Pixel Tree Plots},
  booktitle = {Advances in {{Visual Computing}}},
  author = {Burch, Michael and Raschke, Michael and Weiskopf, Daniel},
  editor = {Bebis, George and Boyle, Richard and Parvin, Bahram and Koracin, Darko and Chung, Ronald and Hammoud, Riad and Hussain, Muhammad and {Kar-Han}, Tan and Crawfis, Roger and Thalmann, Daniel and Kao, David and Avila, Lisa},
  year = 2010,
  series = {Lecture {{Notes}} in {{Computer Science}}},
  volume = {6453},
  pages = {338--349},
  publisher = {Springer},
  address = {Berlin, Heidelberg},
  doi = {10.1007/978-3-642-17289-2_33},
  isbn = {978-3-642-17289-2},
  langid = {english}
}

@article{DyckhoffTypologyCuttingPacking1990,
  title = {A Typology of Cutting and Packing Problems},
  author = {Dyckhoff, Harald},
  year = 1990,
  month = jan,
  journal = {European Journal of Operational Research},
  series = {Cutting and {{Packing}}},
  volume = {44},
  number = {2},
  pages = {145--159},
  issn = {0377-2217},
  doi = {10.1016/0377-2217(90)90350-K},
  urldate = {2026-01-13}
}

@phdthesis{DevkotaVisualizingControlFlow2021,
  title = {Visualizing Control Flow Graphs},
  author = {Devkota, Sabin},
  year = 2021,
  address = {Tucson, Arizona, USA},
  url = {https://repository.arizona.edu/handle/10150/661593},
  urldate = {2026-01-13},
  copyright = {http://rightsstatements.org/vocab/InC/1.0/},
  langid = {english},
  school = {The University of Arizona}
}

@article{SnyderSpecifyingTextualGraphical1991,
  title = {Specifying Textual to Graphical Conversion},
  author = {Snyder, Robin M.},
  year = 1991,
  month = sep,
  journal = {Journal of Systems and Software},
  volume = {16},
  number = {1},
  pages = {17--28},
  issn = {0164-1212},
  doi = {10.1016/0164-1212(91)90028-5},
  urldate = {2026-01-14}
}

@misc{PieperEtAlSyntaxdi2020,
  title = {{{syntaxdi}}},
  author = {Pieper, Johannes and Hilbig, Andr{\'e} and Kuhaupt, Johannes and Spittank, Daniel and Humbert, Ludger and Salamon, Adrian},
  year = 2020,
  month = oct,
  url = {https://www.ctan.org/pkg/syntaxdi}
}

@misc{WoodingSyntaxmdw1996,
  title = {{{syntax-mdw}}},
  author = {Wooding, Mark},
  year = 1996,
  month = may,
  url = {https://ctan.org/pkg/syntax-mdw}
}

@misc{KlockerSyngen1996,
  title = {{{syngen}}},
  author = {Kl{\"o}cker, Jens},
  year = 1996,
  month = nov,
  url = {https://www.ctan.org/pkg/syngen},
  urldate = {2026-02-02}
}

@misc{BarthelmannRooijakkersRail1998,
  title = {{{rail}}},
  author = {Barthelmann, Klaus Georg and Rooijakkers, Luc},
  year = 1998,
  month = may,
  url = {https://www.ctan.org/pkg/rail},
  urldate = {2026-02-03}
}

@misc{DundalekGrammKit2023,
  title = {{{GrammKit}}},
  author = {Dundalek, Jakub},
  year = 2023,
  month = apr,
  url = {https://dundalek.com/GrammKit/},
  urldate = {2026-02-11}
}

@misc{JacquesDrawGrammar2018,
  title = {{{DrawGrammar}}},
  author = {Jacques, Vincent},
  year = 2018,
  month = nov,
  url = {https://jacquev6.github.io/DrawGrammar/},
  urldate = {2026-02-11}
}

@misc{FlavelKatesGrammarTool2025,
  title = {{{KGT}}: {{Kate}}'s {{Grammar Tool}}},
  author = {Flavel, Katherine},
  year = 2025,
  month = apr,
  url = {https://github.com/katef/kgt},
  urldate = {2026-02-11}
}

@misc{chevrotaincontributorsChevrotain2026,
  title = {{{chevrotain}}},
  author = {{chevrotain contributors}},
  year = 2026,
  month = mar,
  url = {https://github.com/Chevrotain/chevrotain},
  urldate = {2026-04-09}
}

@inproceedings{ChengParreauxSimpleRecipeWriting2026,
  title = {A Simple Recipe for Writing Decent Recursive Descent Parsers (Pearl)},
  booktitle = {40th {{European Conference}} on {{Object-Oriented Programming}} ({{ECOOP}} 2026)},
  author = {Cheng, Luyu and Parreaux, Lionel},
  year = 2026,
  series = {Leibniz International Proceedings in Informatics ({{LIPIcs}})},
  volume = {372},
  publisher = {Schloss Dagstuhl -- Leibniz-Zentrum f\"ur Informatik},
  address = {Dagstuhl, Germany}
}

@misc{KenningaOptimalLineBreak2003b,
  title = {Optimal Line Break Determination},
  author = {Kenninga, Eric A.},
  year = 2003,
  month = jan,
  howpublished = {U.S. Patent 6,510,441 B1, assigned to Adobe Systems, Inc.}
}

@misc{BrissonMethodsSystemConverting1997a,
  title = {Methods and System for Converting a Text-Based Grammar to a Compressed Syntax Diagram},
  author = {Brisson, James Paul},
  year = 1997,
  month = oct,
  urldate = {2026-01-08},
  howpublished = {U.S. Patent 5,678,052 A, assigned to IBM Corporation}
}

@misc{dagstuhl-artifact-25789,
  title = {{{LibRRD}}},
  author = {Chiplunkar, Shardul and Pit-Claudel, Clément},
  howpublished = {Software, version ecoop2026-artifact (visited on 2026-05-05)},
  url = {https://github.com/epfl-systemf/librrd/releases/tag/ecoop2026-artifact},
  doi = {10.4230/artifacts.25789}
}

\end{document}